\documentclass[12pt,a4paper,reqno,final]{amsart}
\usepackage{amsmath,amsfonts,amssymb,amsthm}
\usepackage{eucal,bbold}
\usepackage[notref,notcite,color]{showkeys}
\usepackage{array}
\usepackage[T1]{fontenc}
\usepackage[latin1]{inputenc}
\usepackage{jmarvosym}
\usepackage{jbox}
\usepackage{subfigure}
\usepackage{graphicx}
\DeclareMathOperator{\E}{E}
\DeclareMathOperator{\GL}{GL}
\DeclareMathOperator{\SL}{SL}
\DeclareMathOperator{\SO}{SO}
\DeclareMathOperator{\SU}{SU}
\DeclareMathOperator{\Spin}{Spin}
\DeclareMathOperator{\Tr}{Tr}
\DeclareMathOperator{\STr}{STr}
\DeclareMathOperator{\U}{U}
\DeclareMathOperator{\ad}{ad}
\DeclareMathOperator{\dalem}{\square}
\DeclareMathOperator{\dslash}{\partial\!\!\!/}
\DeclareMathOperator{\fsl}{\mathfrak{sl}}
\DeclareMathOperator{\sign}{sign}

\DeclareMathOperator{\su}{\mathfrak{su}}
\DeclareMathSymbol{\aok}{\mathord}{AMSa}{"58}%
\DeclareMathOperator{\ads}{adS}
\newcommand{\exref}[1]{Exercise \ref{#1}}
\newcommand{\probref}[1]{Problem \ref{#1}}

\newcommand{\Dslash}{\eD\!\!\!\!/\,}
\newcommand{\bra}[1]{\mathord{\langle#1|}}
\newcommand{\ket}[1]{\mathord{|#1\rangle}}
\newcommand{\vac}{\ket{\text{vac}}}
\newcommand{\cav}{\bra{\text{vac}}}
\newcommand{\half}{{\tfrac{1}{2}}}
\renewcommand{\d}{\partial}
\newcommand{\be}{{\boldsymbol e}}
\newcommand{\bPhi}{{\boldsymbol\Phi}}
\newcommand{\bphi}{{\boldsymbol\phi}}
\newcommand{\bz}{{\boldsymbol z}}
\newcommand{\eD}{{\mathcal D}}
\newcommand{\eF}{{\mathcal F}}
\newcommand{\eG}{{\mathcal G}}
\newcommand{\eH}{{\mathcal H}}
\newcommand{\eL}{{\mathcal L}}
\newcommand{\eM}{{\mathcal M}}
\newcommand{\eV}{{\mathcal V}}
\newcommand{\fS}{{\mathfrak S}}
\newcommand{\fg}{{\mathfrak g}}
\newcommand{\sA}{{\mathsf A}}

\newcommand{\sD}{{\mathsf D}}
\newcommand{\sK}{{\mathsf K}}
\newcommand{\sM}{{\mathsf M}}
\newcommand{\sP}{{\mathsf P}}
\newcommand{\sQ}{{\mathsf Q}}
\newcommand{\sR}{{\mathsf R}}
\newcommand{\sS}{{\mathsf S}}
\newcommand{\sW}{{\mathsf W}}
\newcommand{\sX}{{\mathsf X}}
\newcommand{\sY}{{\mathsf Y}}
\newcommand{\sZ}{{\mathsf Z}}
\newcommand{\sq}{{\mathsf q}}
\newcommand{\sz}{{\mathsf z}}
\newcommand{\1}{{\mathbb 1}}
\newcommand{\CC}{{\mathbb C}}
\newcommand{\RR}{{\mathbb R}}
\newcommand{\VV}{{\mathbb V}}
\newcommand{\WW}{{\mathbb W}}
\newcommand{\ZZ}{{\mathbb Z}}
\theoremstyle{plain}  \newtheorem{ex}{Exercise}[section]
\theoremstyle{definition} \newtheorem{tut}{Problem}
\newenvironment{exercise}%
{\begin{flushright}{\Huge\Writinghand~}%
    \begin{minipage}[t]{0.92\textwidth}\small\begin{ex}}
 {\end{ex}\end{minipage}\end{flushright}}
\newenvironment{scholium}%
{\begin{flushright}{\Huge\Pointinghand~}%
    \begin{minipage}[t]{0.92\textwidth}\scriptsize}
 {\end{minipage}\end{flushright}}
\newenvironment{caveat}%
{\begin{flushright}{\Huge\Stopsign~}%
    \begin{minipage}[t]{0.92\textwidth}\small}
 {\end{minipage}\end{flushright}}
\newenvironment{amusing}%
{\begin{flushright}{\Huge\Smiley~}%
    \begin{minipage}[t]{0.92\textwidth}\small}
 {\end{minipage}\end{flushright}}
\begin{document}
%
%
\title{BUSSTEPP Lectures on Supersymmetry}
\author[JM Figueroa-O'Farrill]{José Miguel Figueroa-O'Farrill}
\thanks{{\Large\Letter} $\left<\texttt{jmf@maths.ed.ac.uk}\right>$}
\begin{abstract}
  This is the written version of the supersymmetry lectures delivered 
  at the 30th and 31st British Universities Summer Schools in
  Theoretical Elementary Particle Physics (BUSSTEPP) held in Oxford in
  September 2000 and in Manchester in August-September 2001.\\
  \begin{center}
    (Version of 21 September 2001)
  \end{center}
\end{abstract}
\maketitle
%
%
\tableofcontents
\section*{Introduction}

The aim of these lectures is to introduce supersymmetry to graduate
students in Physics having completed the first year of their PhD
studies in a British university.  No previous exposure to
supersymmetry is expected, but familiarity with topics normally
covered in an introductory course in relativistic field theory will be
assumed.  These include, but are not restricted to, the following:
lagrangian formulation of relativistic field theories, Lie symmetries,
representations of the Poincaré group, gauge theories and spontaneous
symmetry breaking.  I have adopted a conservative approach to the
subject, discussing exclusively four-dimensional rigid $N{=}1$
supersymmetry.

The lecture notes are accompanied by a series of Exercises and
Problems.  Exercises are meant to fill in the details of the lectures.
They are relatively easy and require little else than following the
logic flow of the lectures.  Problems are more involved (although none
are really difficult) and make good topics for tutorials.

The written version of the lectures contains more material than can be
comfortably covered during the School and certainly more exercises and
problems than can possibly be completed by the student during this
time.  It is my hope, however, that the interested student can
continue working on the problems and exercises after the School has
ended and that the written version of these notes can be of help in
this task.

\begin{caveat}
  Throughout the written version of the lectures you will find
  paragraphs like this one with one of the following signs:
  \begin{center}
    \begin{tabular}{cccc}
      {\Huge\Stopsign} & {\Huge\Writinghand} & {\Huge\Pointinghand} &
      {\Huge\Smiley}
    \end{tabular}
  \end{center}
  indicating, respectively, caveats, exercises, scholia and the (very)
  occasional amusing comment.
\end{caveat}

These notes are organised as follows.

In Lecture~\ref{lec:WZmodel} we will introduce the simplest field
theoretical model exhibiting (linearly realised) supersymmetry: the
Wess--Zumino model.  It will serve to illustrate many of the
properties found in more phenomenologically realistic models.  We will
prove that the Wess--Zumino model is invariant under a ``super''
extension of the Poincaré algebra, known as the $N{=}1$ Poincaré
superalgebra.  The tutorial problem for this lecture investigates the
superconformal invariance of the massless Wess--Zumino model.

In Lecture~\ref{lec:SYM} we will study another simple four-dimensional 
supersymmetric field theory: supersymmetric Yang--Mills.  This is
obtained by coupling pure Yang--Mills theory to adjoint fermions.  We
will show that the action is invariant under the Poincaré
superalgebra, and that the algebra closes on-shell and up to gauge
transformations.  This theory is also classically superconformal
invariant, and this is the topic of the tutorial problem for this
lecture.

In Lecture~\ref{lec:Reps} we will study the representations of the
$N{=}1$ Poincaré superalgebra.  We will see that representations of
this superalgebra consist of mass-degenerate multiplets of irreducible
representations of the Poincaré algebra.  We will see that unitary
representations of the Poincaré superalgebra have non-negative energy
and that they consist of an equal number of bosonic and fermionic
fields.  We will discuss the most important multiplets: the chiral
multiplet, the gauge multiplet and the supergravity multiplet.
Constructing supersymmetric field theories can be understood as
finding field-theoretical realisations of these multiplets.  The
tutorial problem introduces the extended Poincaré superalgebra, the
notion of central charges and the ``BPS'' bound on the mass of any
state in a unitary representation.

In Lecture~\ref{lec:superspace} we will introduce superspace and
superfields.  Superspace does for the Poincaré superalgebra what
Minkowski space does for the Poincaré algebra; namely it provides a
natural arena in which to discuss the representations and in which to
build invariant actions.  We will learn how to construct invariant
actions and we will recover the Wess--Zumino model as the simplest
possible action built out of a chiral superfield.  The tutorial
problem discusses more general models built out of chiral superfields:
we will see that the most general renormalisable model consists of $N$
chiral multiplets with a cubic superpotential and the most general
model consists of a supersymmetric sigma model on a Kähler manifold
and a holomorphic function on the manifold (the superpotential).

In Lecture~\ref{lec:SYM2} we continue with our treatment of
superspace, by studying supersymmetric gauge theories in superspace.
We will see that supersymmetric Yang--Mills is the natural theory
associated to a vector superfield.  We start by discussing the abelian 
theory, which is easier to motivate and then generalise to the
nonabelian case after a brief discussion of the coupling of gauge
fields to matter (in the form of chiral superfields).  This is all
that is needed to construct the most general renormalisable
supersymmetric lagrangian in four dimensions.  In the tutorial problem 
we introduce the Kähler quotient in the simple context of the
$\CC P^N$ model.

In Lecture~\ref{lec:SB} we will discuss the spontaneous breaking of
supersymmetry. We will discuss the relation between spontaneous
supersymmetry breaking and the vacuum energy and the vacuum
expectation values of auxiliary fields.  We discuss the O'Raifeartaigh
model, Fayet--Iliopoulos terms and the Witten index.  In the tutorial
problem we discuss an example of Higgs mechanism in an $\SU(5)$
supersymmetric gauge theory.

Lastly, there are two appendices.  Appendix~\ref{app:conventions}
includes the basic mathematical definitions needed in the lectures.
More importantly, it also includes our conventions.  It should
probably be skimmed first for notation and then revisited as needed.
It is aimed to be self-contained.  Appendix~\ref{app:formulas} is a
``reference card'' containing formulas which I have found very useful
in calculations.  I hope you do too.

Enjoy!

\section*{Notes for lecturers}

The format of the School allocated six one-hour lectures to this
topic.  With this time constraint I was forced to streamline the
presentation.  This meant among other things that many of the
Exercises were indeed left as exercises; although I tried to do enough
to illustrate the different computational techniques.

The six lectures in the School did not actually correspond to the six
lectures in the written version of the notes.  (In fact, since the
conventions must be introduced along the way, the written version
really has seven lectures.)  The first lecture was basically
Lecture~\ref{lec:WZmodel}, only that there was only enough time to do
the kinetic term in detail.  The second lecture did correspond to
Lecture~\ref{lec:SYM} with some additional highlights from
Lecture~\ref{lec:Reps}: the notion of supermultiplet, the balance
between bosonic and fermionic degrees of freedom, and the positivity
of the energy in a unitary representation.  This allowed me to devote
the third lecture to introducing superspace, roughly speaking the
first three sections in Lecture~\ref{lec:superspace}, which was then
completed in the fourth lecture.  The fifth lecture covered the
abelian part of Lecture~\ref{lec:SYM2} and all too briefly mentioned
the extension to nonabelian gauge theories.  The sixth and final
lecture was devoted to Lecture~\ref{lec:SB}.

It may seem strange to skip a detailed analysis of the representation
theory of the Poincaré superalgebra, but this is in fact not strictly
speaking necessary in the logical flow of the lectures, which are
aimed at supersymmetric field theory model building.  Of course, they
are an essential part of the topic itself, and this is why they have
been kept in the written version.


\section*{Acknowledgements}

Supersymmetry is a vast subject and trying to do justice to it in just
a few lectures is a daunting task and one that I am not sure I have
accomplished to any degree.  In contrast, I had the good fortune to
learn supersymmetry from Peter van Nieuwenhuizen and at a much more
leisurely pace.  His semester-long course on supersymmetry and
supergravity was memorable.  At the time I appreciated the course from
the perspective of the student.  Now, from that of the lecturer, I
appreciate the effort that went into it.  I can think of no better
opportunity than this one to thank him again for having given it.

The first version of these lectures were started while on a visit to
CERN and finished, shortly before the School, while on a visit to the
Spinoza Institute.  I would like to extend my gratitude to both
institutions, and Bernard de Wit in particular, for their hospitality.

I would like to thank Sonia Stanciu for finding several critical minus
signs that I had misplaced; although I remain solely responsible for
any which have yet to find their right place.

On a personal note, I would like to extend a warm ``Mul{\c t}umesc
frumos'' to Veronica R{\u a}dulescu for providing a very pleasant
atmosphere and for so gracefully accommodating my less-than-social
schedule for two weeks during the writing of the first version of
these lectures.

Last, but by no means least, I would also like to thank John Wheater
and Jeff Forshaw for organising the Summer Schools, and all the other
participants (students, tutors and lecturers alike) for their
questions and comments, some of which have been incorporated in the
updated version of these notes.  Thank you all!


\section*{References}

The amount of literature in this topic can be overwhelming to the
beginner.  Luckily there are not that many books to choose from and I
have found the following references to be useful in the preparation of
these notes:
\begin{itemize}
\item \textsl{Supersymmetry and Supergravity}\\
   J~Wess and J~Bagger\\
  (Princeton University Press, 1983) (Second Edition, 1992)
\item \textsl{Supersymmetric gauge field theory and string theory}\\
  D~Bailin and A~Love\\
  (IoP Publishing, 1994)
\item \textsl{Superspace}\\
  SJ~Gates, Jr., MT~Grisaru, M~Ro\v{c}ek and W~Siegel\\
  (Benjamin/Cummings, 1983)
\item \textsl{Fields}\\
  W~Siegel\\
  (\texttt{arXiv:hep-th/9912205})
\item \textsl{Dynamical breaking of supersymmetry}\\
  E~Witten\\
  Nucl.Phys. \textbf{B188} (1981) 513--554.
\end{itemize}

Of course, they use different conventions to the ones in these notes
and thus care must be exercised when importing/exporting any formulas.
As mentioned in the Introduction, my aim has been to make these notes
self-contained, at least as far as calculations are concerned.  Please
let me know if I have not succeeded in this endeavour so that I can
correct this in future versions.  Similarly, I would appreciate any
comments or suggestions, as well as pointers to errors of any kind:
computational, conceptual, pedagogical,...  (My email is written in
the cover page of these notes.)

The latest version of these notes can always be found at the following
URL:
\begin{center}
  \texttt{http://www.maths.ed.ac.uk/\~{}jmf/BUSSTEPP.html}
\end{center}


\section{The Wess--Zumino model}
\label{lec:WZmodel}

We start by introducing supersymmetry in the context of a simple
four-dimensional field theory: the Wess--Zumino model.  This is
arguably the simplest supersymmetric field theory in four dimensions.
We start by discussing the free massless Wess--Zumino model and then
we make the model more interesting by adding masses and interactions.

\subsection{The free massless Wess--Zumino model}

The field content of the Wess--Zumino model consists of a real scalar
field $S$, a real pseudoscalar field $P$ and a real (i.e., Majorana)
spinor $\psi$.  (See the Appendix for our conventions.)  Of course,
$\psi$ is anticommuting.  The (free, massless) lagrangian for these
fields is:
\begin{equation}
  \label{eq:WZkin}
  \eL_{\text{kin}} = -\half \left(\d S\right)^2  - \half \left(\d
    P\right)^2 - \half  \bar\psi \dslash \psi~,
\end{equation}
where $\dslash = \d_\mu \gamma^\mu$ and $\bar\psi = \psi^t C =
\psi^\dagger i \gamma^0$.  The signs have been chosen in order to make
the hamiltonian positive-semidefinite in the chosen (mostly plus)
metric.  The action is defined as usual by
\begin{equation}
  I_{\text{kin}} = \int d^4 x \, \eL_{\text{kin}}
\end{equation}
To make the action have the proper dimension, the bosonic fields $S$
and $P$ must have dimension $1$ and the fermionic field $\psi$ must
have dimension $\frac32$, in units where $\d_\mu$ has dimension $1$.

\begin{caveat}
  You may wonder why it is that $P$ is taken to be a pseudoscalar,
  since the above action is clearly symmetric in $S$ and $P$.  The
  pseudoscalar nature of $P$ will manifest itself shortly when we
  discuss supersymmetry, and at the end of the lecture when we
  introduce interactions: the Yukawa coupling between $P$ and $\psi$
  will have a $\gamma_5$.  Since changing the orientation changes the
  sign in $\gamma_5$, the action would not be invariant unless $P$
  also changed sign.  This means that it is a pseudoscalar.
\end{caveat}

\begin{exercise}
  Check that the action $I_{\text{kin}}$ is real and that the
  equations of motion are
  \begin{equation}
    \dalem S = \dalem P = \dslash \psi = 0~,
  \end{equation}
  where $\dalem = \d_\mu \d^\mu$.
\end{exercise}

We now discuss the symmetries of the action $I_{\text{kin}}$.  It will
turn out that the action is left invariant by a ``super'' extension of
the Poincaré algebra, so we briefly remind ourselves of the Poincaré
invariance of the above action.  The Poincaré algebra is the Lie
algebra of the group of isometries of Minkowski space.  As such it is
isomorphic to the semidirect product of the algebra of Lorentz
transformations and the algebra of translations.  Let
$\sM_{\mu\nu}=-\sM_{\nu\mu}$ be a basis for the (six-dimensional)
Lorentz algebra and let $\sP_\mu$ be a basis for the
(four-dimensional) translation algebra.  The form of the algebra in
this basis is recalled in \eqref{eq:poincare} in the Appendix.

Let $\tau^\mu$ and $\lambda^{\mu\nu} = -\lambda^{\nu\mu}$ be constant
parameters.  Then for any field $\varphi = S$, $P$ or $\psi$ we define 
infinitesimal Poincaré transformations by
\begin{equation}
  \begin{split}
    \delta_{\tau} \varphi &= \tau^\mu \sP_\mu \cdot \varphi\\
    \delta_{\lambda} \varphi &= \half \lambda^{\mu\nu} \sM_{\mu\nu}
    \cdot \varphi~,
\end{split}
\end{equation}
where
\begin{equation}
  \label{eq:PoincareTransf}
  \begin{aligned}
    \sP_\mu \cdot S &= - \d_\mu S \\
    \sP_\mu \cdot P &= - \d_\mu P\\
    \sP_\mu \cdot \psi &= - \d_\mu \psi
\end{aligned}
\qquad
\begin{aligned}
  \sM_{\mu\nu} \cdot S &= - (x_\mu \d_\nu - x_\nu \d_\mu) S\\
  \sM_{\mu\nu} \cdot P &= - (x_\mu \d_\nu - x_\nu \d_\mu) P\\
  \sM_{\mu\nu} \cdot \psi &= - (x_\mu \d_\nu - x_\nu \d_\mu) \psi -
  \Sigma_{\mu\nu}\psi~,
\end{aligned}
\end{equation}
and $\Sigma_{\mu\nu} = \half \gamma_{\mu\nu}$.

\begin{scholium}
  The reason for the minus signs is that the action on functions is
  inverse to that on points.  More precisely, let $G$ be a group of
  transformations on a space $X$: every group element $g\in G$ sends a
  point $x\in X$ to another point $g \cdot x \in X$.  Now suppose that
  $f : X \to \RR$ is a function.  How does the group act on it?  The
  physically meaningful quantity is the value $f(x)$ that the function
  takes on a point; hence this is what should be invariant.  In other
  words, the transformed function on the transformed point $(g \cdot
  f)(g \cdot x)$ should be the same as the original function on the
  original point $f(x)$.  This means that $(g \cdot f)(x) =
  f(g^{-1}\cdot x)$ for all $x\in X$.
  
  As an illustration, let's apply this to the translations on
  Minkowski space, sending $x^\mu$ to $x^\mu + \tau^\mu$.  Suppose
  $\varphi$ is a scalar field.  Then the action of the translations is
  $\varphi \mapsto \varphi'$ where $\varphi'(x^\mu) = \varphi(x^\mu -
  \tau^\mu)$.  For infinitesimal $\tau^\mu$ we have $\varphi'(x^\mu) =
  \varphi(x^\mu) - \tau^\mu \d_\mu \varphi(x^\mu)$, or equivalently
  \begin{equation*}
    \tau^\mu \sP_\mu \cdot \varphi = \varphi' - \varphi = - \tau^\mu
    \d_\mu \varphi~,
  \end{equation*}
  which agrees with the above definition.
\end{scholium}

\begin{exercise}
  Show that the above operators satisfy the Poincaré algebra
  \eqref{eq:poincare} and show that
  \begin{equation}
    \begin{aligned}
      \delta_\tau \eL_{\text{kin}} &= \d_\mu \left( - \tau^\mu
      \eL_{\text{kin}}\right)\\
      \delta_\lambda \eL_{\text{kin}} &= \d_\mu \left(\lambda^{\mu\nu}
      x_\nu \eL_{\text{kin}} \right)~.
\end{aligned}
\end{equation}
  Conclude that the action $I_{\text{kin}}$ is Poincaré invariant.
\end{exercise}

\begin{caveat}
  I should issue a word of warning when computing the algebra of
  operators such as $\sP_\mu$ and $\sM_{\mu\nu}$.  These operators are
  defined \emph{only on fields}, where by ``fields'' we mean products
  of $S$, $P$ and $\psi$.  For instance, $\sP_\mu \cdot (x^\nu S) =
  x^\nu \sP_\mu \cdot S$: it does \emph{not} act on the coordinate
  $x^\nu$.  Similarly, $\sM_{\mu\nu} \cdot \d_\rho S = \d_\rho
  (\sM_{\mu\nu} \cdot S)$, and of course the $\d_\rho$ \emph{does} act
  on the coordinates which appear in $\sM_{\mu\nu} \cdot S$.
\end{caveat}

\subsection{Invariance under supersymmetry}

More interestingly the action is invariant under the following
``supersymmetry'' transformations:
\begin{equation}
  \label{eq:WZkinsusy}
  \begin{aligned}
    \delta_\varepsilon S &=  \bar \varepsilon \psi\\
    \delta_\varepsilon P &=  \bar \varepsilon \gamma_5 \psi\\
    \delta_\varepsilon \psi &= \dslash (S + P\gamma_5)
    \varepsilon~,
\end{aligned}
\end{equation}
where $\varepsilon$ is a constant Majorana spinor.  Notice that
because transformations of any kind should not change the Bose--Fermi
parity of a field, we are forced to take $\varepsilon$ anticommuting,
just like $\psi$.  Notice also that for the above transformations to
preserve the dimension of the fields, $\varepsilon$ must have
dimension $-\half$.  Finally notice that they preserve the reality
properties of the fields.

\begin{exercise}
  Show that under the above transformations the free lagrangian
  changes by a total derivative:
  \begin{equation}
    \delta_\varepsilon \eL_{\text{kin}} = \d_\mu \left(- \half
      \bar\varepsilon \gamma^\mu \dslash\left(S +
      P\gamma_5\right)\psi\right)~,
  \end{equation}
  and conclude that the action is invariant.
\end{exercise}

The supersymmetry transformations are generated by a spinorial
\emph{supercharge} $\sQ$ of dimension $\half$ such that for all fields
$\varphi$,
\begin{equation}
  \delta_\varepsilon \varphi = \bar\varepsilon \sQ \cdot \varphi~.
\end{equation}
The action of $\sQ$ on the bosonic fields is clear:
\begin{equation}
  \sQ \cdot S =  \psi \qquad\text{and}\qquad \sQ \cdot P  =  \gamma_5\,
  \psi~.
\end{equation}
To work out the action of $\sQ$ on $\psi$ it is convenient to
introduce indices.

First of all notice that $\bar\varepsilon \sQ = \varepsilon^b C_{ba}
\sQ^a = \varepsilon_a \sQ^a = - \varepsilon^a \sQ_a$, whereas
\begin{equation}
  \delta_\varepsilon \psi^a = \left( (\gamma^\mu)^a{}_b \d_\mu S +
  (\gamma^\mu\gamma_5)^a{}_b \d_\mu P \right) \varepsilon^b~.
\end{equation}
Equating the two, and taking into account that $\psi_a = \psi^b
C_{ba}$ and similarly for $\sQ$, one finds that
\begin{equation}
  \sQ_a \cdot \psi_b = -\left(\gamma^\mu\right)_{ab} \d_\mu S +
  \left(\gamma^\mu\gamma_5\right)_{ab} \d_\mu P~,
\end{equation}
where we have lowered the indices of $\gamma^\mu$ and
$\gamma^\mu\gamma_5$ with $C$ and used respectively the symmetry and
antisymmetry of the resulting forms.

\subsection{On-shell closure of the algebra}

We now check the closure of the algebra generated by $\sP_\mu$,
$\sM_{\mu\nu}$ and $\sQ_a$.  We have already seen that $\sP_\mu$ and
$\sM_{\mu\nu}$ obey the Poincaré algebra, so it remains to check the
brackets involving $\sQ_a$.  The supercharges $\sQ_a$ are spinorial
and hence transform nontrivially under Lorentz transformations.  We
therefore expect their bracket with the Lorentz generators
$\sM_{\mu\nu}$ to reflect this.  Also the dimension of $\sQ_a$ is
$\half$ and the dimension of the translation generators $\sP_\mu$ is
$1$, whence their bracket would have dimension $\frac32$.  Since there
is no generator with the required dimension, we expect that their
bracket should vanish.  Indeed, we have the following.

\begin{exercise}
  Show that
\begin{equation}
  \begin{aligned}
    \left[\sP_\mu, \sQ_a\right] \cdot \varphi &= 0\\
    \left[\sM_{\mu\nu}, \sQ_a\right] \cdot \varphi &= -
    \left(\Sigma_{\mu\nu}\right)_a{}^b \sQ_b \cdot \varphi~,
\end{aligned}
\end{equation}
where $\varphi$ is any of the fields $S$, $P$ or $\psi$.
\end{exercise}

We now compute the bracket of two supercharges.  The first thing we
notice is that, because $\sQ_a$ anticommutes with the parameter
$\varepsilon$, it is the \emph{anticommutator} of the generators which
appears in the commutator of transformations.  More precisely,
\begin{equation}
  \begin{split}
    \left[ \delta_{\varepsilon_1}, \delta_{\varepsilon_2} \right]
    \cdot \varphi &= \left[ -\varepsilon_1^a \sQ_a , - \varepsilon_2^b
      \sQ_b \right] \cdot \varphi\\
    &= \varepsilon_1^a \sQ_a \cdot \varepsilon_2^b \sQ_b \cdot \varphi
    - \varepsilon_2^b \sQ_b \cdot \varepsilon_1^a \sQ_a \cdot \varphi\\
    &= - \varepsilon_1^a \varepsilon_2^b \left( \sQ_a \cdot \sQ_b +
      \sQ_b \cdot \sQ_a\right) \cdot \varphi\\
    &= - \varepsilon_1^a \varepsilon_2^b \left[ \sQ_a, \sQ_b \right]
    \cdot \varphi~,
\end{split}
\end{equation}
where we use the \emph{same} notation $[-,-]$ for the bracket of
any two elements in a Lie superalgebra.  On dimensional grounds, the
bracket of two supercharges, having dimension $1$, must be a
translation.  Indeed, one can show the following.

\begin{exercise}
  Show that
  \begin{equation*}
    \left[\sQ_a, \sQ_b\right] \cdot S =  2 
    \left(\gamma^\mu\right)_{ab} \sP_\mu \cdot S
  \end{equation*}
  and similarly for $P$, whereas for $\psi$ one has instead
  \begin{equation*}
    \left[\sQ_a, \sQ_b\right] \cdot \psi = 2 
    \left(\gamma^\mu\right)_{ab} \sP_\mu \cdot \psi + 
    (\gamma^\mu)_{ab}\, \gamma_\mu \dslash \psi~. 
  \end{equation*}
\end{exercise}

If we use the classical equations of motion for $\psi$, the second
term in the right-hand side of the last equation vanishes and we
obtain an \emph{on-shell} realisation of the extension of the Poincaré
algebra \eqref{eq:poincare} defined by the following extra brackets:
\begin{equation}
  \label{eq:superpoincare}
  \begin{aligned}
    \left[\sP_\mu, \sQ_a\right] &= 0\\
    \left[\sM_{\mu\nu}, \sQ_a\right] &= -
    \left(\Sigma_{\mu\nu}\right)_a{}^b \sQ_b\\
    \left[\sQ_a, \sQ_b\right] &= 2 \left(\gamma^\mu\right)_{ab}
    \sP_\mu~.
\end{aligned}
\end{equation}
These brackets together with \eqref{eq:poincare} define the
\emph{($N{=}1$) Poincaré superalgebra}.

\begin{scholium}
  The fact that the commutator of two supersymmetries is a translation
  has a remarkable consequence.  In theories where supersymmetry is
  local, so that the spinor parameter is allowed to depend on the
  point, the commutator of two local supersymmetries is an
  infinitesimal translation whose parameter is allowed to depend on
  the point; in other words, it is an infinitesimal general coordinate
  transformation or, equivalently, an infinitesimal diffeomorphism.
  This means that theories with local supersymmetry automatically
  incorporate gravity.  This is why such theories are called
  supergravity theories.
\end{scholium}

A ($N{=}1$) supersymmetric field theory is by definition any field
theory which admits a realisation of the ($N{=}1$) Poincaré
superalgebra on the space of fields (maybe on-shell and up to gauge
equivalence) which leaves the action invariant.  In particular this
means that supersymmetry transformations take solutions to solutions.

\begin{scholium}
  It may seem disturbing to find that supersymmetry is only realised
  on-shell, since in computing perturbative quantum corrections, it is
  necessary to consider virtual particles running along in loops.
  This problem is of course well-known, e.g., in gauge theories where
  the BRST symmetry is only realised provided the antighost equation
  of motion is satisfied.  The solution, in both cases, is the
  introduction of non-propagating auxiliary fields.  We will see the
  need for this when we study the representation theory of the
  Poincaré superalgebra.  In general finding a complete set of
  auxiliary fields is a hard (sometimes unsolvable) problem; but we
  will see that in the case of $N{=}1$ Poincaré supersymmetry, the
  superspace formalism to be introduced in
  Lecture~\ref{lec:superspace} will automatically solve this problem.
\end{scholium}

\subsection{Adding masses and interactions}

There are of course other supersymmetric actions that can be built out 
of the same fields $S$, $P$ and $\psi$ by adding extra terms to the
free action $I_{\text{kin}}$.  For example, we could add mass terms:
\begin{equation}
  \eL_{\text{m}} = -\half m_1^2 S^2 - \half m_2^2 P^2 - \half  m_3
  \bar\psi \psi~,
\end{equation}
where $m_i$ for $i=1,2,3$ have units of mass.

\begin{exercise}
  Show that the action
  \begin{equation}
    \label{eq:massiveaction}
   \int d^4x \, \left( \eL_{\text{kin}} + \eL_{\text{m}}\right)
  \end{equation}
  is invariant under a modified set of supersymmetry transformations
  \begin{equation}
    \begin{aligned}
      \delta_\varepsilon S &=  \bar\varepsilon\psi\\
      \delta_\varepsilon P &=  \bar\varepsilon\gamma_5\psi\\
      \delta_\varepsilon \psi &= (\dslash - m)(S +
      P\gamma_5)\varepsilon~,
  \end{aligned}
\end{equation}
  provided that $m_1=m_2=m_3=m$.  More concretely, show that with
  these choices of $m_i$,
  \begin{equation}
    \delta_\varepsilon \left( \eL_{\text{kin}} + \eL_{\text{m}}\right) 
    = \d_\mu X^\mu~,
  \end{equation}
  where
  \begin{equation}
    X^\mu = -\half \bar\varepsilon \gamma^\mu \left(S -
    P\gamma_5\right) \left(\overleftarrow{\dslash} - m\right) \psi~,
  \end{equation}
  where for any $\zeta$, $\zeta \overleftarrow{\dslash} =
  \d_\mu\zeta\gamma^\mu$.  Moreover show that the above supersymmetry
  transformations close, up to the equations of motion of the
  fermions, to realise the Poincaré superalgebra.
\end{exercise}

This result illustrates an important point: irreducible
representations of the Poincaré superalgebra are mass degenerate; that
is, all fields have the same mass. This actually follows easily from
the Poincaré superalgebra itself.  The (squared) mass is up to a sign
the eigenvalue of the operator $\sP^2 = \eta^{\mu\nu}\sP_\mu\sP_\nu$
which, from equations \eqref{eq:poincare} and the first equation in
\eqref{eq:superpoincare}, is seen to be a Casimir of the Poincaré
superalgebra.  Therefore on an irreducible representation $\sP^2$ must
act as a multiple of the identity.

The action \eqref{eq:massiveaction} is still free, hence physically
not very interesting.  It is possible to add interacting terms in such 
a way that Poincaré supersymmetry is preserved.

Indeed, consider the following interaction terms
\begin{equation}
  \eL_{\text{int}} = - \lambda \left(  \bar\psi \left(S -
      P\gamma_5\right)\psi + \half \lambda \left(S^2 + P^2\right)^2 +
    m S \left(S^2+P^2\right)\right)~.
\end{equation}

The \emph{Wess--Zumino model} is defined by the action
\begin{equation}
  \label{eq:WZmodel}
  I_{\text{WZ}} = \int d^4 x \left( \eL_{\text{kin}} + \eL_{\text{m}} +
  \eL_{\text{int}} \right)~.
\end{equation}

\begin{exercise}
  Prove that $I_{\text{WZ}}$ is invariant under the following modified 
  supersymmetry transformations:
  \begin{equation}
    \label{eq:modsusy}
    \begin{aligned}
      \delta_\varepsilon S &= \bar\varepsilon\psi\\
      \delta_\varepsilon P &= \bar\varepsilon\gamma_5\psi\\
      \delta_\varepsilon \psi &= \left[\dslash - m - \lambda
        \left(S+P\gamma_5\right)\right] \left(S +
        P\gamma_5\right)\varepsilon ~,
  \end{aligned}
\end{equation}
  and verify that these transformations close on-shell to give a
  realisation of the Poincaré superalgebra.  More concretely, show
  that
  \begin{equation}
    \delta_\varepsilon \eL_{\text{WZ}} = \d_\mu Y^\mu~,
  \end{equation}
  where
  \begin{equation}
    Y^\mu = - \half \bar\varepsilon \gamma^\mu (S-P\gamma_5)
    \left(\overleftarrow{\dslash} - m - \lambda (S-P\gamma_5)\right)
    \psi~.
  \end{equation}
\end{exercise}

For future reference, we notice that the supersymmetry transformations
in \eqref{eq:modsusy} can be rewritten in terms of the generator
$\sQ_a$ as follows:
\begin{equation}
  \label{eq:susyWZ}
  \begin{aligned}
    \sQ_a \cdot S &=  \psi_a\\
    \sQ_a \cdot P &= -  (\gamma_5)_a{}^b \psi_b\\
    \sQ_a \cdot \psi_b &= -\d_\mu S (\gamma^\mu)_{ab} + \d_\mu P
    (\gamma^\mu\gamma_5)_{ab} - m S C_{ab} - m P (\gamma_5)_{ab}\\
    &\quad - \lambda(S^2-P^2) C_{ab} -2 \lambda S P
    (\gamma_5)_{ab}~.
  \end{aligned}
\end{equation}

\begin{tut}[\textsc{Superconformal invariance, Part I}]\indent\par
  \label{pr:superconformalWZ}
  In this problem you are invited to show that the massless
  Wess--Zumino model is classically invariant under a larger symmetry
  than the Poincaré superalgebra: the conformal superalgebra.
  
  The conformal algebra of Minkowski space contains the Poincaré
  algebra as a subalgebra, and in addition it has five other
  generators: the dilation $\sD$ and the special conformal
  transformations $\sK_\mu$.  The conformal algebra has the following
  (nonzero) brackets in addition to those in \eqref{eq:poincare}:
  \begin{equation}
    \label{eq:conformal}
    \begin{aligned}
      \left[\sP_\mu,\sD\right] &= \sP_\mu\\
      \left[\sK_\mu,\sD\right] &= -\sK_\mu\\
      \left[\sP_\mu,\sK_\nu\right] &= 2 \eta_{\mu\nu} \sD - 2
      \sM_{\mu\nu}\\
      \left[\sM_{\mu\nu}, \sK_\rho\right] &= \eta_{\nu\rho} \sK_\mu -
      \eta_{\mu\rho} \sK_\nu~.
    \end{aligned}
  \end{equation}
  Any supersymmetric field theory which is in addition conformal
  invariant will be invariant under the smallest superalgebra
  generated by these two Lie (super)algebras.  This superalgebra is
  called the conformal superalgebra.  We will see that the massless
  Wess--Zumino model is classically conformal invariant.  This will
  then show that it is also classically superconformal invariant.  In
  the course of the problem you will also discover the form of the
  conformal superalgebra.

  \begin{enumerate}
  \item Prove that the following, together with
    \eqref{eq:PoincareTransf}, define a realisation of the conformal
    algebra on the fields in the Wess--Zumino model:
    \begin{equation*}
      \begin{aligned}
        \sD \cdot S &= -x^\mu\d_\mu S - S\\
        \sD \cdot P &= -x^\mu\d_\mu P - P\\
        \sD \cdot \psi &= -x^\mu\d_\mu \psi - \tfrac32 \psi\\
        \sK_\mu \cdot S & = - 2 x_\mu x^\nu \d_\nu S + x^2 \d_\mu S -
        2
        x_\mu S\\
        \sK_\mu \cdot P & = - 2 x_\mu x^\nu \d_\nu P + x^2 \d_\mu P -
        2
        x_\mu P\\
        \sK_\mu \cdot \psi &= - 2 x_\mu x^\nu \d_\nu \psi + x^2 \d_\mu
        \psi - 3 x_\mu \psi + x^\nu \gamma_{\nu\mu} \psi
    \end{aligned}
\end{equation*}
  \item Prove that the massless Wess--Zumino action with lagrangian
    \begin{multline}
      \label{eq:masslessWZ}
      \eL_{\text{mWZ}} = - \half \left(\d S\right)^2 - \half \left(\d
        P\right)^2 - \half \bar\psi \dslash \psi\\
      - \lambda \bar\psi \left(S - P\gamma_5\right)\psi - \half
        \lambda^2 \left(S^2 + P^2\right)^2
    \end{multline}
    is conformal invariant.  More precisely, show that
    \begin{equation*}
      \sD \cdot \eL_{\text{mWZ}} = \d_\mu ( - x^\mu \eL_{\text{mWZ}} )
    \end{equation*}
    and that
    \begin{equation*}
      \sK_\mu \cdot \eL_{\text{mWZ}} = \d_\nu \left[ \left(-2 x_\mu
      x^\nu + x^2 \delta_\mu^\nu \right) \eL_{\text{mWZ}} \right]~,
    \end{equation*}
    and conclude that the action is invariant.
  \end{enumerate}
  
  \begin{scholium}
    It is actually enough to prove that the action is invariant under
    $\sK_\mu$ and $\sP_\mu$, since as can be easily seen from the
    explicit form of the algebra, these two sets of elements generate
    the whole conformal algebra.
  \end{scholium}
  
  We now know that the massless Wess--Zumino model is invariant both
  under supersymmetry and under conformal transformations.  It
  follows that it is also invariant under any transformation
  obtained by taking commutators of these and the resulting
  transformations until the algebra closes (at least on-shell).  We
  will now show that this process results in an on-shell realisation
  of the conformal superalgebra.  In addition to the conformal and
  superPoincaré generators, the conformal superalgebra has also a
  second spinorial generator $\sS_a$, generating conformal
  supersymmetries and a further bosonic generator $\sR$ generating
  the so-called R-symmetry to be defined below.
  
  Let $\kappa_\mu$ be a constant vector and let $\delta_\kappa$ denote
  an infinitesimal special conformal transformation, defined on fields
  $\varphi$ by $\delta_\kappa \varphi = \kappa^\mu \sK_\mu \cdot
  \varphi$.  The commutator of an infinitesimal supersymmetry and an
  infinitesimal special conformal transformation is, by definition, a
  conformal supersymmetry.  These are generated by a spinorial
  generator $\sS_a$ defined by
  \begin{equation*}
   \left[\sK_\mu,\sQ_a\right] = + (\gamma_\mu)_a{}^b \sS_b~.
  \end{equation*}
  Let $\zeta$ be an anticommuting Majorana spinor and define an
  infinitesimal conformal supersymmetry $\delta_\zeta$ as
  $\delta_\zeta \varphi = \bar\zeta \sS \cdot \varphi$.

  \begin{enumerate}
  \item[3.] Show that the infinitesimal conformal supersymmetries take
    the following form:
    \begin{equation}
      \label{eq:conformalsusies}
      \begin{aligned}
        \delta_\zeta S &= \bar\zeta x^\mu\gamma_\mu \psi\\
        \delta_\zeta P &= \bar\zeta x^\mu\gamma_\mu \gamma_5\psi\\
        \delta_\zeta \psi &= -\left[\left(\dslash - \lambda
            (S+P\gamma_5) \right) (S+P\gamma_5)\right] x^\mu\gamma_\mu
        \zeta - 2 (S-P\gamma_5)\zeta~.
      \end{aligned}
    \end{equation}
    \item[4.] Show that the action of $\sS_a$ on fields is given
      by:
    \begin{equation}
      \begin{aligned}
        \sS_a \cdot S &= x^\mu(\gamma_\mu)_{ab}\, \psi^b\\
        \sS_a \cdot P &= x^\mu(\gamma_\mu \gamma_5)_{ab}\, \psi^b\\
        \sS_a \cdot \psi_b &= - (x^\mu \d_\mu + 2) S C_{ab} + (x^\mu
        \d_\mu + 2) P (\gamma_5)_{ab})\\
        & \quad {} - (x_\mu \d_\nu S + \half
        \epsilon_{\mu\nu}{}^{\rho\sigma}
        x_\rho \d_\sigma P) (\gamma^{\mu\nu})_{ab}\\
        & \quad {} - \lambda(S^2-P^2) x^\mu (\gamma_\mu)_{ab} - 2
        \lambda S P x^\mu(\gamma_\mu\gamma_5)_{ab}~.
      \end{aligned}
    \end{equation}
  \item[5.] Show that the action of $\sP_\mu$, $\sM_{\mu\nu}$,
    $\sK_\mu$, $\sD$, $\sR$, $\sQ_a$ and $\sS_a$ on the fields $S$,
    $P$ and $\psi$ defines an on-shell realisation of the
    \emph{conformal superalgebra} defined by the following (nonzero)
    brackets in addition to those in \eqref{eq:poincare},
    \eqref{eq:superpoincare} and \eqref{eq:conformal}:
    \begin{equation}
      \label{eq:superconformal}
      \begin{aligned}
        \left[\sM_{\mu\nu}, \sS_a\right] &= -
        \left(\Sigma_{\mu\nu}\right)_a{}^b \sS_b\\
        \left[\sR, \sQ_a\right] &= + \half (\gamma_5)_a{}^b \sQ_b\\
        \left[\sR, \sS_a\right] &= - \half (\gamma_5)_a{}^b \sS_b\\
        \left[\sD, \sQ_a\right] &= - \half \sQ_a\\
        \left[\sD, \sS_a\right] &= + \half \sS_a\\
        \left[\sK_\mu,\sQ_a\right] &= + (\gamma_\mu)_a{}^b \sS_b\\
        \left[\sP_\mu,\sS_a\right] &= -(\gamma_\mu)_a{}^b \sQ_b\\
        \left[\sS_a,\sS_b\right] &= - 2 (\gamma^\mu)_{ab} \sK_\mu\\
        \left[\sQ_a,\sS_b\right] &= + 2 C_{ab} \sD - 2 (\gamma_5)_{ab}
        \sR + (\gamma^{\mu\nu})_{ab} \sM_{\mu\nu}~,
      \end{aligned}
    \end{equation}
    where the action of the R-symmetry on fields is
  \begin{equation}
    \label{eq:Rsymmetry}
    \begin{aligned}
      \sR \cdot S &=  P\\
      \sR \cdot P &= -S\\
      \sR \cdot \psi &= \half \gamma_5\, \psi~.
    \end{aligned}
  \end{equation}
  \end{enumerate}
  
  \begin{scholium}
    This shows that the massless Wess--Zumino model is classically
    superconformal invariant.  However several facts should alert us
    that this symmetry will be broken by quantum effects.  First of
    all the R-symmetry acts via $\gamma_5$ and this sort of symmetries
    are usually quantum-mechanically anomalous.  Similarly, we expect
    that the trace and conformal anomalies should break invariance
    under $\sD$ and $\sK_\mu$ respectively.  This is in fact the case.
    What is remarkable is that the Wess--Zumino model (with or without
    mass) is actually quantum mechanically supersymmetric to all
    orders in perturbation theory.  Moreover the model only requires
    wave function renormalisation: the mass $m$ and the coupling
    constant $\lambda$ do not renormalise.  This sort of
    \emph{nonrenormalisation theorems} are quite common in
    supersymmetric theories.  We will be able to explain why this is
    the case in a later lecture, although we will not have the time to
    develop the necessary formalism to prove it.
  \end{scholium}

\end{tut}


\section{Supersymmetric Yang--Mills theory}
\label{lec:SYM}

In this section we introduce another simple model exhibiting
supersymmetry: supersymmetric Yang--Mills.  This model consists of
ordinary Yang--Mills theory coupled to adjoint fermions.  We will see
that this model admits an on-shell realisation of the Poincaré
superalgebra which however only closes up to gauge transformations.
More is true and in the tutorial you will show that the theory is
actually superconformal invariant, just like the massless Wess--Zumino
model.

\subsection{Supersymmetric Yang--Mills}

The existence of a supersymmetric extension of Yang--Mills theory
could be suspected from the study of the representations of the
Poincaré superalgebra (see Lecture~\ref{lec:Reps}), but this does not
mean that it is an obvious fact.  Indeed, the existence of
supersymmetric Yang--Mills theories depends on the dimensionality and
the signature of spacetime.  Of course one can always write down the
Yang--Mills action in any dimension and then couple it to fermions,
but as we will see in the next lecture, supersymmetry requires a
delicate balance between the bosonic and fermionic degrees of freedom.
Let us consider only lorentzian spacetimes.  A gauge field in $d$
dimensions has $d-2$ physical degrees of freedom corresponding to the
transverse polarisations.  The number of degrees of freedom of a
fermion field depends on what kind fermion it is, but it always a
power of $2$.  An unconstrained Dirac spinor in $d$ dimensions has
$2^{d/2}$ or $2^{(d-1)/2}$ real degrees of freedom, for $d$ even or
odd respectively: a Dirac spinor has $2^{d/2}$ or $2^{(d-1)/2}$
complex components but the Dirac equation cuts this number in half.
In even dimensions, one can further restrict the spinor by imposing
that it be chiral (Weyl).  This cuts the number of degrees of freedom
by two.  Alternatively, in some dimensions (depending on the signature
of the metric) one can impose a reality (Majorana) condition which
also halves the number of degrees of freedom.  For a lorentzian metric
of signature $(1,d-1)$, Majorana spinors exist for $d\equiv 1,2,3,4
\mod 8$.  When $d\equiv 2 \mod 8$ one can in fact impose that a spinor
be both Majorana and Weyl, cutting the number of degrees of freedom in
four.  The next exercise asks you to determine in which dimensions can
supersymmetric Yang--Mills theory exist based on the balance between
bosonic and fermionic degrees of freedom.

\begin{exercise}
  Verify via a counting of degrees of freedom that ($N{=}1$)
  supersymmetric Yang--Mills can exist only in the following
  dimensions and with the following types of spinors:
\begin{center}
\renewcommand{\arraystretch}{1.2}
\begin{tabular}{r|l}
  $d$&Spinor\\
  \hline
  3&Majorana\\
  4&Majorana (or Weyl)\\
  6&Weyl\\
  10&Majorana--Weyl
\end{tabular}
\end{center}
\end{exercise}

\begin{scholium}
  The fact that these dimensions are of the form $2+n$, where
  $n=1,2,4,8$ are the dimensions of the real division algebras is not
  coincidental.  It is a further curious fact that these are precisely
  the dimensions in which the classical superstring exists.  Unlike
  superstring theory, in which only the ten-dimensional theory
  survives quantisation, it turns out that supersymmetric Yang--Mills
  theory exists in each of these dimensions.  Although we are mostly
  concerned with four-dimensional field theories in these notes, the
  six-dimensional and ten-dimensional theories are useful tools since
  upon dimensional reduction to four dimensions they yield $N{=}2$ and
  $N{=}4$ supersymmetric Yang--Mills, respectively.
\end{scholium}

We will now write down a supersymmetric Yang--Mills theory in four
dimensions.  We will show that the action is invariant under a
supersymmetry algebra which closes on-shell and up to gauge
transformations to a realisation of the Poincaré superalgebra.

\subsection{A brief review of Yang--Mills theory}

Let us start by reviewing Yang--Mills theory.  We pick a gauge group
$G$ which we take to be a compact Lie group.  We let $\fg$ denote its
Lie algebra.  We must also make the choice of an invariant inner
product in the Lie algebra, which we will call $\Tr$.  Fix a basis
$\{T_i\}$ for $\fg$ and let $G_{ij} = \Tr T_i T_j$ be the invariant
inner product and $f_{ij}{}^k$ be the structure constants.

The gauge field is a one-form in Minkowski space with values in $\fg$:
$A_\mu = A_\mu^i T_i$.  Geometrically it represents a connection in a
principal $G$ bundle on Minkowski space.  The field-strength
$F_{\mu\nu} = F_{\mu\nu}^i T_i$ is the curvature two-form of that
connection, and it is defined as
\begin{equation*}
  F_{\mu\nu} = \d_\mu A_\nu - \d_\nu A_\mu + g [A_\mu,A_\nu]~,
\end{equation*}
or relative to the basis $\{T_i\}$:
\begin{equation*}
  F_{\mu\nu}^i = \d_\mu A^i_\nu - \d_\nu A^i_\mu + g f_{jk}{}^i
  A^j_\mu A^k_\nu~,
\end{equation*}
where $g$ is the Yang--Mills coupling constant.  The lagrangian
is then given by
\begin{equation*}
  \eL_{\text{YM}} = - \tfrac14 \Tr F_{\mu\nu} F^{\mu\nu}~,
\end{equation*}
and the action is as usual the integral
\begin{equation*}
  I_{\text{YM}} = - \tfrac14 \int d^4 x \, \Tr F_{\mu\nu}
  F^{\mu\nu}~.
\end{equation*}
The sign has been chosen so that with our choice of spacetime metric,
the hamiltonian is positive-semidefinite.

\begin{exercise}
  Show that the action is invariant under the natural action of the
  Poincaré algebra:
  \begin{equation}
    \begin{aligned}
      \sP_\mu \cdot A_\nu &= -\d_\mu A_\nu\\
      \sM_{\mu\nu} \cdot A_\rho &= - (x_\mu \d_\nu - x_\nu \d_\mu)
      A_\rho - \eta_{\nu\rho} A_\mu + \eta_{\mu\rho} A_\nu~.
  \end{aligned}
\end{equation}
\end{exercise}

The action is also invariant under gauge transformations.  Let $U(x)$
be a $G$-valued function on Minkowski space.  The gauge field $A_\mu$
transforms in such a way that the covariant derivative $\eD_\mu = \d_\mu
+ g A_\mu$ transforms covariantly:
\begin{equation*}
  \eD_\mu^U = \d_\mu + g A_\mu^U = U \eD_\mu U^{-1} = U \left( \d_\mu +
    g A_\mu \right) U^{-1}~,
\end{equation*}
whence the transformed gauge field is
\begin{equation*}
  A_\mu^U = U A_\mu U^{-1} - \tfrac{1}{g}(\d_\mu U) U^{-1}~.
\end{equation*}
The field-strength transforms covariantly
\begin{equation*}
  F_{\mu\nu}^U = U F_{\mu\nu} U^{-1}~,
\end{equation*}
which together with the invariance of the inner product (or
equivalently, cyclicity of the trace) implies that the Lagrangian is
invariant.

Suppose that $U(x) = \exp \omega(x)$ where $\omega(x) = \omega(x)^i
T_i$ is a $\fg$-valued function.  Keeping only terms linear in
$\omega$ in the gauge transformation of the gauge field, we arrive at
the infinitesimal gauge transformations:
\begin{equation}
  \label{eq:infgaugetrans}
  \delta_\omega A_\mu = - \tfrac{1}{g} \eD_\mu \omega \implies
  \delta_\omega F_{\mu\nu} = [\omega, F_{\mu\nu}]~,
\end{equation}
which is easily verified to be an invariance of the Yang--Mills
lagrangian.

\subsection{Supersymmetric extension}

We will now find a supersymmetric extension of this action. Because
supersymmetry exchanges bosons and fermions, we will add some
fermionic fields.  Since the bosons $A_\mu$ are $\fg$-valued,
supersymmetry will require that so be the fermions.  Therefore we will 
consider an adjoint Majorana spinor $\Psi = \Psi^i T_i$.  The
natural gauge invariant interaction between the spinors and the gauge
fields is the minimally coupled lagrangian
\begin{equation}
  \label{eq:minimalcoupling}
  - \half \Tr \bar\Psi \Dslash \Psi~,
\end{equation}
where $\bar\Psi = \Psi^t C$, $\Dslash = \gamma^\mu \eD_\mu$ and
\begin{equation*}
  \eD_\mu \Psi = \d_\mu \Psi + g [A_\mu,\Psi] \implies
  \eD_\mu \Psi^i = \d_\mu \Psi^i + g f_{jk}{}^i A^j_\mu \Psi^k~.
\end{equation*}

\begin{exercise}
  Prove that the minimal coupling interaction
  \eqref{eq:minimalcoupling} is invariant under the infinitesimal
  gauge transformations \eqref{eq:infgaugetrans} and
  \begin{equation}
    \label{eq:infgaugetransPsi}
    \delta_\omega \Psi = [\omega, \Psi]~.
  \end{equation}
\end{exercise}

The action
\begin{equation}
  I_{\text{SYM}} = \int d^4x \, \eL_{\text{SYM}}~,
\end{equation}
with
\begin{equation}
  \label{eq:symlag}
  \eL_{\text{SYM}} = - \tfrac14 \Tr F_{\mu\nu} F^{\mu\nu} - \half
  \Tr \bar\Psi \Dslash \Psi
\end{equation}
is therefore both Poincaré and gauge invariant.  One can also verify
that it is real.  In addition, as we will now show, it is also
invariant under supersymmetry.

Taking into account dimensional considerations, Bose--Fermi parity,
and equivariance under the gauge group (namely that gauge
transformations should commute with supersymmetry) we arrive at the
following supersymmetry transformations rules:
\begin{equation*}
  \begin{aligned}
    \delta_\varepsilon A_\mu = \bar\varepsilon \gamma_\mu \Psi
    &\implies \sQ_a \cdot A_\mu = - (\gamma_\mu)_a{}^b
    \Psi_b\\
    \delta_\varepsilon \Psi = \half \alpha F_{\mu\nu} \gamma^{\mu\nu}
    \varepsilon &\implies \sQ_a \cdot \Psi_b = -\half \alpha
    F_{\mu\nu} (\gamma^{\mu\nu})_{ab}~,
\end{aligned}
\end{equation*}
where $\alpha$ is a parameter to be determined and $\varepsilon$ in
again an anticommuting Majorana spinor.

\begin{scholium}
  The condition on gauge equivariance is essentially the condition
  that we should only have rigid supersymmetry.  Suppose that
  supersymmetries and gauge transformations would not commute.  Their
  commutator would be another type of supersymmetry (exchanging bosons 
  and fermions) but the parameter of the transformation would be
  local, since gauge transformations have local parameters.  This
  would imply the existence of a local supersymmetry.  Since we are
  only considering rigid supersymmetries, we must have that
  supersymmetry transformations and gauge transformations commute.
\end{scholium}

\begin{exercise}
  Prove that the above ``supersymmetries'' commute with infinitesimal
  gauge transformations:
  \begin{equation*}
    [\delta_\epsilon, \delta_\omega] \varphi = 0~,
  \end{equation*}
  on any field $\varphi = A_\mu, \Psi$.
\end{exercise}

Now let us vary the lagrangian $\eL_{\text{SYM}}$.  This task is made
a little easier after noticing that for any variation $\delta A_\mu$
of the gauge field---including, of course, supersymmetries---the field
strength varies according to
\begin{equation*}
  \delta F_{\mu\nu} = \eD_\mu \delta A_\nu - \eD_\nu \delta A_\mu~.
\end{equation*}
Varying the lagrangian we notice that there are two types of terms in
$\delta_\varepsilon \eL_{\text{SYM}}$: terms linear in $\Psi$ and
terms cubic in $\Psi$.  Invariance of the action demands that they
should vanish separately.

It is easy to show that the terms linear in $\Psi$ cancel up a total
derivative provided that $\alpha =-1$.  This result uses equation
\eqref{eq:gammagammagamma} and the Bianchi identity
\begin{equation*}
  \eD_{[\mu} F_{\nu\rho]} = 0~.
\end{equation*}
On the other hand, the cubic terms vanish on their own using the Fierz
identity \eqref{eq:specialfierz} and the identities \eqref{eq:useful}.

\begin{exercise}
  Prove the above claims; that is, prove that under the supersymmetry
  transformations:
  \begin{equation}
    \label{eq:superpoincareYM}
    \begin{aligned}
      \delta_\varepsilon A_\mu = \bar\varepsilon \gamma_\mu \Psi\\
      \delta_\varepsilon \Psi = -\half F_{\mu\nu} \gamma^{\mu\nu}
      \varepsilon
    \end{aligned}
  \end{equation}
  the lagrangian $\eL_{\text{SYM}}$ transforms into a total
  derivative
  \begin{equation*}
    \delta_\varepsilon \eL_{\text{SYM}} = \d_\mu
    \left( - \tfrac14 \bar\varepsilon \gamma^\mu \gamma^{\rho\sigma}
      F_{\rho\sigma} \Psi\right)~,
  \end{equation*}
  and conclude that the action $I_{\text{SYM}}$ is invariant.
\end{exercise}

\subsection{Closure of the supersymmetry algebra}

We have called the above transformations ``supersymmetries'' but we
have still to show that they correspond to a realisation of the
Poincaré superalgebra \eqref{eq:superpoincare}.  We saw in the
Wess--Zumino model that the algebra only closed up to the equations of
motion of the fermions.  In this case we will also have to allow for
gauge transformations.  The reason is the following: although
supersymmetries commute with gauge transformations, it is easy to see
that translations do not.  Therefore the commutator of two
supersymmetries could not simply yield a translation.  Instead, and
provided the equations of motion are satisfied, it yields a
translation and an infinitesimal gauge transformation.

\begin{exercise}
  Prove that 
  \begin{equation*}
    [\sQ_a, \sQ_b] \cdot A_\mu =  2 (\gamma^\rho)_{ab} \sP_\rho \cdot 
    A_\mu + 2 (\gamma^\rho)_{ab} \eD_\mu A_\rho~.
  \end{equation*}
\end{exercise}

Notice that the second term in the above equation has the form of an
infinitesimal gauge transformation with (field-dependent) parameter
$-2 g \gamma^\rho A_\rho$, whereas the first term agrees with the
Poincaré superalgebra \eqref{eq:superpoincare}.

\begin{exercise}
  Prove that up to terms involving the equation of motion of the
  fermion ($\Dslash \Psi = 0$),
  \begin{equation*}
    [\sQ_a, \sQ_b] \cdot \Psi = 2 (\gamma^\rho)_{ab} \sP_\rho \cdot 
    \Psi - 2 g (\gamma^\rho)_{ab} [A_\rho, \Psi]~.
  \end{equation*}
\end{exercise}

Again notice that the second term has the form of an infinitesimal
gauge transformation with the \emph{same} parameter $-2 g \gamma^\rho 
A_\rho$, whereas again the first term agrees with the Poincaré
superalgebra \eqref{eq:superpoincare}.

The fact that the gauge transformation is the same one in both cases
allows us to conclude that the Poincaré superalgebra is realised
on-shell and up to gauge transformations on the fields $A_\mu$ and
$\Psi$.

\begin{scholium}
  There is a geometric picture which serves to understand the above
  result.  One can understand infinitesimal symmetries as vector
  fields on the (infinite-dimensional) space of fields $\eF$.  Each
  point in this space corresponds to a particular field configuration
  $(A_\mu, \Psi)$.  An infinitesimal symmetry $(\delta A_\mu,
  \delta\Psi)$ is a particular kind of vector field on $\eF$; in other
  words, the assignment of a small displacement (a tangent vector
  field) to every field configuration.
  
  Now let $\eF_0\subset\eF$ be the subspace corresponding to those
  field configurations which obey the classical equations of motion.
  A symmetry of the action preserves the equations of motion, and
  hence sends solutions to solutions.  Therefore symmetries preserve
  $\eF_0$ and infinitesimal symmetries are vector fields which are
  tangent to $\eF_0$.
  
  The group $\eG$ of gauge transformations, since it acts by
  symmetries, preserves the subspace $\eF_0$ and in fact foliates it
  into gauge orbits: two configurations being in the same orbit if
  there is a gauge transformation that relates them.  Unlike other
  symmetries, gauge-related configurations are physically
  indistinguishable.  Therefore the space of physical configurations
  is the space $\eF_0/\eG$ of gauge orbits.
  
  Now, any vector field on $\eF_0$ defines a vector field on
  $\eF_0/\eG$: one simply throws away the components tangent to the
  gauge orbits.  The result we found above can be restated as saying
  that in the space of physical configurations we have a realisation
  of the Poincaré superalgebra.
\end{scholium}

We have proven that the theory defined by the lagrangian
\eqref{eq:symlag} is a supersymmetric field theory.  It is called
\emph{($N{=}1$) (pure) supersymmetric Yang--Mills}.  This is the
simplest four-dimensional supersymmetric gauge theory, but by no means
the only one.  One can add matter coupling in the form of Wess--Zumino
multiplets.  Some of these theories have extended supersymmetry (in
the sense of \probref{pr:extended}).  Extended supersymmetry
constrains the dynamics of the gauge theory.  In the last five years
there has been much progress made on gauge theories with extended
supersymmetry, including for the first time the exact
(nonperturbative) solution of nontrivial interacting four-dimensional
quantum field theories.

\begin{tut}[\textsc{Superconformal invariance, Part II}]\indent\par
  \label{pr:superconformalYM}
  This problem does for supersymmetric Yang--Mills what
  \probref{pr:superconformalWZ} did for the Wess--Zumino model:
  namely, it invites you to show that supersymmetric Yang--Mills is
  classically invariant under the conformal superalgebra.  As with the
  Wess--Zumino model the strategy will be to show that the theory is
  conformal invariant and hence that it is invariant under the
  smallest superalgebra generated by the Poincaré supersymmetry and
  the conformal transformations.  This superalgebra will be shown to
  be (on-shell and up to gauge transformations) the conformal
  superalgebra introduced in \probref{pr:superconformalWZ}.

  \begin{enumerate}
  \item Show that supersymmetric Yang--Mills theory described by
    the action $I_{\text{SYM}}$ with lagrangian \eqref{eq:symlag} is
    invariant under the conformal transformations:
    \begin{equation*}
      \begin{aligned}
        \sD \cdot A_\rho &= -x^\mu\d_\mu A_\rho - A_\rho\\
        \sD \cdot \Psi &= -x^\mu\d_\mu \Psi - \tfrac32 \Psi\\
        \sK_\mu \cdot A_\rho & = - 2 x_\mu x^\nu \d_\nu A_\rho + x^2
        \d_\mu A_\rho - 2 x_\mu A_\rho + 2 x_\rho A_\mu - 2
        \eta_{\mu\rho} x^\nu A_\nu\\
        \sK_\mu \cdot \Psi &= - 2 x_\mu x^\nu \d_\nu \Psi + x^2 \d_\mu
        \Psi - 3 x_\mu \Psi + x^\nu \gamma_{\nu\mu} \Psi~.
      \end{aligned}
    \end{equation*}
    More precisely show that
    \begin{equation*}
      \begin{aligned}
        \sD \cdot \eL_{\text{SYM}} &= \d_\mu \left( - x^\mu
          \eL_{\text{SYM}} \right)\\
        \sK_\mu \cdot \eL_{\text{SYM}} &= \d_\nu \left[\left( -2 x_\mu
            x^\nu + x^2 \delta_\mu^\nu\right) \eL_{\text{SYM}}
        \right]~,
      \end{aligned}
    \end{equation*}
    and conclude that the action $I_{\text{SYM}}$ is invariant.
  \item Show that $I_{\text{SYM}}$ is invariant under the
    R-symmetry:
    \begin{equation*}
      \sR \cdot A_\mu = 0  \qquad\text{and}\qquad \sR \cdot \Psi = \half
      \gamma_5 \Psi~.
    \end{equation*}
  \item Referring to the discussion preceding Part~3 in
    \probref{pr:superconformalWZ}, show that the infinitesimal
    conformal supersymmetry of supersymmetric Yang--Mills takes the
    form:
    \begin{equation*}
      \begin{aligned}
        \delta_\zeta A_\mu &= \bar\zeta x^\nu \gamma_\nu\gamma_\mu
        \Psi\\
        \delta_\zeta \Psi &= \half x^\rho F_{\mu\nu} \gamma^{\mu\nu}
        \gamma^\rho \zeta~.
      \end{aligned}
    \end{equation*}
    \item Defining the generator $\sS_a$ by
    \begin{equation*}
      \delta_\zeta \varphi = \bar\zeta \sS \cdot \varphi = - \zeta^a
      \sS_a \cdot \varphi
    \end{equation*}
    show that the action of $\sS_a$ is given by
    \begin{equation*}
      \begin{aligned}
        \sS_a \cdot A_\mu &= \left(x^\nu \gamma_\nu
          \gamma_\mu\right)_{ab} \Psi^b\\
        \sS_a \cdot \Psi &= -\half x_\rho F_{\mu\nu}
        \left(\gamma^{\mu\nu}\gamma^\rho\right)_{ba}\\
        &= \half \epsilon^{\mu\nu\rho\sigma} x_\rho F_{\mu\nu}
        \left(\gamma_\sigma\gamma_5\right)_{ab} + x^\mu F_{\mu\nu}
        \left(\gamma^\nu\right)_{ab}~.
      \end{aligned}
    \end{equation*}
  \item Finally show that $\sS_a$, together with $\sM_{\mu\nu}$,
    $\sP_\mu$, $\sK_\mu$, $\sD$, $\sR$ and $\sQ_a$, define an on-shell
    (and up to gauge transformations) realisation of the conformal
    superalgebra defined by the brackets \eqref{eq:poincare},
    \eqref{eq:superpoincare}, \eqref{eq:conformal} and
    \eqref{eq:superconformal}.
  \end{enumerate}
  
  \begin{scholium}
    Again we expect that the classical superconformal symmetry of
    supersymmetric Yang--Mills will be broken by quantum effects:
    again the R-symmetry acts by chiral transformations which are
    anomalous, and as this theory has a nonzero beta function,
    conformal invariance will also fail at the quantum level.
    Nevertheless Poincaré supersymmetry will be preserved at all
    orders in perturbation theory.
    
    Remarkably one can couple supersymmetric Yang--Mills to
    supersymmetric matter in such a way that the resulting quantum
    theory is still superconformal invariant.  One such theory is the
    so-called $N{=}4$ supersymmetric Yang--Mills.  This theory has
    vanishing beta function and is in fact superconformally invariant
    to all orders.  It is not a realistic quantum field theory for
    phenomenological purposes, but it has many nice properties: it is
    maximally supersymmetric (having 16 supercharges), it exhibits
    electromagnetic (Montonen--Olive) duality and it has been
    conjectured (Maldacena) to be equivalent at strong coupling to
    type IIB string theory on a ten-dimensional background of the form 
    $\ads_5 \times S^5$, where $S^5$ is the round $5$-sphere and
    $\ads_5$, five-dimensional anti-de~Sitter space, is the lorentzian 
    analogue of hyperbolic space in that dimension.
  \end{scholium} 
  
\end{tut}


\section{Representations of the Poincaré superalgebra}
\label{lec:Reps}

In the first two lectures we met the Poincaré superalgebra and showed
that it is a symmetry of the Wess--Zumino model (in
Lecture~\ref{lec:WZmodel}) and of Yang--Mills theories with adjoint
fermions (in Lecture~\ref{lec:SYM}).  In the present lecture we will
study the representations of this algebra.  We will see that
irreducible representations of the Poincaré superalgebra consist of
multiplets of irreducible representations of the Poincaré algebra
containing fields of different spins (or helicities) but of the same
mass.  This degeneracy in the mass is not seen in nature and hence
supersymmetry, if a symmetry of nature at all, must be broken.  In
Lecture~\ref{lec:SB} we will discuss spontaneous supersymmetry
breaking.

\subsection{Unitary representations}

It will prove convenient both in this lecture and in later ones, to
rewrite the Poincaré superalgebra in terms of two-component spinors.
(See the Appendix for our conventions.)  The supercharge
$\sQ^a$, being a Majorana spinor decomposes into two Weyl spinors
\begin{equation}
  \sQ^a = 
  \begin{pmatrix}
    \sQ^\alpha\\
    \bar\sQ_{\dot\alpha}
  \end{pmatrix}~,
\end{equation}
in terms of which, the nonzero brackets in \eqref{eq:superpoincare}
now become
\begin{equation}
  \label{eq:superpoincare2}
  \begin{aligned}
    \left[\sM_{\mu\nu}, \sQ_\alpha\right] &= - \half
    \left(\sigma_{\mu\nu}\right)_\alpha{}^\beta \sQ_\beta\\
    \left[\sM_{\mu\nu}, \bar\sQ_{\dot\alpha}\right] &= \half
    \left(\bar\sigma_{\mu\nu}\right)_{\dot\alpha}{}^{\dot\beta}
    \bar\sQ_{\dot\beta}\\
    \left[\sQ_\alpha, \bar\sQ_{\dot\beta}\right] &= 2 i
    \left(\sigma^\mu\right)_{\alpha\dot\beta} \sP_\mu~.
\end{aligned}
\end{equation}

For the purposes of this lecture we will be interested in unitary
representations of the Poincaré superalgebra.  This means that
representations will have a positive-definite invariant hermitian
inner product and the generators of the algebra will obey the
following hermiticity conditions:
\begin{equation}
  \label{eq:hermiticity}
  \sM_{\mu\nu}^\dagger = - \sM_{\mu\nu} \qquad
  \sP_\mu^\dagger = - \sP_\mu \qquad
  \sQ_\alpha^\dagger = \bar\sQ_{\dot\alpha}~.
\end{equation}

\begin{exercise}
  Show that these hermiticity conditions are consistent with the
  Poincaré superalgebra.
\end{exercise}

Notice that $\sP_\mu$ is antihermitian, hence its eigenvalues will be
imaginary.  Indeed, we have seen that $\sP_\mu$ acts like $-\d_\mu$ on
fields.  For example, acting on a plane wave $\varphi = \exp(i p\cdot
x)$, $\sP_\mu \cdot \varphi = - i p_\mu\, \varphi$.  Therefore on a
momentum eigenstate $\ket{p}$, the eigenvalue of $\sP_\mu$ is
$-ip_\mu$.

A remarkable property of supersymmetric theories is that the energy is 
positive-semidefinite in a unitary representation.  Indeed, acting on
a momentum eigenstate $\ket{p}$ the supersymmetry algebra becomes
\begin{equation*}
  \left[\sQ_\alpha, \bar\sQ_{\dot\beta}\right] \ket{p} = 
  2 \begin{pmatrix}
    - p_0 + p_3 & p_1 - i p_2\\
    p_1 + i p_2 & - p_0 - p_3
\end{pmatrix} \ket{p}~.
\end{equation*}
Recalling that the energy is given by $p^0=-p_0$, we obtain
\begin{equation*}
  p^0 \ket{p} = \tfrac14 \left([\sQ_1, \sQ_1^\dagger] + [\sQ_2,
  \sQ_2^\dagger]\right) \ket{p}~.
\end{equation*}
In other words, the hamiltonian can be written in the following
manifestly positive-semidefinite way:
\begin{equation}
  \label{eq:Hsusy}
  H = \tfrac14 \left(\sQ_1 \sQ_1^\dagger + \sQ_1^\dagger \sQ_1 +
  \sQ_2 \sQ_2^\dagger + \sQ_2^\dagger \sQ_2\right)~.
\end{equation}
This shows that energy of any state is positive unless the state is
annihilated by all the supercharges, in which case it is zero.
Indeed, if $\ket{\psi}$ is \emph{any} state, we have that the
expectation value of the hamiltonian (the energy) is given by a sum of
squares:
\begin{equation*}
  \bra{\psi} H \ket{\psi} = \tfrac14 \left( \|\sQ_1\ket{\psi}\|^2 +
  \|\sQ_1^\dagger\ket{\psi}\|^2 + \|\sQ_2\ket{\psi}\|^2 +
  \|\sQ_2^\dagger\ket{\psi}\|^2 \right)~.
\end{equation*}
This is a very important fact of supersymmetry and one which plays a
crucial role in many applications, particularly in discussing the
spontaneous breaking of supersymmetry.

\subsection{Induced representations in a nutshell}

The construction of unitary representations of the Poincaré
superalgebra can be thought of as a mild extension of the construction
of unitary representations of the Poincaré algebra.  This method is
originally due to Wigner and was greatly generalised by Mackey.  The
method consists of inducing the representation from a
finite-dimensional unitary representation of some compact subgroup.
Let us review this briefly as it will be the basis for our
construction of irreducible representations of the Poincaré
superalgebra.

The Poincaré algebra has two Casimir operators: $\sP^2$ and $\sW^2$,
where $\sW^\mu = \half \epsilon^{\mu\nu\lambda\rho}\sP_\nu
\sM_{\lambda\rho}$ is the Pauli--Lubansky vector.  By Schur's lemma,
on an irreducible representation they must both act as multiplication
by scalars.  Let's focus on $\sP^2$.  On an irreducible representation
$\sP^2 = m^2$, where $m$ is the ``rest-mass'' of the particle
described by the representation.  Remember that on a state with
momentum $p$, $\sP_\mu$ has eigenvalue $-i p_\mu$, hence $\sP^2$ has
eigenvalues $-p^2$, which equals $m^2$ with our choice of metric.
Because physical masses are real, we have $m^2\geq 0$, hence we can
distinguish two kinds of physical representations: \emph{massless} for
which $m^2=0$ and \emph{massive} for which $m^2>0$.

Wigner's method starts by choosing a nonzero momentum $p$ on the
mass-shell: $p^2 = - m^2$.  We let $G_p$ denote the subgroup of the
Lorentz group (or rather of $\SL(2,\CC)$) which leaves $p$ invariant.
$G_p$ is known as the \emph{little group}.  Wigner's method, which we
will not describe in any more detail than this, consists in the
following.  First one chooses a unitary finite-dimensional irreducible
representation of the little group.  Doing this for every $p$ in the
mass shell defines a family of representations indexed by $p$.  The
representation is carried by functions assigning to a momentum $p$ in
the mass shell, a state $\phi(p)$ in this representation.  Finally,
one Fourier transforms to obtain fields on Minkowski spacetime subject
to their classical equations of motion.

\begin{scholium}
  In more mathematical terms, the construction can be described as
  follows.  The mass shell $\eM_{m^2} = \{p^\mu \mid p^2 = -m^2\}$ is
  acted on transitively by the Lorentz group $L$.  Fix a vector $p \in
  \eM_{m^2}$ and let $G_p$ be the little group.  Then $\eM_{m^2}$ can
  be seen as the space of right cosets of $G_p$ in $L$; that is, it is
  a homogeneous space $L/G_p$.  Any representation $\VV$ of $G_p$
  defines a homogeneous vector bundle on $\eM_{m^2}$ whose space of
  sections carries a representation of the Poincaré group.  This
  representation is said to be \emph{induced} from $\VV$.  If $\VV$ is
  unitary and irreducible, then so will be the induced representation.
  The induced representation naturally lives in momentum space, but
  for field theoretical applications we would like to work with fields
  in Minkowski space.  This is easily achieved by Fourier transform,
  but since the momenta on the mass-shell obey $p^2 = - m^2$, it
  follows that the Fourier transform $\varphi(x)$ of a function
  $\tilde\varphi(p)$ automatically satisfies the Klein--Gordon
  equation.  More is true, however, and the familiar classical
  relativistic equations of motion: Klein--Gordon, Dirac,
  Rarita--Schwinger,... can be understood group theoretically simply
  as projections onto irreducible representations of the Poincaré
  group.
\end{scholium}

In extending this method to the Poincaré superalgebra all that happens
is that now the Lie algebra of the little group gets extended by the
supercharges, since these commute with $\sP_\mu$ and hence stabilise
the chosen 4-vector $p_\mu$.  Therefore we now induce from a unitary
irreducible representations of the little (super)group.  This
representation will be reducible when restricted to the little group
and will at the end of the day generate a \emph{supermultiplet} of
fields.

Before applying this procedure we will need to know about the
structure of the little groups.  The little group happens to be
different for massive and for massless representations, as the next
exercise asks you to show.

\begin{exercise}
  Let $p_\mu$ be a momentum obeying $p^0 >0$, $p^2 = - m^2$.
  Prove that the little group of $p_\mu$ is isomorphic to:
  \begin{itemize}
  \item $\SU(2)$, for $m^2>0$;
  \item $\widetilde \E_2$, for $m^2=0$,
  \end{itemize}
  where $\E_2 \cong \SO(2)\ltimes\RR^2$, is the two-dimensional
  euclidean group and $\widetilde \E_2\cong \Spin(2)\ltimes\RR^2$ is
  its double cover.\\
  {\rm (Hint: Argue that two momenta which are Lorentz-related have
    isomorphic little groups and then choose a convenient $p_\mu$ in
    each case.)}
  \label{ex:ltlgrp}
\end{exercise}

\subsection{Massless representations}

Let us start by considering massless representations.  As shown in
\exref{ex:ltlgrp}, the little group for the momentum $p^\mu$ of a
massless particle is noncompact.  Therefore its finite-dimensional
unitary representations must all come from its maximal compact
subgroup $\Spin(2)$ and be trivial on the translation subgroup
$\RR^2$. The unitary representations of $\Spin(2)$ are one-dimensional
and indexed by a number $\lambda\in\half\ZZ$ called the
\emph{helicity}.  Since $CPT$ reverses the helicity, it may be
necessary to include both helicities $\pm\lambda$ in order to obtain a
$CPT$-self-conjugate representation.

Let's choose $p^\mu = (E,0,0,-E)$, with $E>0$.  Then the algebra of the 
supercharges
\begin{equation*}
  \left[\sQ_\alpha, \bar\sQ_{\dot\beta}\right] = 
  4 E
  \begin{pmatrix}
    0 & 0\\
    0 & 1
  \end{pmatrix}~.
\end{equation*}

\begin{exercise}
  Show that as a consequence of the above algebra, $\sQ_1=0$ in any
  unitary representation.
\end{exercise}

Let us now define $\sq \equiv (1/2\sqrt{E})\, \sQ_2$, in terms
of which the supersymmetry algebra becomes the fermionic oscillator
algebra:
\begin{equation*}
  \sq \, \sq^\dagger + \sq^\dagger \, \sq = 1~.
\end{equation*}
This algebra has a unique irreducible representation of dimension
$2$.  If $\ket{\Omega}$ is a state annihilated by $\sq$, then the
representation has as basis $\{\ket{\Omega},
\sq^\dagger\ket{\Omega}\}$.  Actually, $\ket{\Omega}$ carries quantum
numbers corresponding to the momentum $p$ and also to the helicity
$\lambda$, so that $\ket{\Omega} = \ket{p,\lambda}$.

\begin{exercise}
  Paying close attention to the helicity of the supersymmetry charges,
  prove that $\sq$ lowers the helicity by $\half$, and that
  $\sq^\dagger$ raises it by the same amount.  Deduce that the
  massless supersymmetry multiplet of helicity $\lambda$ contains two
  irreducible representations of the Poincaré algebra with helicities
  $\lambda$ and $\lambda+\half$.
\end{exercise}

For example, if we take $\lambda=0$, then we have two irreducible
representations of the Poincaré algebra with helicities $0$ and
$\half$.  This representation cannot be realised on its own in a
quantum field theory, because of the CPT invariance of quantum field
theories.  Since CPT changes the sign of the helicity, if a
representation with helicity $s$ appears, so will the representation
with helicity $-s$.  That means that representations which are not
CPT-self-conjugate appear in CPT-conjugate pairs.  The CPT-conjugate
representation to the one discussed at the head of this paragraph has
helicities $-\half$ and $0$.  Taking both representations into account
we find two states with helicity $0$ and one state each with helicity
$\pm\half$.  This is precisely the helicity content of the massless
Wess--Zumino model: the helicity $0$ states are the scalar and the
pseudoscalar fields and the states of helicities $\pm\half$ correspond
to the physical degrees of freedom of the spinor.

If instead we start with helicity $\lambda=\half$, then the
supermultiplet has helicities $\half$ and $1$ and the $CPT$-conjugate
supermultiplet has helicities $-1$ and $-\half$.  These are precisely
the helicities appearing in supersymmetric Yang--Mills.  The multiplet
in question is therefore called the \emph{gauge multiplet}.

Now take the $\lambda=\tfrac32$ supermultiplet and add its
$CPT$-conjugate.  In this way we obtain a $CPT$-self-conjugate
representation with helicities $-2,-\tfrac32,\tfrac32,2$.  This has
the degrees of freedom of a graviton (helicities $\pm 2$) and a
\emph{gravitino} (helicities $\pm \frac32$).  This multiplet is
realised field theoretically in supergravity, and not surprisingly it
is called the \emph{supergravity multiplet}.

\subsection{Massive representations}

Let us now discuss massive representations.  As shown in
\exref{ex:ltlgrp}, the little group for the momentum $p_\mu$ of a
massive particle is $\SU(2)$.  Its finite-dimensional irreducible
unitary representations are well-known: they are indexed by the
\emph{spin} $s$, where $2s$ is a non-negative integer, and have
dimension $2s+1$.

A massive particle can always be boosted to its rest frame, so that we
can choose a momentum $p^\mu = (m,0,0,0)$ with $m>0$.  The
supercharges now obey
\begin{equation*}
  \left[\sQ_\alpha, \bar\sQ_{\dot\beta}\right] = 
  2 m \begin{pmatrix}
    1 & 0\\
    0 & 1
\end{pmatrix}~.
\end{equation*}
Thus we can introduce $\sq_\alpha \equiv (1/\sqrt{2m}) \sQ_\alpha$, in
terms of which the supersymmetry algebra is now the algebra of two
identical fermionic oscillators:
\begin{equation}
  \sq_\alpha \, (\sq_\beta)^\dagger + (\sq_\beta)^\dagger\, \sq_\alpha
  = \delta_{\alpha\beta}~.
\end{equation}
This algebra has a unique irreducible representation of dimension
$4$ with basis
\begin{equation*}
\{\ket{\Omega}, (\sq_1)^\dagger \ket{\Omega}, (\sq_2)^\dagger
\ket{\Omega}, (\sq_1)^\dagger (\sq_2)^\dagger \ket{\Omega}\}~,
\end{equation*}
where $\ket{\Omega}$ is a nonzero state obeying
\begin{equation*}
  \sq_1 \ket{\Omega} = \sq_2 \ket{\Omega} = 0~.
\end{equation*}

However unlike the case of massless representations, $\ket{\Omega}$ is
now degenerate, since it carries spin: for spin $s$, $\ket{\Omega}$ is
really a $(2s+1)$-dimensional $\SU(2)$ multiplet.  Notice that
$(\sq_\alpha)^\dagger$ transforms as an $\SU(2)$-doublet of spin
$\half$.  This must be taken into account when determining the spin
content of the states in the supersymmetry multiplet.  Instead of
simply adding the helicities like in the massless case, now we must
use the Clebsch--Gordon series to add the spins.  On the other hand,
massive representations are automatically $CPT$-self-conjugate so we
don't have to worry about adding the CPT-conjugate representation.

For example, if we take $s=0$, then we find the following spectrum:
$\ket{p,0}$ with spin 0, $(\sq_\alpha)^\dagger\ket{p,0}$ with spin
$\half$ and $(\sq_1)^\dagger (\sq_2)^\dagger \ket{p,0}$ which has spin 0
too.  The field content described by this multiplet is then a scalar
field, a pseudo-scalar field, and a Majorana fermion, which is
precisely the field content of the Wess--Zumino model.  The multiplet
is known as the \emph{scalar} or \emph{Wess--Zumino multiplet}.

\begin{exercise}
  What is the spin content of the massive supermultiplet with
  $s=\half$?  What would be the field content of a theory admitting
  this representation of the Poincaré superalgebra?
\end{exercise}

All representations of the Poincaré superalgebra share the property
that the number of fermionic and bosonic states match.  For the
massless representations this is clear because the whatever the
Bose--Fermi parity of $\ket{p,\lambda}$, it is opposite that of
$\ket{p,\lambda+\half}$.

For the massive representations we see that whatever the Bose--Fermi
parity of the $2s+1$ states $\ket{p,s}$, it is opposite that of the 
$2(2s+1)$ states $(\sq_\alpha)^\dagger \ket{p,s}$ and the same as that 
of the $2s+1$ states $(\sq_1)^\dagger (\sq_2)^\dagger \ket{p,s}$.
Therefore there are $2(2s+1)$ bosonic and $2(2s+1)$ fermionic states.

\begin{tut}[\textsc{Supersymmetry and the BPS bound}]\indent\par
  \label{pr:extended}
  Here we introduce the extended Poincaré superalgebra and study its
  unitary representations.  In particular we will see the emergence of
  central charges, the fact that the mass of a unitary representation
  satisfies a bound, called the BPS bound, and that the sizes of
  representations depends on whether the bound is or is not saturated.
  
  The extended Poincaré superalgebra is the extension of the Poincaré
  algebra by $N$ supercharges $\sQ_I$ for $I=1,2,\dots,N$.  The
  nonzero brackets are now
  \begin{equation}
    \begin{aligned}
      \left[\sQ_{\alpha\,I},\sQ_{\beta\,J}\right] &= 2
      \epsilon_{\alpha\beta} \sZ_{IJ}\\
      \left[\sQ_{\alpha\,I},\bar\sQ_{\dot\alpha}^J\right] &= 2 i
      \delta_I^J\, (\sigma^\mu)_{\alpha\dot\alpha} \sP_\mu~,
  \end{aligned}
\end{equation}
  where $\sZ_{IJ}$ commute with all generators of the algebra and are
  therefore known as the \emph{central charges}.  Notice that
  $\sZ_{IJ} = - \sZ_{JI}$, whence central charges requires $N\geq2$.
  The hermiticity condition on the supercharges now says that
  \begin{equation*}
    \left(\sQ_{\alpha\, I}\right)^\dagger = \bar\sQ_{\dot\alpha}^I~.
  \end{equation*}
  
  We start by considering massless representations.  Choose a
  lightlike momentum $p^\mu = (E,0,0,-E)$ with $E>0$.  The
  supercharges obey
  \begin{equation*}
    [\sQ_{\alpha\, I}, (\sQ_{\beta\, J})^\dagger] = 4 E \delta_I^J 
    \begin{pmatrix}
      0 & 0 \\ 0 & 1
    \end{pmatrix}~.
  \end{equation*}
  
  \begin{enumerate}
  \item Prove that all $\sQ_{1\, I}$ must act trivially on any unitary
    representation, and conclude that the central charges must vanish
    for massless unitary representations.
  \item Consider a massless multiplet with lowest helicity $\lambda$.
    Which helicities appear and with what multiplicities?
  \item Prove that $CPT$-self-conjugate multiplets exist only for even
    $N$.  Discuss the $CPT$-self-conjugate multiplets for $N=2$, $N=4$
    and $N=8$.  These are respectively the $N{=}2$
    \emph{hypermultiplet}, the $N{=}4$ \emph{gauge multiplet} and
    $N{=}8$ \emph{supergravity multiplet}.
  \end{enumerate}
  
  Now we consider massive representations without central charges.
  The situation is very similar to the $N{=}1$ case discussed in
  lecture.
  
  \begin{enumerate}
  \item[4.] Work out the massive $N{=}2$ multiplets without central
    charges and with spin $s{=}0$ and $s{=}\half$.  Show that for
    $s{=}0$ the spin content is $(0^5, \half^4, 1)$ in the obvious
    notation, and for $s{=}\half$ it is given by $(\tfrac32, 1^4,
    \half^6,0^4)$.
  \end{enumerate}
  
  Now consider massive $N{=}2$ multiplets with central charges.  In
  this case $\sZ_{IJ} = \sz \epsilon_{IJ}$, where there is only one
  central charge $\sz$.  Since $\sz$ is central, it acts as a multiple
  of the identity, say $z$, in any irreducible representation.  The
  algebra of supercharges is now:
  \begin{equation*}
    \begin{aligned}
      \left[\sQ_{\alpha\, I}, \sQ_{\alpha\, J}\right] &= z\,
      \epsilon_{IJ}
      \begin{pmatrix}
        0 & 1\\ -1 & 0
      \end{pmatrix}\\
      \left[\sQ_{\alpha\, I}, (\sQ_{\alpha\, J})^\dagger\right] &=
      2m\, \delta_I^J
      \begin{pmatrix}
        1 & 0\\ 0 & 1
      \end{pmatrix}~.
    \end{aligned}
  \end{equation*}
  
  \begin{enumerate}
  \item[5.] Show that for a unitary massive representation of mass
    $m$, the following bound is always satisfied: $m \geq |z|$.
    \textsl{(Hint: Consider the algebra satisfied by the linear
      combination of supercharges $\sQ_{\alpha\, 1} \pm
      \bar\epsilon^{\dot\alpha\dot\beta} (\sQ_{\beta\, 2})^\dagger$.)}
  \item[6.] Show that representations where the bound is not
    saturated---that is, $m > |z|$---have the same multiplicities as
    massive representations without central charge.
  \item[7.] Show that massive representations where the bound is
    saturated have the same multiplicities as massless
    representations.
  \end{enumerate}
  \begin{scholium}
    The bound in Part~5 above is called the \emph{BPS bound} since it
    generalises the Bogomol'nyi bound for the Prasad--Sommerfield
    limit of Yang--Mills--Higgs theory.  In fact, in the context of
    $N{=}2$ supersymmetric Yang--Mills it is \emph{precisely} the
    Bogomol'nyi bound.
    
    The result in Part~7 above explains why BPS saturated multiplets
    are also called \emph{short multiplets}.  The difference in
    multiplicity between ordinary massive multiplets and those which
    are BPS saturated underlies the rigidity of the BPS-saturated
    condition under deformation: either under quantum corrections or
    under other continuous changes in the parameters of the model.
  \end{scholium}
\end{tut}


\section{Superspace and Superfields}
\label{lec:superspace}

In the previous lectures we have studied the representations of the
Poincaré superalgebra and we have seen some of its field theoretical
realisations.  In both the Wess--Zumino model and supersymmetric
Yang--Mills, proving the supersymmetry of the action was a rather
tedious task, and moreover the superalgebra was only realised on-shell
and, in the case of supersymmetric Yang--Mills, up to gauge
transformations.

It would be nice to have a formulation in which supersymmetry was
manifest, just like Poincaré invariance is in usual relativistic field
theories.  Such theories must have in addition to the physical fields,
so-called auxiliary fields in just the right number to reach the
balance between bosonic and fermionic fields which supersymmetry
demands.  For example, in the Wess--Zumino model this balance is
present on the physical degree of freedoms: $2$ bosonic and $2$
fermionic.  In order to have a manifestly supersymmetric formulation
this balance in the degrees of freedom must be present without the
need to go on-shell.  For example, in the Wess--Zumino model, the
bosons are defined by $2$ real functions $S$ and $P$, whereas the
fermions are defined by $4$: $\psi^a$.  We conclude therefore that a
manifestly supersymmetric formulation must contain at least two
additional bosonic fields.  The superfield formulation will do just
that.

Superfields are fields in superspace, and superspace is to the
Poincaré superalgebra what Minkowski space is to the Poincaré algebra.
Just like we can easily write down manifestly Poincaré invariant
models as theories of fields on Minkowski space, we will be able to
(almost) effortlessly write down models invariant under the Poincaré
superalgebra as theories of superfields in superspace.

In this lecture we will introduce the notions of superspace and
superfields.  We will discuss the scalar superfields and will rewrite
the Wess--Zumino model in superspace.  Unpacking the superspace
action, we will recover a version of this model with the requisite
number of auxiliary fields for the off-shell closure of the Poincaré
superalgebra.  The auxiliary fields are essential not only in the
manifestly supersymmetric formulation of field theories but, as we
will see in Lecture~\ref{lec:SB}, also play an important role in the
breaking of supersymmetry.

\subsection{Superspace}

For our purposes the most important characteristic of Minkowski space
is that, as discussed in the Appendix, it is acted upon transitively
by the Poincaré group.  We would now like to do something similar with
the ``Lie supergroup'' corresponding to the Poincaré superalgebra.

\begin{caveat}
  We will not give the precise mathematical definition of a Lie
  supergroup in these lectures.  Morally speaking a Lie supergroup is
  what one obtains by exponentiating elements of a Lie superalgebra.
  We will formally work with exponentials of elements of the
  superalgebra keeping in mind that the parameters associated to odd
  elements are themselves anticommuting.
\end{caveat}

By analogy with the treatment of Minkowski space in the Appendix, we
will define Minkowski superspace (or superspace for short) as the
space of right cosets of the Lorentz group.  Notice that the Poincaré
superalgebra has the structure of a semidirect product, just like the
Poincaré algebra, where the translation algebra is replaced by the
superalgebra generated by $\sP_\mu$ and $\sQ_a$.  Points in superspace
are then in one-to-one correspondence with elements of the Poincaré
supergroup of the form
\begin{equation*}
  \exp(x^\mu \sP_\mu) \exp(\bar\theta \sQ)~,
\end{equation*}
where $\theta$ is an anticommuting Majorana spinor and $\bar\theta\sQ
= -\theta^a \sQ_a$ as usual.

The Poincaré group acts on superspace by left multiplication with the
relevant group element.  However as we discussed in the Appendix, this
action generates an antirepresentation of the Poincaré superalgebra.
In order to generate a representation of the Poincaré superalgebra we
must therefore start with the opposite superalgebra---the superalgebra
where all brackets are multiplied by $-1$.  In the case of the
Poincaré superalgebra, the relevant brackets are now
\begin{equation}
  \label{eq:opposite}
  \begin{aligned}
    \left[\sP_\mu, \sQ_a\right] &= 0\\
    \left[\sQ_a, \sQ_b\right] &= - 2 \left(\gamma^\mu\right)_{ab}
    \sP_\mu~.
\end{aligned}
\end{equation}

Translations act as expected:
\begin{equation*}
  \exp(\tau^\mu \sP_\mu) \exp(x^\mu \sP_\mu) \exp(\bar\theta \sQ) = 
  \exp\left((x^\mu+\tau^\mu) \sP_\mu\right) \exp(\bar\theta \sQ)~,
\end{equation*}
so that the point $(x,\theta)$ gets sent to the point $(x+\tau,
\theta)$.

The action of the Lorentz group is also as expected: $x^\mu$
transforms as a vector and $\theta$ as a Majorana spinor.  In
particular, Lorentz transformations do not mix the coordinates.

On the other hand, the noncommutativity of the superalgebra generated
by $\sP_\mu$ and $\sQ_a$ has as a consequence that a supertranslation
does not just shift $\theta$ but also $x$, as the next exercise asks
you to show.  This is the reason why supersymmetry mixes bosonic and
fermionic fields.

\begin{exercise}
  With the help of the Baker--Campbell--Hausdorff formula
  \eqref{eq:BCH}, show that
  \begin{equation*}
    \exp(\bar\varepsilon \sQ) \exp(\bar\theta \sQ) = 
    \exp(-\bar\varepsilon\gamma^\mu\theta \sP_\mu)
    \exp\left((\overline{\theta+\varepsilon}) \sQ\right)~.
  \end{equation*}
  \label{ex:BCH}
\end{exercise}

It follows that the action of a supertranslation on the point $(x^\mu, 
\theta)$ is given by $(x^\mu - \bar\varepsilon\gamma^\mu\theta, \theta +
\varepsilon)$.

\begin{caveat}
  We speak of points in superspace, but in fact, as in noncommutative
  geometry, of which superspace is an example (albeit a mild one), one
  is supposed to think of $x$ and $\theta$ are coordinate functions.
  There are no points corresponding to $\theta$, but rather nilpotent
  elements in the (noncommutative) algebra of functions.  For
  simplicity of exposition we will continue to talk of $(x,\theta)$ as
  a point, although it is good to keep in mind that this is an
  oversimplification.  Doing so will avoid ``koans'' like
  \begin{center}
    \emph{What is the point with coordinates $x^\mu -
    \bar\varepsilon\gamma^\mu\theta$?}
  \end{center}
  This question has no answer because whereas (for fixed $\mu$)
  $x^\mu$ is an ordinary function assigning a real number to each
  point, the object $x^\mu - \bar\varepsilon\gamma^\mu\theta$ is quite
  different, since $\bar\varepsilon\gamma^\mu\theta$ is certainly not
  a number.  What it is, is an even element in the ``coordinate ring''
  of the superspace, which is now a Grassmann algebra: with generators
  $\theta$ and $\bar\theta$ and coefficients which are honest
  functions of $x^\mu$.  This is to be understood in the sense of
  noncommutative geometry, as we now briefly explain.
  
  Noncommutative geometry starts from the observation that in many
  cases the (commutative) algebra of functions of a space determines
  the space itself, and moreover that many of the standard geometric
  concepts with which we are familiar, can be rephrased purely in
  terms of the algebra of functions, without ever mentioning the
  notion of a point.  (This is what von~Neumann called ``pointless
  geometry''.)  In noncommutative geometry one simply starts with a
  noncommutative algebra and interprets it as the algebra of functions
  on a ``noncommutative space.''  Of course, this space does not
  really exist.  Any question for which this formalism is appropriate
  should be answerable purely in terms of the noncommutative algebra.
  Luckily this is the case for those applications of this formalism to
  supersymmetry with which we are concerned in these lectures.
  
  In the case of superspace, the noncommutativity is mild.  There are
  commuting coordinates, the $x^\mu$, but also (mildly) noncommuting
  coordinates $\theta$ and $\bar\theta$.  More importantly, these
  coordinates are nilpotent: big enough powers of them vanish.  In
  some sense, superspace consists of ordinary Minkowski space with
  some ``nilpotent fuzz'' around each point.
\end{caveat}

\subsection{Superfields}

A superfield $\Phi(x,\theta)$ is by definition a (differentiable)
function of $x$ and $\theta$.  By linearising the geometric action on
points, and recalling that the action on functions is inverse to that
on points, we can work out the infinitesimal actions of $\sP_\mu$ and
$\sQ_a$ on superfields:
\begin{equation}
  \label{eq:derivations}
  \begin{aligned}
    \sP_\mu \cdot \Phi &= -\d_\mu \Phi\\
    \sQ_a \cdot \Phi &= \left( \d_a + (\gamma^\mu)_{ab}\theta^b \d_\mu
    \right) \Phi~,
\end{aligned}
\end{equation}
where by definition $\d_a \theta^b = \delta_a^b$.

\begin{exercise}
  Verify that the above derivations satisfy the opposite superalgebra
  \eqref{eq:opposite}.
\end{exercise}

Since both $\sP_\mu$ and $\sQ_a$ act as derivations, they obey the
Leibniz rule and hence products of superfields transform under
(super)translations in the same way as a single superfield.  Indeed,
if $f$ is any differentiable function, $f(\Phi)$ transforms under
$\sP_\mu$ and $\sQ_a$ as in equation \eqref{eq:derivations}.
Similarly, if $\Phi^i$ for $i=1,2,\dots, n$ transform as in
\eqref{eq:derivations}, so will any differentiable function
$f(\Phi^i)$.

The derivations $-\d_\mu$ and $Q_a := \d_a + (\gamma^\mu)_{ab}\theta^b
\d_\mu$ are the vector fields generating the infinitesimal \emph{left}
action of the Poincaré supergroup.  The infinitesimal \emph{right}
action is also generated by vector fields which, because left and
right multiplications commute, will (anti)commute with them.  Since
ordinary translations commute, right translations are also generated
by $-\d_\mu$.  On the other hand, the noncommutativity of the
supertranslations means that the expression for the right action of
$\sQ_a$ is different.  In fact, from \exref{ex:BCH} we read off
\begin{equation*}
  \exp(\bar\theta \sQ) \exp(\bar\varepsilon \sQ) = 
    \exp(+\bar\varepsilon\gamma^\mu\theta \sP_\mu)
    \exp\left((\overline{\theta+\varepsilon}) \sQ\right)~,
\end{equation*}
whence the infinitesimal generator (on superfields) is given by the
\emph{supercovariant derivative}
\begin{equation*}
  D_a := \d_a - (\gamma^\mu)_{ab}\theta^b \d_\mu~.
\end{equation*}

\begin{exercise}
  Verify that the derivations $Q_a$ and $D_a$ anticommute and that
  \begin{equation}
    \label{eq:Dalgebra}
    [D_a, D_b] = -2 (\gamma^\mu)_{ab} \d_\mu~.
  \end{equation}
\end{exercise}

We are almost ready to construct supersymmetric lagrangians.  Recall
that a lagrangian $\eL$ is supersymmetric if it is Poincaré
invariant and such that its supersymmetric variation is a total
derivative:
\begin{equation*}
  \delta_\varepsilon \eL = \d_\mu \left( \bar\varepsilon K^\mu
  \right)~.
\end{equation*}
It is very easy to construct supersymmetric lagrangians using
superfields.

To explain this let us make several crucial observations.  First of
all notice that because the odd coordinates $\theta$ are
anticommuting, the dependence on $\theta$ is at most polynomial, and
because $\theta$ has four real components, the degree of the
polynomial is at most $4$.

\begin{exercise}
  Show that a superfield $\Phi(x,\theta)$ has the following
  $\theta$-expansion
  \begin{multline*}
    \Phi(x,\theta) = \phi(x) + \bar\theta \chi(x) + \bar\theta\theta\, 
    F(x) + \bar\theta\gamma_5\theta\, G(x)\\
    + \bar\theta \gamma^\mu\gamma_5\theta\,  v_\mu(x) +
      \bar\theta\theta\, \bar\theta \xi(x) + \bar\theta\theta\,
      \bar\theta\theta\, E(x)~,
  \end{multline*}
  where $\phi$, $E$, $F$, $G$, $v_\mu$, $\chi$ and $\xi$ are fields in
  Minkowski space.\\
  {\rm (Hint: You may want to use the Fierz-like identities
  \eqref{eq:thetapowers}.)}
  \label{ex:thetaexp4}
\end{exercise}

Now let $L(x,\theta)$ be any Lorentz-invariant function of $x$ and
$\theta$ which transforms under supertranslations according to
equation \eqref{eq:derivations}.  For example, any function built out
of superfields, their derivatives and their supercovariant derivatives
transforms according to equation \eqref{eq:derivations}.  The next
exercise asks you to show that under a supertranslation, the component
of $L$ with the highest power of $\theta$ transforms into a total
derivative.  Its integral is therefore invariant under
supertranslations, Lorentz invariant (since $L$ is) and, by the
Poincaré superalgebra, also invariant under translations.  In other
words, it is invariant under supersymmetry!

\begin{exercise}
  Let $\Phi(x,\theta)$ be a superfield and let $E(x)$ be its
  $(\bar\theta\theta)^2$ component, as in \exref{ex:thetaexp4}.  Show
  that $E(x)$ transforms into a total derivative under
  supertranslations:
  \begin{equation*}
    \delta_\varepsilon E = \d_\mu \left( -\tfrac14 \bar\varepsilon
      \gamma^\mu \xi\right)~.
  \end{equation*}
  {\rm (Hint: As in \exref{ex:thetaexp4}, you may want to use the
  identities \eqref{eq:thetapowers}.)}
\end{exercise}

We will see how this works in practice in two examples: the
Wess--Zumino model presently and in the next lecture the case of
supersymmetric Yang--Mills.

\subsection{Superfields in two-component formalism}

The cleanest superspace formulation of the Wess--Zumino model requires
us to describe superspace in terms of two-component spinors.  Since
$\theta$ is a Majorana spinor, it can be written as $\theta^a =
(\theta^\alpha, \bar\theta_{\dot\alpha})$.  Taking into account
equation \eqref{eq:4to2innerproduct}, a point in superspace can be
written as
\begin{equation*}
  \exp(x^\mu \sP_\mu) \, \exp\left(-(\theta\sQ +
  \bar\theta\bar\sQ)\right)~.
\end{equation*}

The two-component version of the opposite superalgebra
\eqref{eq:opposite} is now
\begin{equation}
  \label{eq:opposite2}
  \left[\sQ_\alpha, \bar\sQ_{\dot\beta}\right] = - 2 i
  (\sigma^\mu)_{\alpha\dot\beta} \sP_\mu~,
\end{equation}
with all other brackets vanishing.

\begin{exercise}
  Show that under left multiplication by $\exp(\varepsilon\sQ)$ the
  point $(x^\mu,\theta,\bar\theta)$ gets sent to the point $(x^\mu - i 
  \varepsilon \sigma^\mu \bar\theta, \theta - \varepsilon,
  \bar\theta)$.  Similarly, show that under left multiplication by
  $\exp(\bar\varepsilon\bar\sQ)$, $(x^\mu,\theta,\bar\theta)$ gets
  sent to $(x^\mu - i \bar\varepsilon \bar\sigma^\mu \theta, \theta,
  \bar\theta - \bar\varepsilon)$.
\end{exercise}

This means that action on superfields (recall that the action on
functions is inverse to that on points) is generated by the following
derivations:
\begin{equation}
  Q_\alpha = \d_\alpha + i (\sigma^\mu)_{\alpha\dot\beta}
  \bar\theta^{\dot\beta} \d_\mu \qquad\text{and}\qquad
  \bar Q_{\dot\alpha} = \bar\d_{\dot\alpha} + i
  (\bar\sigma^\mu)_{\dot\alpha\beta} \theta^\beta\d_\mu~.
\end{equation}

Repeating this for the right action, we find the following expressions 
for the supercovariant derivatives:
\begin{equation}
  D_\alpha = \d_\alpha - i (\sigma^\mu)_{\alpha\dot\beta}
  \bar\theta^{\dot\beta} \d_\mu \qquad\text{and}\qquad
  \bar D_{\dot\alpha}= \bar\d_{\dot\alpha} - i
  (\bar\sigma^\mu)_{\dot\alpha\beta} \theta^\beta\d_\mu~.
\end{equation}

\begin{exercise}
  Verify that $\bar Q_{\dot \alpha} = (Q_\alpha)^*$ and $\bar
  D_{\dot\alpha} = (D_\alpha)^*$.  Also show that any of $Q_\alpha$
  and $\bar Q_{\dot\alpha}$ anticommute with any of $D_\alpha$ and
  $\bar D_{\dot\alpha}$, and that they obey the following brackets:
  \begin{equation}
    [Q_\alpha,\bar Q_{\dot\beta}] = + 2 i
    \sigma^\mu_{\alpha\dot\beta}\d_\mu\qquad\text{and}\qquad
    [D_\alpha,\bar D_{\dot\beta}] = - 2 i
    \sigma^\mu_{\alpha\dot\beta}\d_\mu~.
  \end{equation}
\end{exercise}

\subsection{Chiral superfields}

Let $\Phi(x,\theta,\bar\theta)$ be a complex superfield.  Expanding it
as a series in $\theta$ we obtain
\begin{multline}
  \label{eq:scalarsuperfield}
  \Phi(x,\theta,\bar\theta) = \phi(x) + \theta\chi(x) +
  \bar\theta\bar\chi'(x) + \bar\theta\bar\sigma^\mu\theta v_\mu(x)\\
  + \theta^2 F(x) + \bar\theta^2 \bar F'(x) + \bar\theta^2 \theta
    \xi(x) + \theta^2  \bar\theta\bar\xi'(x)  + \theta^2
    \bar\theta^2 D(x)~,
\end{multline}
where $\phi$, $\chi$, $\bar\chi'$, $v_\mu$, $\xi$, $\bar\xi'$, $F$,
$F'$ and $D$ are all different complex fields.

Therefore an unconstrained superfield $\Phi$ gives rise to a large
number of component fields.  Taking $\phi$, the lowest component of
the superfield, to be a complex scalar we see that the superfield
contains too many component fields for it to yield an irreducible
representation of the Poincaré superalgebra.  Therefore we need to
impose constraints on the superfield in such a way as to cut down the
size of the representation.  We now discuss one such constraint and in 
the following lecture will discuss another.

Let us define a \emph{chiral superfield} as a superfield $\Phi$ which
satisfies the condition
\begin{equation}
  \label{eq:chiral}
  \bar D_{\dot\alpha} \Phi = 0~.
\end{equation}
Similarly we define an \emph{antichiral superfield} as one satisfying
\begin{equation}
  \label{eq:antichiral}
  D_\alpha \Phi = 0~.
\end{equation}
Chiral superfields behave very much like holomorphic functions.
Indeed, notice that a real (anti)chiral superfield is necessarily
constant.  Indeed, the complex conjugate of a chiral field is
antichiral.  If $\Phi$ is real and chiral, then it also antichiral,
whence it is annihilated by both $D_\alpha$ and $\bar D_{\dot\alpha}$
and hence by their anticommutator, which is essentially $\d_\mu$,
whence we would conclude that $\Phi$ is constant.

It is very easy to solve for the most general (anti)chiral
superfield.  Indeed, notice that the supercovariant derivatives admit
the following operatorial decompositions
\begin{equation}
  \label{eq:twist}
  D_\alpha = e^{i U} \d_\alpha e^{-i U} \qquad\text{and}\qquad
  \bar D_{\dot\alpha} = e^{-i U} \bar\d_{\dot\alpha} e^{i U}~,
\end{equation}
where $U = \theta \sigma^\mu \bar \theta \d_\mu$ is real.

\begin{exercise}
  Use this result to prove that the most general chiral superfield
  takes the form
  \begin{equation*}
    \Phi(x,\theta,\bar\theta) = \phi(y) + \theta\chi(y) + \theta^2
    F(y)~,
  \end{equation*}
  where $y^\mu = x^\mu - i \theta\sigma^\mu\bar\theta$.  Expand this
  to obtain
  \begin{multline}
    \label{eq:chiralcomps}
    \Phi(x,\theta,\bar\theta) = \phi(x) + \theta\chi(x)
    + \theta^2 F(x) + i \bar\theta\bar\sigma^\mu\theta \d_\mu\phi(x)\\
    - \tfrac{i}{2} \theta^2 \bar\theta \bar\sigma^\mu \d_\mu\chi(x) +
    \tfrac14 \theta^2 \bar\theta^2 \dalem\phi(x)~.
  \end{multline}
  \label{ex:chiral}
\end{exercise}

It is possible to project out the different component fields in a
chiral superfields by taking derivatives.  One can think of this as
Taylor expansions in superspace.

\begin{exercise}
  Let $\Phi$ be a chiral superfield.  Show that
  \begin{equation*}
    \begin{aligned}
      \phi(x) &= \Phi\bigr|\\
      \chi_\alpha(x) &= D_\alpha \Phi\bigr|\\
      F(x) &= -\tfrac14 D^2 \Phi\bigr|~,
  \end{aligned}
\end{equation*}
  where $D^2 = D^\alpha D_\alpha$ and where $\bigr|$ denotes the
  operation of setting $\theta=\bar\theta=0$ in the resulting
  expressions.
  \label{ex:chiralprojectors}
\end{exercise}

\subsection{The Wess--Zumino model revisited}

We will now recover the Wess--Zumino model in superspace.  The
lagrangian couldn't be simpler.

Let $\Phi$ be a chiral superfield.  Its dimension is equal to that of
its lowest component $\Phi\bigr|$, which in this case, being a complex
scalar, has dimension $1$.

Since $\theta$ has dimension $-\half$, the highest component of any
superfield (the coefficient of $\theta^2\bar\theta^2$) has dimension 
two more than that of the superfield.  Therefore if we want to build a 
lagrangian out of $\Phi$ we need to take a quadratic expression.
Since $\Phi$ is complex and but the action should be real, we have
essentially one choice: $\bar\Phi\Phi$, where $\bar\Phi = (\Phi)^*$.
The highest component of $\bar\Phi\Phi$ is real, has dimension $4$, is
Poincaré invariant and transforms into a total derivative under
supersymmetry.  It therefore has all the right properties to be a
supersymmetric lagrangian.

\begin{exercise}
  Let $\Phi$ be a chiral superfield and let $\bar\Phi = (\Phi)^*$ be
  its (antichiral) complex conjugate.  Show that the highest component
  of $\bar\Phi\Phi$ is given by
  \begin{equation*}
    - \d_\mu\phi \d^\mu\bar\phi + F \bar F + \tfrac{i}{4} \left( \chi
      \sigma^\mu \d_\mu\bar\chi + \bar\chi \bar\sigma^\mu
      \d_\mu\chi\right) + \tfrac14 \d_\mu \left( \phi 
    \d^\mu\bar\phi + \bar\phi\d^\mu\phi \right)~.
  \end{equation*}
  Rewrite the lagrangian of the free massless Wess--Zumino model
  (given in \eqref{eq:WZkin}) in terms of two-component spinors and
  show that it agrees (up to total derivatives and after using the
  equation of motion of $F$) with $2\bar\Phi\Phi$ where $\phi =
  \half(S + i P)$ and $\psi^a=(\chi^\alpha, \bar\chi_{\dot\alpha})$.
  \label{ex:phibarphicomps}
\end{exercise}

\begin{scholium}
  A complex scalar field is not really a scalar field in the strict
  sense.  Because of CPT-invariance, changing the orientation in
  Minkowski space complex conjugates the complex scalar.  This means
  that the real part is indeed a scalar, but that the imaginary part
  is a pseudoscalar.  This is consistent with the identification $\phi
  = \half (S + i P)$ in the above exercise.
\end{scholium}

Let us now recover the supersymmetry transformations of the component
fields from superspace.  By definition, $\delta_\varepsilon \Phi =
-(\varepsilon Q + \bar\varepsilon \bar Q) \Phi$.  In computing the
action of $Q_\alpha$ and $\bar Q_{\dot\alpha}$ on a chiral superfield
$\Phi$, it is perhaps easier to write $\Phi$ as
\begin{equation*}
  \Phi = e^{-i U} \left( \phi + \theta\chi + \theta^2 F\right)~,
\end{equation*}
and the supercharges as
\begin{equation*}
  Q_\alpha = e^{-i U} \d_\alpha e^{i U} \qquad\text{and}\qquad
  \bar Q_{\dot\alpha} = \bar D_{\dot\alpha} + 2 i
  (\bar\sigma^\mu)_{\dot\alpha\beta}\theta^\beta\d_\mu~,
\end{equation*}
with $U = \theta\sigma^\mu\bar\theta \d_\mu$.

\begin{exercise}
  Doing so, or the hard way, show that
  \begin{equation}
    \label{eq:susychiral}
    \begin{aligned}
      \delta_\varepsilon \phi &= - \varepsilon\chi\\
      \delta_\varepsilon \chi_\alpha &= -2 \varepsilon_\alpha F + 2 i
      \bar\varepsilon^{\dot\alpha} (\bar\sigma^\mu)_{\dot\alpha\alpha}
      \d_\mu\phi\\
      \delta_\varepsilon F &= i \bar\varepsilon \bar\sigma^\mu \d_\mu
      \chi~.
  \end{aligned}
\end{equation}
  Now rewrite the supersymmetry transformations \eqref{eq:WZkinsusy}
  of the free massless Wess--Zumino model in terms of two-component
  spinors and show that they agree with the ones above after using the 
  $F$ equations of motion and under the identification $\phi=\half(S + 
  i P)$ and $\psi^a=(\chi^\alpha, \bar\chi_{\dot\alpha})$.
\end{exercise}

The above result illustrates why in the formulation of the
Wess--Zumino model seen in Lecture~\ref{lec:WZmodel}, the Poincaré
superalgebra only closes on-shell.  In that formulation the auxiliary
field $F$ has been eliminated using its equation of motion $F=0$.
However for this to be consistent, its variation under supersymmetry
has to vanish as well, and as we have just seen $F$ varies into the
equation of motion of the fermion.

Let us introduce the following notation:
\begin{equation*}
 \int d^2\theta d^2\bar\theta \leftrightarrow \text{the coefficient of 
 $\theta^2\bar\theta^2$.}
\end{equation*}

\begin{caveat}
  The notation is supposed to be suggestive of integration in
  superspace.  Of course this integral is purely formal and has not
  measure-theoretic content.  It is an instance of the familiar
  Berezin integral in the path integral formulation of theories with
  fermions; only that in this case the definition is not given in this
  way, since the Grassmann algebra in quantum field theory has to be
  infinitely generated so that correlation functions of an arbitrary
  number of fermions are not automatically zero.  Therefore it makes
  no sense to extract the ``top'' component of an element of the
  Grassmann algebra.
\end{caveat}

In this notation, the (free, massless) Wess--Zumino model is described
by the following action:
\begin{equation}
  \label{eq:WZkinsuper}
   \int d^4 x d^2\theta d^2\bar\theta\, 2 \bar\Phi \Phi~.
\end{equation}

A convenient way to compute superspace integrals of functions of
chiral superfields is to notice that
\begin{equation}
   \int d^4 x d^2\theta d^2\bar\theta K(\Phi,\bar\Phi) =  \int d^4 x
   \tfrac{1}{16} D^2 \bar D^2 K(\Phi,\bar\Phi)\bigr|~.
\end{equation}
This is true even if $\Phi$ is not a chiral superfield, but it becomes
particularly useful if it is, since we can use chirality and
\exref{ex:chiralprojectors} to greatly simplify the computations.

\begin{exercise}
  Take $K(\Phi,\bar\Phi) = \bar\Phi\Phi$ and, using the above
  expression for $\int d^4 x d^2 \theta d^2\bar\theta
  K(\Phi,\bar\Phi)$, rederive the result in the first part of
  \exref{ex:phibarphicomps}.
\end{exercise}

\begin{scholium}
  In Problem~\ref{pr:superconformalWZ} we saw that the free massless
  Wess--Zumino model is invariant under the R-symmetry
  \eqref{eq:Rsymmetry}.  This symmetry can also be realised
  geometrically in superspace.  Notice that the infinitesimal
  R-symmetry acts on the component fields of the superfield as
  \begin{equation*}
    \sR \cdot \phi = i \phi \qquad \sR \cdot \chi = - \tfrac{i}{2} \chi 
    \qquad\text{and}\qquad \sR \cdot \bar\chi = \tfrac{i}{2} \bar\chi~.
  \end{equation*}
  Since $\phi = \Phi|$ we are forced to set $\sR \cdot \Phi = i \Phi$,
  which is consistent with the R-symmetry transformation properties of
  the fermions provided that $\theta$ and $\bar\theta$ transform
  according to
  \begin{equation}
    \label{eq:Rsymmetrysuperspace}
    \sR \cdot \theta = \tfrac{3i}{2} \theta \qquad\text{and}\qquad \sR
    \cdot \bar\theta = - \tfrac{3i}{2} \bar\theta~.
  \end{equation}
  This forces the superspace ``measures'' $d^2\theta$ and
  $d^2\bar\theta$ to transform as well:
  \begin{equation}
    \label{eq:Rsymmetrymeasure}
    \sR \cdot d^2\theta = -3i d^2 \theta \qquad\text{and}\qquad \sR
    \cdot d^2 \bar\theta = 3i d^2\bar\theta~,
  \end{equation}
  and this shows that the lagrangian $\int d^2\theta
  d^2\bar\theta \Phi\bar\Phi$ is manifestly invariant under the
  R-symmetry.
\end{scholium}

\subsection{The superpotential}

We now add masses and interactions to the theory with superspace
lagrangian $\bar\Phi\Phi$.

The observation that allows us to do this is the following.  It
follows from the supersymmetry transformation properties
\eqref{eq:susychiral} of a chiral superfield, that its $\theta^2$
component transforms as a total derivative.  Now suppose that $\Phi$
is a chiral superfield.  Then so is any power of $\Phi$ and in fact
any differentiable function $W(\Phi)$.  Therefore the $\theta^2$
component of $W(\Phi)$ is supersymmetric.  However it is not real, so
we take its real part.  The function $W(\Phi)$ is called the
\emph{superpotential}.  In the case of the Wess--Zumino model it is
enough to take $W$ to be a cubic polynomial.  In fact, on dimensional
grounds, a renormalisable superpotential is at most cubic.  This
follows because the $\theta^2$ component of $W(\Phi)$ has dimension
$1$ more than that of $W(\Phi)$.  Since the dimension of a lagrangian
term must be at most four, the dimension of $W(\Phi)$ must be at most
three.  Since $\Phi$ has dimension $1$ and renormalisability does not allow
coupling constants of negative dimension, we see that $W(\Phi)$ must
be at most cubic.

Let us introduce the notation
\begin{gather*}
 \int d^2\theta \leftrightarrow \text{the coefficient of $\theta^2$}\\
 \int d^2\bar\theta \leftrightarrow \text{the coefficient of $\bar\theta^2$,}
\end{gather*}
with the same caveat about superspace integration as before.  A
convenient way to compute such chiral superspace integrals is again to 
notice that
\begin{equation}
  \label{eq:intisder}
  \begin{aligned}
    \int d^4 x d^2\theta\, W(\Phi) &= - \int d^4 x\, \tfrac14 D^2
    W(\Phi)\bigr|~,\\
    \int d^4 x d^2\bar\theta\, \overline W(\bar\Phi) &= - \int d^4 x\, 
    \tfrac14 \bar D^2 \overline W(\bar\Phi)\bigr|~.
 \end{aligned}
\end{equation}

\begin{exercise}
  Let $W(\Phi)$ be given by
  \begin{equation*}
    W(\Phi) = \mu \Phi^2 + \nu \Phi^3~.
  \end{equation*}
  Determine $\mu$ and $\nu$ in such a way that the action obtained by
  adding to the action \eqref{eq:WZkinsuper} the superpotential term
  \begin{equation}
    \label{eq:superpotential}
    \int d^2\theta\, W(\Phi) +  \int d^2\bar\theta\,
    \overline{W(\Phi)}
  \end{equation}
  and eliminating the auxiliary field via its equation of motion we
  recover the Wess--Zumino model, under the identification
  $\phi=\half(S+i P)$ and $\psi^a=(\chi^\alpha,
  \bar\chi_{\dot\alpha})$.\\
  {\rm (Hint: I get $\mu = m$ and $\nu = \tfrac43 \lambda$.)}
\end{exercise}

\begin{scholium}
  R-symmetry can help put constraints in the superpotential.  Notice
  that the R-symmetry transformation properties of the superspace
  measures $d^2\theta$ and $d^2\bar\theta$ in
  \eqref{eq:Rsymmetrymeasure} says that an R-invariant superpotential
  must transform as $\sR \cdot W(\Phi) = 3 i W(\Phi)$.  This means
  that only the cubic term is invariant and in particular that the
  model must be massless.  This is consistent with the results of
  \probref{pr:superconformalWZ}: the conformal superalgebra contains
  the R-symmetry, yet it is not a symmetry of the model unless the
  mass is set to zero.
  
  It is nevertheless possible to redefine the action of the R-symmetry
  on the fields in such a way that the mass terms are R-invariant.
  For example, we could take $\sR \cdot \Phi = \frac{3i}{2} \Phi$, but
  this then prohibits the cubic term in the superpotential and renders
  the theory free.  Of course the massive theory, even if free, is not 
  (super)conformal invariant.
\end{scholium}

In other words, we see that the Wess--Zumino model described by the
action \eqref{eq:WZmodel} can be succinctly written in superspace as
\begin{equation}
  \label{eq:WZmodelSuperspace}
  \int d^4 x\, d^2\theta\, d^2\bar\theta\,  2 \bar\Phi\Phi +
  \left[\int d^4 x\, d^2\theta \left( m \Phi^2 + \tfrac43 \lambda
  \Phi^3 \right) + \text{c.c.}\right]~.
\end{equation}

Using equation \eqref{eq:intisder} and \exref{ex:chiralprojectors} it
is very easy to read off the contribution of the superpotential to the 
the lagrangian:
\begin{equation*}
  \frac{d W(\phi)}{d\phi} F - \tfrac14 \frac{d^2 W(\phi)}{d\phi^2}
  \chi\chi + \text{c.c}~,
\end{equation*}
and hence immediately obtain the Yukawa couplings and the fermion
mass.  The scalar potential (including the masses) is obtained after
eliminating the auxiliary field.

We leave the obvious generalisations of the Wess--Zumino model to the
tutorial problem.  It is a pleasure to contemplate how much simpler it
is to write these actions down in superspace than in components, and
furthermore the fact that we know a priori that the resulting theories
will be supersymmetric.

The power of superfields is not restricted to facilitating the
construction of supersymmetric models.  There is a full-fledged
superspace approach to supersymmetric quantum field theories, together
with Feynman rules for ``supergraphs'' and manifestly supersymmetry
regularisation schemes.  This formalism has made it possible to prove
certain powerful ``nonrenormalisation'' theorems which lie at the
heart of the attraction of supersymmetric theories.  A simple
consequence of superspace perturbation theory is that in a theory of
chiral superfields, any counterterm is of the form of an integral over
all of superspace (that is, of the form $\int d^4 x\, d^2 \theta
d^2\bar\theta$).  This means that in a renormalisable theory, the
superpotential terms---being integrals over chiral superspace (that
is, $\int d^4 x d^2 \theta$ or $\int d^4 x d^2 \bar\theta$)---are not
renormalised.  Since the superpotential contains both the mass and the
couplings of the chiral superfields, it means that the tree level
masses and couplings receive no perturbative loop corrections.  In
fact, ``miraculous cancellations'' at the one-loop level were already
observed in the early days of supersymmetry, which suggested that
there was only need for wave-function renormalisation.  The
nonrenormalisation theorem (for chiral superfields) is the statement
that this persists to all orders in perturbation theory.  More
importantly, the absence of mass renormalisation provides a solution
of the gauge hierarchy problem, since a hierarchy of masses fixed at
tree-level will receive no further radiative corrections.  From a
phenomenological point of view, this is one of the most attractive
features of supersymmetric theories.

\begin{tut}[\textsc{Models with chiral superfields}]\indent\par
  \label{pr:chiralmodels}
  In this tutorial problem we discuss the most general supersymmetric
  models which can be constructed out of chiral superfields.  Let
  $\Phi^i$, for $i=1,2,\dots,N$, be chiral superfields, and let
  $(\Phi^i)^* = \bar\Phi^{\bar\imath}$ be the conjugate antichiral
  fields.
  \begin{enumerate}
  \item Show that the most general supersymmetric renormalisable
    lagrangian involving these fields is given by the sum of a kinetic 
    term
    \begin{equation*}
      \int d^2 \theta d^2\bar\theta\, K_{i\bar\jmath} \Phi^i
      \bar\Phi^{\bar\jmath}
    \end{equation*}
    and a superpotential term \eqref{eq:superpotential} with
    \begin{equation*}
      W(\Phi) = a_i \Phi^i + \half m_{ij} \Phi^i\Phi^j + \tfrac13
      \lambda_{ijk} \Phi^i\Phi^j\Phi^k~,
    \end{equation*}
    where $a_i$, $m_{ij}$ and $\lambda_{ijk}$ are totally symmetric
    real constants, and $K_{i\bar\jmath}$ is a constant hermitian
    matrix.  Moreover unitarity of the model forces $K_{i\bar\jmath}$
    to be positive definite.
  \item Argue that via a complex change of variables $\Phi^i \mapsto
    M^i{}_j \Phi^j$, where $M$ is a matrix in $\GL(N,\CC)$, we can
    take $K_{i\bar\jmath} = \delta_{i\bar\jmath}$ without loss of
    generality.  Moreover we we still have the freedom to make a
    unitary transformation $\Phi^i \mapsto U^i{}_j \Phi^j$, where $U$
    is a matrix in $\U(N)$ with which to diagonalise the mass matrix
    $m_{ij}$.  Conclude that the most general supersymmetric
    renormalisable lagrangian involving $N$ chiral superfields is
    given by the sum of a kinetic term
    \begin{equation*}
      \int d^2\theta d^2\bar\theta \sum_{i=1}^N \Phi^i \bar\Phi_i~,
    \end{equation*}
    where $\bar\Phi_i = \delta_{i\bar\jmath} \bar\Phi^{\bar\jmath}$,
    and a superpotential term \eqref{eq:superpotential} with
    \begin{equation*}
      W = a_i \Phi^i + \sum_{i=1}^N m_i (\Phi^i)^2 + \tfrac13
      \lambda_{ijk} \Phi^i\Phi^j\Phi^k~.
    \end{equation*}
  \item Expand the above action into components and eliminate the
    auxiliary fields via their equations of motion.
  \end{enumerate}
  
  If we don't insist on renormalisability, we can generalise the above
  model in two ways.  First of all we can consider more general
  superpotentials, but we can also contemplate more complicated
  kinetic terms.  Let $K(\Phi,\bar\Phi)$ be a real function of
  $\Phi^i$ and $\bar\Phi^{\bar\imath}$ and consider the kinetic term
  \begin{equation}
    \label{eq:ssmkin}
      \int d^4 x\, d^2 \theta d^2\bar\theta\, K(\Phi,\bar\Phi)~.
  \end{equation}
  \begin{enumerate}
  \item[4.] Show that the above action is invariant under the
  transformations
  \begin{equation}
    \label{eq:klrgt}
      K(\Phi,\bar\Phi) \mapsto K(\Phi,\bar\Phi) + \Lambda(\Phi) +
      \overline{\Lambda(\Phi)}~.
    \end{equation}
  \item[5.] Expand the above kinetic term and show that it gives rise
    to a supersymmetric extension of the ``hermitian sigma model''
    \begin{equation*}
      - \int d^4 x\, g_{i\bar\jmath}(\phi,\bar\phi) \d_\mu \phi^i \d^\mu
        \bar\phi^{\bar\jmath}~,
    \end{equation*}
    with metric
    \begin{equation*}
      g_{i\bar\jmath}(\phi,\bar\phi) = \d_i\d_{\bar\jmath}
      K(\phi,\bar\phi)~,
    \end{equation*}
    where $\d_i = \d/\d\phi^i$ and $\d_{\bar\imath} =
    \d/\d\bar\phi^{\bar\imath}$.
  \end{enumerate}
  \begin{scholium}
    Such a metric $g_{i\bar\jmath}$ is called \emph{Kähler}.  Notice
    that it is the metric which is physical even though the superspace
    action is written in terms of the \emph{Kähler potential} $K$.
    This is because the action is invariant under the \emph{Kähler
      gauge transformations} \eqref{eq:klrgt} which leave the metric
    invariant.
  \end{scholium}
\begin{enumerate}
  \item[6.] Eliminate the auxiliary fields via their equations of
    motion and show that the resulting lagrangian becomes (up to a
    total derivative)
    \begin{equation*}
      - g_{i\bar\jmath} \d_\mu \phi^i \d^\mu
        \bar\phi^{\bar\jmath} + \tfrac{i}{2} g_{i\bar\jmath} \chi^i
        \sigma^\mu\nabla_\mu \bar\chi^{\bar\jmath} + \tfrac{1}{16}
        R_{ij\bar k\bar \ell} \chi^i \chi^j \bar\chi^{\bar k}
        \bar\chi^{\bar \ell}~,
    \end{equation*}
    where
    \begin{equation*}
      \begin{aligned}
        \nabla_\mu\bar\chi^{\bar\imath} &= \d_\mu\bar\chi^{\bar\imath}
        + \Gamma_{\bar\jmath\bar k}{}^{\bar\imath}
        \d_\mu\bar\phi^{\bar\jmath}
        \bar\chi^{\bar k}\\
        \Gamma_{\bar\jmath\bar k}{}^{\bar\imath} &= g^{i\bar\imath}
        \d_{\bar k} g_{i\bar\jmath} \qquad
        \left(\text{and}~\Gamma_{jk}{}^i = g^{i\bar\imath} \d_k
          g_{\bar\imath j}\right)\\
        R_{ij\bar k\bar \ell} &= \d_i \d_{\bar k} g_{j\bar \ell} -
        g^{m\bar m} \d_i g_{j\bar m} \d_{\bar k} g_{m\bar \ell}~,
    \end{aligned}
\end{equation*}
    where $g^{i\bar\jmath}$ is the inverse of $g_{i\bar\jmath}$, which
    is assumed invertible due to the positive-definiteness (or more
    generally, nondegeneracy) of the kinetic term.
  \item[7.] Finally, consider an arbitrary differentiable function
    $W(\Phi)$ and add to the kinetic term \eqref{eq:ssmkin} the
    corresponding superpotential term \eqref{eq:superpotential}.
    Expand the resulting action in components and eliminate the
    auxiliary fields using their field equations to arrive at the most
    general supersymmetric action involving only scalar multiplets:
    \begin{multline}
      \label{eq:ssmodel}
      - g_{i\bar\jmath} \d_\mu \phi^i \d^\mu \bar\phi^{\bar\jmath} +
      \tfrac{i}{2} g_{i\bar\jmath} \chi^i \sigma^\mu\nabla_\mu
      \bar\chi^{\bar\jmath} + \tfrac{1}{16} R_{ij\bar k\bar \ell}
      \chi^i \chi^j \bar\chi^{\bar k}
      \bar\chi^{\bar \ell}\\
      - g^{i \bar\jmath} \d_i W \d_{\bar\jmath}\overline W - \tfrac14
      \chi^i\chi^j H_{ij}(W) - \tfrac14
      \bar\chi^{\bar\imath}\bar\chi^{\bar\jmath}
      H_{\bar\imath\bar\jmath}(\overline W)~,
    \end{multline}
    where
    \begin{equation*}
      \begin{aligned}
        H_{ij}(W) &= \nabla_i \d_j W = \d_i \d_j W - \Gamma_{i
          j}{}^k\d_k
        W\\
        H_{\bar\imath\bar\jmath}(\overline W) &= \nabla_{\bar\imath}
        \d_{\bar\jmath} \overline W = \d_{\bar\imath} \d_{\bar\jmath}
        \overline W - \Gamma_{\bar\imath\bar\jmath}{}^{\bar k}\d_{\bar
          k} \overline W
    \end{aligned}
\end{equation*}
    is the Hessian of $W$.
  \end{enumerate}
  \begin{scholium}
    Models such as \eqref{eq:ssmodel} are known as supersymmetric
    sigma models.  The scalar fields can be understood as maps from
    the spacetime to a riemannian manifold.  Not every riemannian
    manifold admits a supersymmetric sigma model and indeed this
    problem shows that supersymmetry requires the metric to be Kähler.
    The data of a supersymmetric sigma model is thus geometric in
    nature: a Kähler manifold $(M,g)$ and a holomorphic function $W$
    on $M$.  This and similar results underlie the deep connections
    between supersymmetry and geometry.
  \end{scholium}
\end{tut}


\section{Supersymmetric Yang--Mills revisited}
\label{lec:SYM2}

The general supersymmetric renormalisable models in four dimensions
can be built out of the chiral superfields introduced in the previous
lecture and the vector superfields to be introduced presently.  In
terms of components, chiral superfields contain complex scalar fields
(parametrising a Kähler manifold, which must be flat in renormalisable
models) and Majorana fermions.  This is precisely the field content of
the Wess--Zumino model discussed in Lecture~\ref{lec:WZmodel} and in
the previous lecture we saw how to write (and generalise) this model
in superspace.  In contrast, the vector superfield is so called
because it contains a vector boson as well as a Majorana fermion. This
is precisely the field content of the supersymmetric Yang--Mills
theory discussed in Lecture~\ref{lec:SYM} and in the present lecture
we will learn how to write this theory down in superspace.  By the end
of this lecture we will know how to write down the most general
renormalisable supersymmetric theory in four dimensions.  The tutorial
problem will introduce the Kähler quotient, in the context of the $\CC
P^N$ supersymmetric sigma model.  Apart from its intrinsic
mathematical interest, this construction serves to illustrate the fact
that in some cases, the low energy effective theory of a
supersymmetric gauge theory is a supersymmetric sigma model on the
space of vacua.

\subsection{Vector superfields}

In the component expansion \eqref{eq:scalarsuperfield} of a general
scalar superfield one finds a vector field $v_\mu$.  If we wish to
identify this field with a vector boson we must make sure that it is
real.  Complex conjugating the superfield sends $v_\mu$ to its complex
conjugate $\bar v_\mu$, hence reality of $v_\mu$ implies the reality
of the superfield.  I hope this motivates the following definition.

A \emph{vector superfield} $V$ is a scalar superfield which satisfies
the reality condition $\bar V = V$.

\begin{exercise}
  Show that the general vector superfield $V$ has the following
  component expansion:
  \begin{multline}
    \label{eq:realsuperfield}
      V(x,\theta,\bar\theta) = C(x) + \theta\xi(x) +
      \bar\theta\bar\xi(x) + \bar\theta\bar\sigma^\mu\theta v_\mu(x)\\ 
      + \theta^2 G(x) + \bar\theta^2 \bar G(x) + \bar\theta^2 \theta
      \eta(x) + \theta^2  \bar\theta\bar\eta(x)  + \theta^2 \bar\theta^2 E(x)~,
  \end{multline}
  where $C$, $v_\mu$ and $E$ are real fields.
\end{exercise}

The real part of a chiral superfield $\Lambda$ is a particular kind of
vector superfield, where the vector component is actually a
derivative:
\begin{multline}
  \Lambda + \bar\Lambda = (\phi + \bar\phi) + \theta\chi +
  \bar\theta\bar\chi + \theta^2 F + \bar\theta^2\bar F + i
  \bar\theta\bar\sigma^\mu\theta \d_\mu(\phi - \bar\phi)\\
  - \tfrac{i}{2} \theta^2 \bar\theta\bar\sigma^\mu\d_\mu\chi -
    \tfrac{i}{2} \bar\theta^2 \theta\sigma^\mu\d_\mu\bar\chi +
    \tfrac14 \theta^2\bar\theta^2 \dalem(\phi + \bar\phi)~.
\end{multline}

This suggests that the transformation
\begin{equation}
  \label{eq:gaugetransf}
  V \mapsto V - (\Lambda + \bar\Lambda)~,
\end{equation}
where $V$ is a vector superfield and $\Lambda$ is a chiral superfield,
should be interpreted as the superspace version of a $\U(1)$ gauge
transformation.

\begin{exercise}
  Show that the transformation \eqref{eq:gaugetransf} has the
  following effect on the components of the vector superfield:
  \begin{equation*}
    \begin{aligned}
      C &\mapsto C -(\phi + \bar\phi)\\
      \xi &\mapsto \xi - \chi\\
      G &\mapsto G - F\\
      v_\mu &\mapsto v_\mu - i \d_\mu(\phi - \bar\phi)\\
      \eta_\alpha &\mapsto \eta_\alpha + \tfrac{i}{2}
      (\sigma^\mu)_{\alpha\dot\beta} \d_\mu \bar\chi^{\dot\beta}\\
      E &\mapsto E - \tfrac14 \dalem (\phi + \bar\phi)~.
  \end{aligned}
\end{equation*}
\end{exercise}

This result teaches us two things.  First of all, we see that the
combinations
\begin{equation}
  \label{eq:gaugeinvariants}
  \begin{aligned}
    \lambda_\alpha &= \eta_\alpha - \tfrac{i}{2}
    (\sigma^\mu)_{\alpha\dot\beta} \d_\mu \bar\xi^{\dot\beta}\\
    D &:= E - \tfrac14 \dalem C
\end{aligned}
\end{equation}
are gauge invariant.
\begin{caveat}
  I hope that the gauge-invariant field $D$ will not be confused with
  the supercovariant derivative.  This abuse of notation has become
  far too ingrained in the supersymmetry literature for me to even
  attempt to correct it here.
\end{caveat}
Of these gauge-invariant quantities, it is $\lambda_\alpha$ which is
the lowest component in the vector superfield.  This suggests that we
try to construct a gauge-invariant lagrangian out of a superfield
having $\lambda_\alpha$ as its lowest component.  Such a superfield
turns out to be easy to construct, as we shall see in the next
section.

The second thing we learn is that because the fields $C$, $G$ and
$\xi$ transform by shifts, we can choose a special gauge in which they
vanish.  This gauge is called the \emph{Wess--Zumino gauge} and it of
course breaks supersymmetry.  Nevertheless it is a very convenient
gauge for calculations, as we will have ample opportunity to
demonstrate.  For now, let us merely notice that in the Wess--Zumino
gauge the vector superfield becomes
\begin{equation}
  \label{eq:WZgauge}
  V = \bar\theta\bar\sigma^\mu\theta v_\mu + \bar\theta^2 \theta\lambda + 
  \theta^2 \bar\theta\bar\lambda + \theta^2\bar\theta^2 D~,
\end{equation}
and that powers are very easy to compute:
\begin{equation*}
  V^2 = -\half \theta^2\bar\theta^2 v_\mu v^\mu~,
\end{equation*}
with all higher powers vanishing.  This is not a gratuitous comment.
We will see that in coupling to matter and indeed already in the
nonabelian case, it will be necessary to compute the exponential of
the vector superfield $e^V$, which in the Wess--Zumino gauge becomes
simply
\begin{equation}
  \label{eq:eVWZgauge}
  e^V = 1 + \bar\theta\bar\sigma^\mu\theta v_\mu + \bar\theta^2
  \theta\lambda + \theta^2 \bar\theta\bar\lambda +
  \theta^2\bar\theta^2 \left(D - \tfrac14 v_\mu v^\mu\right)~.
\end{equation}
Furthermore gauge transformations with imaginary parameter $\phi = -
\bar\phi$ and $\chi=F=0$ still preserve the Wess--Zumino gauge and
moreover induce in the vector field $v_\mu$ the expected $\U(1)$ gauge
transformations
\begin{equation}
  \label{eq:residualgaugetrans}
  v_\mu \mapsto v_\mu - i \d_\mu (\phi - \bar\phi)~.
\end{equation}

\subsection{The gauge-invariant action}

Define the following spinorial superfields
\begin{equation}
  \label{eq:superfieldstrengths}
  W_\alpha := -\tfrac14 \bar D^2 D_\alpha V \qquad\text{and}\qquad
  \overline W_{\dot\alpha} := -\tfrac14 D^2 \bar D_{\dot\alpha} V~.
\end{equation}
Notice that reality of $V$ implies that $(W_\alpha)^* = \overline
W_{\dot\alpha}$.  To show that the lowest component of $W_\alpha$ is
$\lambda_\alpha$ it will be convenient to compute it in the
Wess--Zumino gauge \eqref{eq:WZgauge}.  This is allowed because
$W_\alpha$ is actually gauge invariant, so it does not matter in which
gauge we compute it.

\begin{exercise}
  Prove that the supercovariant derivatives satisfy the following
  identities:
  \begin{equation}
    \label{eq:Didentities}
    \begin{aligned}
      \left[\bar D_{\dot\alpha},\left[\bar D_{\dot\beta},
          D_\gamma\right]\right] &= 0\\
      \bar D_{\dot\alpha} \bar D^2 &= 0~,
    \end{aligned}
  \end{equation}
  and use them to prove that $W_\alpha$ is both chiral:
  \begin{equation*}
    \bar D_{\dot\beta} W_\alpha = 0~,
  \end{equation*}
  and gauge invariant.  Use complex conjugation to prove that
  $\overline W_{\dot\alpha}$ is antichiral and gauge invariant.
  Finally, show that the following ``real'' equation is satisfied:
  \begin{equation}
    \label{eq:real}
    D^\alpha W_\alpha = \bar D^{\dot\alpha} \overline
    W_{\dot\alpha}~.
  \end{equation}
\end{exercise}

In the Wess--Zumino gauge, the vector superfield $V$ can be written as
\begin{equation}
  \label{eq:dressedvector}
  V = e^{-iU}\left[ \bar\theta\bar\sigma^\mu\theta v_\mu +
  \bar\theta^2 \theta\lambda + \theta^2\bar\theta\bar\lambda +
  \theta^2\bar\theta^2 \left( D + \tfrac{i}{2} \d^\mu v_\mu \right)
  \right]~,
\end{equation}
where as usual $U = \theta\sigma^\mu\bar\theta\d_\mu$.

\begin{exercise}
  Using this fact show that
  \begin{equation*}
      \bar D^{\dot\alpha} V = e^{-iU} \left[ - \theta_\alpha
        (\sigma^\mu)^{\alpha\dot\alpha} v_\mu + 2
        \bar\theta^{\dot\alpha} \theta\lambda + \theta^2
        \bar\lambda^{\dot\alpha} + 2 \theta^2 \bar\theta^{\dot\alpha}
        \left( D + \tfrac{i}{2} \d^\mu v_\mu
        \right) \right]
  \end{equation*}
  and that
  \begin{equation*}
      -\tfrac14 \bar D^2 V  = e^{-iU} \left[ \theta\lambda + \theta^2
        \left( D + \tfrac{i}{2} \d^\mu v_\mu \right) \right]~,
  \end{equation*}
  and conclude that $W_\alpha$ takes the following expression
  \begin{equation}
    \label{eq:abelianW}
    W_\alpha = e^{-iU} \left[ \lambda_\alpha + 2 \theta_\alpha D +
      \tfrac{i}{2} \theta_\beta (\sigma^{\mu\nu})^\beta{}_\alpha
      f_{\mu\nu} + i \theta^2 (\bar\sigma^\mu)_{\dot\beta\alpha} \d_\mu
      \bar\lambda^{\dot\beta}\right]~,
  \end{equation}
  where $f_{\mu\nu} = \d_\mu v_\nu - \d_\nu v_\mu$ is the
  field-strength of the vector $v_\mu$.\\
  {\rm (Hint: You may want to use the expressions \eqref{eq:twist} for the
  supercovariant derivatives.)}
\end{exercise}

Since $W_\alpha$ is chiral, so is $W^\alpha W_\alpha$, which is
moreover Lorentz invariant.  The $\theta^2$ component is also Lorentz
invariant and transforms as a total derivative under supersymmetry.
Its real part can therefore be used as a supersymmetric lagrangian.

\begin{exercise}
  Show that
  \begin{equation}
    \int d^2 \theta \, W^\alpha W_\alpha = 2 i \lambda \sigma^\mu \d_\mu
    \bar\lambda + 4 D^2 - \tfrac12 f_{\mu\nu} f^{\mu\nu} +
    \tfrac{i}{4} \epsilon^{\mu\nu\rho\sigma} f_{\mu\nu}
    f_{\rho\sigma}~,
  \end{equation}
  and hence that its real part is given by
  \begin{equation}
    \label{eq:symlagsuperspace}
    i \left( \lambda \sigma^\mu \d_\mu \bar\lambda +
    \bar\lambda\bar\sigma^\mu \d_\mu \lambda \right) - \tfrac12
    f_{\mu\nu} f^{\mu\nu} + 4 D^2~.
  \end{equation}
  \label{ex:trw2}
\end{exercise}

\begin{caveat}
  It may seem from this expression that the supersymmetric Yang--Mills
  lagrangian involves an integral over chiral superspace, and perhaps
  that a similar nonrenormalisation theorem to the one for chiral
  superfields would prevent the Yang--Mills coupling constant to
  renormalise.  This is \emph{not} true.  In fact, a closer look at
  the expression for the supersymmetric Yang--Mills reveals that it
  can be written as an integral over all of superspace, since the
  $\bar D^2$ in the definition of $W_\alpha$ acts like a $\int
  d^2\bar\theta$.  In other words, counterterms can \emph{and do}
  arise which renormalise the supersymmetric Yang--Mills action.  
\end{caveat}

Now consider the supersymmetric Yang--Mills action with lagrangian
\eqref{eq:symlag} for the special case of the abelian group
$G=\U(1)$.  The resulting theory is free.  Let $\Psi^a = (\psi^\alpha, 
\bar\psi_{\dot\alpha})$.  Expanding the lagrangian we obtain
\begin{equation}
  \eL_{\text{SYM}} = \tfrac{i}{4} \left( \psi \sigma^\mu \d_\mu
  \bar\psi + \bar\psi\bar\sigma^\mu \d_\mu \psi \right) - \tfrac14
  F_{\mu\nu}
  F^{\mu\nu}~,
\end{equation}
which agrees with half the lagrangian \eqref{eq:symlagsuperspace}
provided that we eliminate the auxiliary field $D$ and identify $A_\mu
= v_\mu$ and $\psi_\alpha = \lambda_\alpha$.  Actually, this last
field identification has a phase ambiguity, and we will fix it by
matching the supersymmetry transformation properties
\eqref{eq:superpoincareYM} with the ones obtained in superspace:
$- (\varepsilon Q + \bar\varepsilon\bar Q) V$.

\subsection{Supersymmetry transformations}

We can (and will) simplify the computation by working in the
Wess--Zumino gauge.  However it should be noticed that this gauge
breaks supersymmetry; that is, the supersymmetry variation of a vector 
superfield in the Wess--Zumino gauge will not remain in the
Wess--Zumino gauge.  In order to get it back to this gauge it will be
necessary to perform a compensating gauge transformation.  This is a
common trick in supersymmetry and it's worth doing it in some detail.

\begin{exercise}
  Compute the supersymmetry transformation of a vector superfield $V$
  in the Wess--Zumino gauge \eqref{eq:WZgauge} to obtain
  \begin{equation}
    \label{eq:susyvectorWZgauge}
    \begin{split}
      - (\varepsilon Q + \bar\varepsilon\bar Q) V = {} & \theta
      \sigma^\mu \bar \varepsilon v_\mu - \bar\theta\bar\sigma^\mu
      \varepsilon v_\mu - \theta^2 \bar\varepsilon\bar\lambda -
      \bar\theta^2 \varepsilon\lambda\\
      & {} + \bar\theta\bar\sigma^\mu\theta
      \left(\bar\varepsilon\bar\sigma_\mu\lambda - \varepsilon
        \sigma_\mu \bar\lambda\right)\\
      & {} - 2 \theta^2 \bar\theta\bar\varepsilon \left(D -
        \tfrac{i}{4} \d^\mu v_\mu \right) - 2 \bar\theta^2
      \theta\varepsilon \left(D + \tfrac{i}{4} \d^\mu v_\mu\right)\\
      & {} - \tfrac{i}{4} \theta^2\bar\theta
      \bar\sigma^{\mu\nu}\bar\varepsilon f_{\mu\nu} - \tfrac{i}{4}
      \bar\theta^2 \theta \sigma^{\mu\nu}\varepsilon f_{\mu\nu}\\
      & {} + \tfrac{i}{2} \theta^2\bar\theta^2 \left(\varepsilon
        \sigma^\mu \d_\mu \bar\lambda + \bar\varepsilon \bar\sigma^\mu
        \d_\mu \lambda\right)~.
    \end{split}
  \end{equation}
  \label{ex:susyvectorWZ}
\end{exercise}

As advertised, the resulting variation is not in the Wess--Zumino
gauge.  Nevertheless we can gauge transform it back to the
Wess--Zumino gauge.  Indeed, we can find a chiral superfield $\Lambda$
with component fields $\phi$, $\chi$ and $F$ such that
\begin{equation}
  \label{eq:compensating}
  \delta_\varepsilon V = - (\varepsilon Q + \bar\varepsilon\bar Q) V -
  (\Lambda + \bar\Lambda)
\end{equation}
is again in the Wess--Zumino gauge.  To do this notice that the first
four terms in the expansion \eqref{eq:susyvectorWZgauge} of
$-(\varepsilon Q + \bar\varepsilon\bar Q) V$ have to vanish in the
Wess--Zumino gauge.  This is enough to fix $\Lambda$ up to the imaginary
part of $\phi$, which simply reflects the gauge invariance of the
component theory.

\begin{exercise}
  Show that the parameters of the compensating gauge transformation
  are given by (where we have chosen the imaginary part of $\phi$ to
  vanish)
  \begin{equation}
    \label{eq:compensator}
    \begin{aligned}
      \phi &= 0\\
      \chi^\alpha &= -(\sigma^\mu)^{\alpha\dot\alpha}
      \bar\varepsilon_{\dot\alpha} v_\mu\\
      F &= - \bar\varepsilon \bar\lambda~,
    \end{aligned}
  \end{equation}
  and hence that
  \begin{equation*}
    \begin{aligned}
      \delta_\varepsilon V &= - (\varepsilon Q + \bar\varepsilon\bar Q)
      V - (\Lambda + \bar\Lambda)\\
      &= \theta\sigma^\mu\bar\theta \delta_\varepsilon v_\mu +
      \bar\theta^2 \theta \delta_\varepsilon \lambda + \theta^2
      \bar\theta\delta_\varepsilon \bar\lambda + \theta^2\bar\theta^2
      \delta_\varepsilon D~,
    \end{aligned}
  \end{equation*}
  with
  \begin{equation}
    \label{eq:susytransvector}
    \begin{aligned}
      \delta_\varepsilon v_\mu &= \varepsilon \sigma_\mu \bar\lambda - 
      \bar\varepsilon\bar\sigma^\mu \lambda\\
      \delta_\varepsilon \lambda_\alpha &= -2 \varepsilon_\alpha D +
      \tfrac{i}{2} (\sigma^{\mu\nu})_{\alpha\beta} \epsilon^\beta
      f_{\mu\nu}\\
      \delta_\varepsilon D &= \tfrac{i}{2} \left( \varepsilon
        \sigma^\mu \d_\mu \bar\lambda + \bar\varepsilon \bar\sigma^\mu
        \d_\mu \lambda\right)~.
    \end{aligned}
  \end{equation}
\end{exercise}

Rewriting the supersymmetry transformations \eqref{eq:superpoincareYM}
of supersymmetric Yang--Mills (for $G=\U(1)$) in terms of $\Psi^a =
(\psi^\alpha, \bar\psi_{\dot\alpha})$ we obtain
\begin{equation*}
  \begin{aligned}
    \delta_\varepsilon A_\mu &= - i (\bar\varepsilon \bar\sigma_\mu
    \psi + \varepsilon \sigma_\mu \bar\psi)\\
    \delta_\varepsilon \psi_\alpha &= -\half F_{\mu\nu}
    (\sigma^{\mu\nu})_{\alpha\beta} \varepsilon^\beta~.
  \end{aligned}
\end{equation*}
Therefore we see that they agree with the transformations
\eqref{eq:susytransvector} provided that as before we identify $v_\mu
= A_\mu$, but now $\psi_\alpha = i \lambda_\alpha$.

In summary, supersymmetric Yang--Mills theory \eqref{eq:symlag} with
gauge group $\U(1)$ can be written in superspace in terms of a vector
superfield $V$ which in the Wess--Zumino gauge has the expansion
\begin{equation*}
  V = \bar\theta\bar\sigma^\mu\theta A_\mu - i \bar\theta^2 \theta\psi
  + i \theta^2 \bar\theta\bar\psi + \theta^2\bar\theta^2 D~,
\end{equation*}
with lagrangian given by
\begin{equation*}
  \eL_{\text{SYM}} = \int d^2 \theta \, \tfrac14 W^\alpha W_\alpha +
  \text{c.c.}~,
\end{equation*}
with $W_\alpha$ given by \eqref{eq:superfieldstrengths}.

\subsection{Coupling to matter}

Let us couple the above theory to matter in the form in one chiral
superfield.  We will postpone discussing more general matter couplings
until we talk about nonabelian gauge theories.

Consider a chiral superfield $\Phi$ in a one-dimensional
representation of the group $\U(1)$ with charge $e$.  That is to say,
if $\exp(i\varphi) \in \U(1)$ then its action on $\Phi$ is given by
\begin{equation*}
  \exp(i\varphi) \cdot \Phi = e^{ie\varphi} \Phi
  \qquad\text{and}\qquad 
  \exp(i\varphi) \cdot \bar\Phi = e^{-ie\varphi} \bar\Phi~.
\end{equation*}
The kinetic term $\bar\Phi\Phi$ is clearly invariant.  If we wish to
promote this symmetry to a gauge symmetry, we need to consider
parameters $\varphi(x)$ which are functions on Minkowski space.
However, $e^{i e \varphi(x)} \Phi$ is not a chiral superfield and
hence this action of the gauge group does not respect supersymmetry.
To cure this problem we need to promote $\varphi$ to a full chiral
superfield $\Lambda$, so that the gauge transformation now reads
\begin{equation}
  \label{eq:chiralGT}
  \Phi \mapsto e^{ie\Lambda} \Phi~.
\end{equation}
Now the gauge transformed superfield remains chiral, but we pay the
price that the kinetic term $\bar\Phi \Phi$ is no longer invariant.
Indeed, it transforms as
\begin{equation*}
  \bar\Phi \Phi \mapsto \bar\Phi \Phi e^{ie(\Lambda - \bar\Lambda)}~.
\end{equation*}
However, we notice that $i(\Lambda - \bar\Lambda)$ is a real
superfield and hence can be reabsorbed in the gauge transformation of
a vector superfield $V$:
\begin{equation}
  \label{eq:vectorGT}
  V \mapsto V - \tfrac{i}{2} (\Lambda - \bar\Lambda)~,
\end{equation}
in such a way that the expression
\begin{equation*}
  \bar\Phi e^{2eV} \Phi
\end{equation*}
is gauge invariant under \eqref{eq:chiralGT} and \eqref{eq:vectorGT}.

The coupled theory is now defined by the lagrangian
\begin{equation}
  \label{eq:scalarQED}
  \int d^2 \theta d^2\bar\theta\, \bar\Phi e^{2eV} \Phi + 
  \left[ \int d^2 \theta\, \tfrac14 W^\alpha W_\alpha + \text{c.c}
  \right]~,
\end{equation}
which can be understood as the supersymmetric version of scalar QED.

The coupling term might look nonpolynomial (and hence
nonrenormalisable), but since it is gauge invariant it can be computed 
in the Wess--Zumino gauge where $V^3=0$.

\begin{exercise}
  Show that the component expansion of the lagrangian
  \eqref{eq:scalarQED}, with $\Phi$ given by \eqref{eq:chiralcomps},
  $V$ in the Wess--Zumino gauge by \eqref{eq:WZgauge} and having
  eliminated the auxiliary fields, is given by 
  \begin{equation}
    \label{eq:scalarQEDcomps}
    \begin{split}
      -\tfrac14 f_{\mu\nu} f^{\mu\nu} + \tfrac{i}{2} \left(\lambda
        \sigma^\mu \d_\mu \bar\lambda + \bar\lambda\bar\sigma^\mu
        \d_\mu \lambda \right) + \tfrac{i}{4} \left( \chi \sigma^\mu
        \overline{\eD_\mu\chi} +
        \bar\chi\bar\sigma^\mu\eD\chi\right)\\
      - \eD^\mu\phi \overline{\eD_\mu\phi} - e
        \left(\bar\phi\lambda\chi + \phi \bar\lambda\bar\chi\right) -
        \half e^2 \left(|\phi|^2\right)^2~,
    \end{split}
  \end{equation}
  where $\eD_\mu \phi = \d_\mu\phi - i e v_\mu\phi$ and
  similarly for $\eD_\mu\chi$.
\end{exercise}

The above model does not allow massive charged matter, since the mass
term in the superpotential is not gauge invariant.  In order to consider
massive matter, and hence supersymmetric QED, it is necessary to
include two oppositely charged chiral superfields $\Phi_\pm$,
transforming under the $\U(1)$ gauge group as
\begin{equation*}
  \Phi_\pm \mapsto e^{\pm i e \Lambda} \Phi_\pm~.
\end{equation*}
Then the supersymmetric QED lagrangian in superspace is given by
\begin{multline}
  \label{eq:sQED}
    \int d^2 \theta d^2\bar\theta\, \left( \bar\Phi_+ e^{2eV} \Phi_+ +
      \bar\Phi_- e^{-2eV} \Phi_- \right)\\
    + \left[ \int d^2 \theta\, \left( \tfrac14 W^\alpha W_\alpha + m
      \Phi_+\Phi_-\right) + \text{c.c.} \right]~.
\end{multline}

\begin{exercise}
  Expand the supersymmetric QED lagrangian in components and verify
  that it describes a massless gauge boson (the \emph{photon}) and a
  charged massive fermion (the \emph{electron}), as well as a a
  massless neutral fermion (the \emph{photino}) and a a massive
  charged scalar (the \emph{selectron}).
\end{exercise}

\begin{amusing}
  Detractors often say, with some sarcasm, that supersymmetry is doing
  well: already half the particles that it predicts have been found.
\end{amusing}

The coupling of supersymmetric gauge fields to supersymmetric matter
suggests that the fundamental object is perhaps not the vector
superfield $V$ itself but its exponential $\exp V$, which in the
Wess--Zumino gauge is not too different an object---compare equations
\eqref{eq:WZgauge} and \eqref{eq:eVWZgauge}.  One might object that
the supersymmetric field-strength $W_\alpha$ actually depends on $V$
and not on its exponential, but this is easily circumvented by
rewriting it thus:
\begin{equation}
  \label{eq:expfieldstrength}
  W_\alpha = -\tfrac14 \bar D^2 e^{-V} D_\alpha e^V~.
\end{equation}
It turns out that this observation facilitates enormously the
construction of nonabelian supersymmetric Yang--Mills theory in
superspace.

\subsection{Nonabelian gauge symmetry}

As in Lecture~\ref{lec:SYM}, let $G$ be a compact Lie group with Lie
algebra $\fg$ and fix an invariant inner product, denoted by $\Tr$ in
the Lie algebra.  The vector superfield $V$ now takes values in $\fg$.
Relative to a fixed basis $\{T_i\}$ for $\fg$ we can write
\begin{equation}
  \label{eq:nonabelianV}
  V = i V^i T_i~,
\end{equation}
where as we will see, the factor of $i$ will guarantee that the
superfields $V^i$ are real.

The expression \eqref{eq:expfieldstrength} for the field-strength
makes sense for a Lie algebra valued $V$, since the only products of
generators $T_i$ appearing in the expression are in the form of
commutators.  The form of the gauge transformations can be deduced by
coupling to matter.

Suppose that $\bPhi$ is a chiral superfield taking values in a unitary
representation of $G$.  This means that under a gauge transformation,
$\bPhi$ transforms as
\begin{equation*}
  \bPhi \mapsto e^{\Lambda} \bPhi~,
\end{equation*}
where $\Lambda$ is an antihermitian matrix whose entries are chiral
superfields.  The conjugate superfield $\bar\bPhi$ takes values in the
conjugate dual representation; this means that now $\bar\Phi$ denotes
the conjugate transpose.  Under a gauge transformation, it transforms
according to
\begin{equation*}
  \bar\bPhi \mapsto \bar\bPhi\, e^{\bar\Lambda}~,
\end{equation*}
where $\bar\Lambda$ is now the \emph{hermitian} conjugate of
$\Lambda$.  Consider the coupling
\begin{equation}
  \label{eq:nonabeliancoupling}  
  \bar\bPhi e^V \bPhi~.
\end{equation}
Reality imposes that $V$ be hermitian,
\begin{equation}
  \label{eq:nonabelianreality}
   \bar V = V
\end{equation}
where $\bar V$ is now the hermitian conjugate of $V$.  Since the $T_i$ 
are antihermitian, this means that the components $V^i$ in
\eqref{eq:nonabelianV} are vector superfields: $\bar V^i = V^i$.
Gauge invariance implies that $V$ should transform according to
\begin{equation}
  \label{eq:nonabelianGT}
  e^V \mapsto e^{-\bar\Lambda} e^V e^{-\Lambda}~.
\end{equation}
We can check that the field-strength \eqref{eq:expfieldstrength}
transforms as expected under gauge transformations.

\begin{exercise}
  Show that the field-strength \eqref{eq:expfieldstrength} transforms
  covariantly under the gauge transformation \eqref{eq:nonabelianGT}:
  \begin{equation*}
    W_\alpha \mapsto e^\Lambda W_\alpha e^{-\Lambda}~,
  \end{equation*}
  and conclude that
  \begin{equation*}
    \int d^2\theta \, \Tr W^\alpha W_\alpha
  \end{equation*}
  is gauge invariant
\end{exercise}

In order to compare this to the component version of supersymmetric
Yang--Mills we would like to argue that we can compute the action in
the Wess--Zumino gauge, but this requires first showing the existence
of this gauge.  The nonabelian gauge transformations
\eqref{eq:nonabelianGT} are hopelessly complicated in terms of $V$,
but using the Baker--Campbell--Hausdorff formula \eqref{eq:BCH} we can 
compute the first few terms and argue that the Wess--Zumino gauge
exists.

\begin{exercise}
  Using the Baker--Campbell--Hausdorff formula \eqref{eq:BCH}, show
  that the nonabelian gauge transformations \eqref{eq:nonabelianGT}
  takes the form
  \begin{equation*}
    V \mapsto V - (\Lambda + \bar\Lambda) - \half
    [V,\Lambda-\bar\Lambda] + \cdots~,
  \end{equation*}
  and conclude that $V$ can be put in the Wess--Zumino gauge
  \eqref{eq:WZgauge} by a judicious choice of $\Lambda + \bar\Lambda$.
\end{exercise}

Notice that in the Wess--Zumino gauge, infinitesimal gauge
transformations simplify tremendously.  In fact, since $V^3=0$, the
gauge transformation formula \eqref{eq:nonabelianGT} for infinitesimal 
$\Lambda$, reduces to
\begin{equation}
  \label{eq:infinitesimalnonabelianGTWZ}
  V \mapsto V - (\Lambda + \bar\Lambda) - \half
  [V,\Lambda-\bar\Lambda] - \tfrac1{12} [V,[V,\Lambda +
  \bar\Lambda]]~.
\end{equation}
Notice that an infinitesimal gauge transformation which preserves the
Wess--Zumino gauge has the form
\begin{equation}
  \label{eq:infGTparameterWZ}
  \Lambda = \omega + i \bar\theta\bar\sigma^\mu\theta \d_\mu \omega
  + \tfrac14 \theta^2 \bar\theta^2 \dalem \omega~,
\end{equation}
for some Lie algebra-valued scalar field $\omega$ obeying $\bar\omega
= - \omega$.  In this case, the term in $V^2$ in the transformation
law \eqref{eq:infinitesimalnonabelianGTWZ} is absent, as it has too
many $\theta$'s.

\begin{exercise}
  Show that the infinitesimal gauge transformation
  \begin{equation*}
    V \mapsto V - (\Lambda + \bar\Lambda) - \half
    [V,\Lambda-\bar\Lambda]
  \end{equation*}
  for $V$ in the Wess--Zumino gauge and with parameter $\Lambda$ given
  by \eqref{eq:infGTparameterWZ}, induces the following transformation
  of the component fields:
  \begin{equation*}
    \begin{aligned}
      \delta_\omega v_\mu &= -2i \d_\mu\omega - [v_\mu,\omega]\\
      \delta_\omega \chi &= - [\chi,\omega]\\
      \delta_\omega D &= -[D,\omega]~.
    \end{aligned}
  \end{equation*}
  Conclude that $A_\mu = \tfrac{1}{2gi} v_\mu$, where $g$ is the
  Yang--Mills coupling constant, obeys the transformation law
  \eqref{eq:infgaugetrans} of a gauge field.
\end{exercise}

This result suggests that in order to identify the fields in the
component formulation of supersymmetric Yang--Mills, we have to
rescale the nonabelian vector superfield by $2g$, with $g$ the
Yang--Mills coupling constant.  In order to obtain a lagrangian with
the correct normalisation for the kinetic term, we also rescale the
spinorial field strength by $1/(2g)$:
\begin{equation}
  \label{eq:nonabelianfieldstrength}
  W_\alpha := -\tfrac{1}{8g} \bar D^2 e^{-2g\, V} D_\alpha e^{2g\,
  V}~.
\end{equation}


\subsection{Nonabelian gauge-invariant action}

We now construct the nonabelian gauge-invariant action.  We will do
this in the Wess--Zumino gauge, but we should realise that the
nonabelian field-strength is no longer gauge invariant.  Nevertheless
we are after the superspace lagrangian $\Tr W^\alpha W_\alpha$, which
is gauge invariant.

\begin{exercise}
  Show that in Wess--Zumino gauge
  \begin{equation}
    \label{eq:WZdressedD}
    e^{-V} D_\alpha e^V = D_\alpha V - \half [V, D_\alpha V]~,
  \end{equation}
  and use this to find the following expression for the nonabelian
  field-strength $W_\alpha$ in \eqref{eq:nonabelianfieldstrength}:
  \begin{equation}
    \label{eq:nonabelianWcomps}
    W_\alpha = e^{-iU} \left[ \lambda_\alpha + 2\theta_\alpha D
    + \tfrac{i}{2} \theta_\beta (\sigma^{\mu\nu})^\beta{}_\alpha
      f_{\mu\nu} + i \theta^2 \eD_\mu \lambda^{\dot\alpha}
      (\bar\sigma^\mu)_{\dot\alpha\alpha}\right]~,
  \end{equation}
  where
  \begin{equation*}
    \begin{aligned}
      f_{\mu\nu} &= \d_\mu v_\nu - \d_\nu v_\mu - i g [v_\mu, v_\nu]\\
      \eD_\mu \lambda &= \d_\mu \lambda - i g [v_\mu, \lambda]~.
    \end{aligned}
  \end{equation*}
\end{exercise}

\begin{caveat}
  The factors of $i$ have to do with the fact that $v_\mu = v_\mu i
  T_i$.  In terms of $A_\mu = - i v_\mu$ these expressions are
  standard:
  \begin{equation*}
    \begin{aligned}
      f_{\mu\nu} &= i \left(\d_\mu A_\nu - \d_\nu A_\mu + g [A_\mu,
      A_\nu]\right)\\
      \eD_\mu \lambda &= \d_\mu \lambda + g [A_\mu, \lambda]~.
    \end{aligned}
  \end{equation*}
\end{caveat}

Comparing \eqref{eq:nonabelianWcomps} with the abelian version
\eqref{eq:abelianW}, we can use the results of \exref{ex:trw2} to
arrive at the component expansion for the lagrangian
\begin{equation}
  \label{eq:nonabelianSYM}
  \eL_{\text{SYM}} = \int d^2\theta \tfrac14 \Tr W^\alpha W_\alpha +
  \text{c.c}
\end{equation}
for (pure, nonabelian) supersymmetric Yang--Mills.  Expanding in
components, we obtain
\begin{equation}
  \label{eq:nonabelianSYMcomps}
  \eL_{\text{SYM}} =  \tfrac{i}{2} \Tr \left( \lambda \sigma^\mu \eD_\mu
  \bar\lambda + \bar\lambda\bar\sigma^\mu \eD_\mu \lambda \right) -
  \tfrac14 \Tr f_{\mu\nu} f^{\mu\nu} + 2 \Tr D^2~.
\end{equation}

In order to fix the correspondence with the component theory discussed
in Lecture~\ref{lec:SYM}, we need again to compare the supersymmetry
transformations.  As in the abelian theory this is once again easiest
to do in the Wess--Zumino gauge, provided that we then perform a
compensating gauge transformation to get the result back to that
gauge.  In other words, we define the supersymmetry transformation of
the nonabelian vector superfield $V$ in the Wess--Zumino gauge by
\begin{equation*}
  \begin{aligned}
    \delta_\varepsilon V &= \theta\sigma^\mu\bar\theta
    \delta_\varepsilon v_\mu + \bar\theta^2 \theta \delta_\varepsilon
    \lambda + \theta^2 \bar\theta\delta_\varepsilon \bar\lambda +
    \theta^2\bar\theta^2 \delta_\varepsilon D\\
    &= - (\varepsilon Q + \bar\varepsilon\bar Q) V - (\Lambda +
    \bar\Lambda) - \half \left[V,\Lambda - \bar\Lambda\right] -
    \tfrac1{12} \left[V,\left[V,\Lambda + \bar\Lambda\right]\right]~,
  \end{aligned}
\end{equation*}
where $\Lambda$ is chosen in such a way that the right hand side in
the second line above is again in the Wess--Zumino gauge.  This
calculation has been done already in the abelian case in
\exref{ex:susyvectorWZ} and we can use much of that result.  The only
difference in the nonabelian case are the commutator terms in the
expression of the gauge transformation: compare the above expression
for $\delta_\varepsilon V$ and equation \eqref{eq:compensating}.

\begin{exercise}
  Let $V$ be a nonabelian vector superfield in the Wess--Zumino
  gauge.  Follow the procedure outlined above to determine the
  supersymmetry transformation laws of the component fields.  In other 
  words, compute
  \begin{equation*}
    \delta_\varepsilon V := - (\varepsilon Q + \bar\varepsilon\bar Q)
    V - (\Lambda + \bar\Lambda) - \half \left[V,\Lambda -
    \bar\Lambda\right] - \tfrac1{12} \left[V,\left[V,\Lambda +
    \bar\Lambda\right]\right]
  \end{equation*}
  for an appropriate $\Lambda$ and show that, after rescaling the
  vector superfield $V \mapsto 2g\, V$, one obtains
  \begin{equation}
    \label{eq:susyNAvectorcomps}
    \begin{aligned}
      \delta_\varepsilon v_\mu &= i \left( \varepsilon \sigma_\mu
      \bar\lambda +  \bar\varepsilon \bar\sigma_\mu \lambda \right)\\
      \delta_\varepsilon \lambda_\alpha &= -2 \varepsilon_\alpha D +
      \tfrac{i}{2} (\sigma^{\mu\nu})_{\alpha\beta} \varepsilon^\beta
      f_{\mu\nu}\\
      \delta_\varepsilon D &= \tfrac{i}{2}\left( \bar\varepsilon
      \bar\sigma^\mu \eD_\mu \lambda - \varepsilon \sigma^\mu
      \overline{\eD_\mu \lambda}\right)~.
    \end{aligned}
  \end{equation}
  Now expand the supersymmetry transformation law
  \eqref{eq:superpoincareYM} with $\Psi=(\psi^\alpha,
  \bar\psi_{\dot\alpha})$ and show that the result agrees with
  \eqref{eq:susyNAvectorcomps} after eliminating the auxiliary field,
  and provided that we identify $A_\mu =  -i v_\mu$ and $\psi^\alpha =
  i \lambda_\alpha$.
\end{exercise}

In summary, the supersymmetric Yang--Mills theory discussed in
Lecture~\ref{lec:SYM} has a superspace description in terms of a
vector superfield
\begin{equation*}
  V = i \bar\theta^2 \bar\sigma^\mu \theta A_\mu - i \bar\theta^2 \theta 
  \psi + i \theta^2 \bar\theta\bar\psi + \theta^2\bar\theta^2 D
\end{equation*}
 with lagrangian
\begin{equation*}
  \int d^2\theta \, \Tr \tfrac14 W^\alpha W_\alpha + \text{c.c.}~,
\end{equation*}
where $W_\alpha$ is given by \eqref{eq:nonabelianfieldstrength}.

To be perfectly honest we have omitted one possible term in the action
which is present whenever the center of the Lie algebra $\fg$ is
nontrivial; that is, whenever there are $\U(1)$ factors in the gauge
group.  Consider the quantity $\Tr \kappa V$ where $\kappa = \kappa^i
T_i$ is a constant element in the center of the Lie algebra.  This
yields a term in the action called a \emph{Fayet--Iliopoulos} term
and, as we will see in Lecture~\ref{lec:SB}, it plays an important
role in the spontaneous breaking of supersymmetry.

\begin{exercise}
  Show that the Fayet--Iliopoulos term
  \begin{equation*}
    \int d^2\theta\, d^2\bar\theta\, \Tr \kappa V = \Tr \kappa D
  \end{equation*}
  is both supersymmetric and gauge-invariant.
\end{exercise}

\subsection{Gauge-invariant interactions}

Having constructed the gauge-invariant action for pure supersymmetric
Yang--Mills and having already seen the coupling to matter
\begin{equation}
  \int d^2\theta d^2\bar\theta \, \bar\bPhi e^{2g V} \bPhi~,
\end{equation}
there remains one piece of the puzzle in order to be able to construct
the most general renormalisable supersymmetric field theory in four
dimensions: a gauge-invariant superpotential.  On dimensional
grounds, we saw that the most general renormalisable superpotential is 
a cubic polynomial
\begin{equation}
  \label{eq:GIsuperpotential}
  W(\bPhi) = a_I \Phi^I + \half m_{IJ} \Phi^I \Phi^J + \tfrac13
  \lambda_{IJK} \Phi^I \Phi^J \Phi^K
\end{equation}
where the $\{\Phi^I\}$ are chiral superfields---the components of
$\bPhi$ relative to some basis $\{\be_I\}$ for the representation.

\begin{exercise}
  Prove that $W(\bPhi)$ is gauge invariant if and only if $a_I$,
  $m_{IJ}$ and $\lambda_{IJK}$ are (symmetric) invariant tensors in the 
  representation corresponding to $\bPhi$.
\end{exercise}

Let us end by summarising what we have learned in this lecture.  The
general renormalisable supersymmetric action is built out of vector
superfields $V$ taking values in the Lie algebra of a compact Lie
group $G$ and a chiral superfield $\bPhi$ taking values in a unitary
representation, not necessarily irreducible.  The lagrangian can be
written as follows:
\begin{multline}
  \label{eq:SYMFIM}
    \int d^2 \theta d^2\bar\theta\, \left( \bar\bPhi e^{2g V} \bPhi +
      \Tr \kappa V \right)\\
    + \left[ \int d^2 \theta\, \left( \tfrac14 \Tr W^\alpha W_\alpha +
      W(\bPhi) \right) + \text{c.c.} \right]~,
\end{multline}
with $W(\bPhi)$ given in \eqref{eq:GIsuperpotential} where $a_I$,
$m_{IJ}$ and $\lambda_{IJK}$ are (symmetric) $G$-invariant tensors in
the matter representation.

\begin{caveat}
  Strictly speaking when the group is not simple, one must then
  restore the Yang--Mills coupling separately in each factor of the
  Lie algebra by rescaling the corresponding vector superfield by
  $2g$, where the coupling constant $g$ can be different for each
  factor, and rescaling the spinorial field-strength accordingly.
  This is possible because the Lie algebra of a compact Lie group
  splits as the direct product of several simple Lie algebras and an
  abelian Lie algebra, itself the product of a number of $\U(1)$'s.
  The Yang--Mills superfield breaks up into the different factors and
  neither the metric nor the Lie bracket couples them.
\end{caveat}

We end this lecture by mentioning the names of the particles
associated with the dynamical fields in the different superfields.  In
the vector superfield, the vector corresponds to the gauge bosons,
whereas its fermionic superpartner is the \emph{gaugino}.  The
supersymmetric partner of the photon and the gluons are called the
\emph{photino} and \emph{gluinos}, respectively.  There are two kinds
of chiral superfield in phenomenological models, corresponding to the
Higgs scalars and the quarks and leptons.  In the former case the
scalars are the Higgs fields and their fermionic partners are the
\emph{Higgsinos}.  In the latter case, the fermions correspond to
either quarks or leptons and their bosonic partners are the
\emph{squarks} and \emph{sleptons}.

\begin{tut}[\textsc{Kähler quotients and the $\CC P^N$ model}]
  \indent\par In this problem we will study the ``moduli space of
  vacua'' of a supersymmetric gauge theory and show that, in the
  absence of superpotential, it is given by a ``Kähler quotient.'' The
  low-energy effective theory is generically a sigma model in the
  moduli space of vacua and we will illustrate this in the so-called
  $\CC P^N$ model.
  
  Let $\Phi^I$, for $I=1,\dots,N$ be $N$ chiral superfields, which we
  will assemble into an $N$-dimensional vector $\bPhi$.  Let
  $\bar\bPhi$ denote the conjugate transpose vector.  It is an
  $N$-dimensional vector of antichiral superfields.

  \begin{enumerate}
  \item Check that the kinetic term
    \begin{equation*}
      \int d^2\theta\, d^2\bar\theta\, \bar\bPhi \bPhi~,
    \end{equation*}
    is invariant under the natural action of $\U(N)$
    \begin{equation*}
      \bPhi \mapsto e^X \bPhi~,
    \end{equation*}
    where $X$ is a constant antihermitian matrix. 
  \end{enumerate}
  
  Let us gauge a subgroup $G \subset \U(N)$ in this model by
  introducing a nonabelian vector superfield $V=V^i (i T_i)$, where
  $\{T_i\}$ is a basis for the Lie algebra $\fg$ of $G$.  Since $G$ is
  a subgroup of the unitary group, the $T_i$ are antihermitian
  matrices.  As we have seen in this lecture, the coupled theory has
  the following lagrangian
  \begin{equation*}
    \int d^2\theta\, d^2\bar\theta\, \left( \bar\bPhi e^{2g\, V} \bPhi -
    2g \kappa^2 \Tr V \right) + \left[
    \int d^2\theta\, \Tr \tfrac14 W^\alpha W_\alpha + \text{c.c.}
    \right]~,
  \end{equation*}
  where $W_\alpha$ is given in \eqref{eq:nonabelianfieldstrength}, and
  where we have introduced a conveniently normalised Fayet--Iliopoulos
  term, since $G$ may have an abelian factor.
  
  A choice of vacuum expectation values of the dynamical scalars in
  the chiral superfield yields a point $z^I = \left< \phi^I \right>$
  in $\CC^N$.  Let $\eM_0 \subset \CC^N$ correspond to those points
  $\bz = (z^I)$ which minimise the potential of the theory.

  \begin{enumerate}
  \item[2.] Show that $\eM_0$ consists of those points $\bz$ in
    $\CC^N$ such that
    \begin{equation*}
      \bar\bz T_i \bz = \kappa^2 \Tr T_i \qquad\text{for all
        $i$,}
    \end{equation*}
    and that the potential is identically zero there.
  \end{enumerate}
  
  \textbf{Notation:} Let $\fg^*$ denote the dual vector space of the
  Lie algebra $\fg$.  Let us define a \emph{momentum map} $\mu: \CC^N
  \to \fg^*$ as follows.  If $\bz \in \CC^N$ then $\mu(\bz)$ is the
  linear functional on $\fg$ which sends $X\in\fg$ to the \emph{real}
  number
  \begin{equation*}
    \left< \mu(\bz), X \right> := i \left(\bar\bz X \bz - \kappa^2 \Tr
    X\right)~.
  \end{equation*}

  \begin{enumerate}
  \item[3.] Show that $\eM_0$ agrees with $\mu^{-1}(0)$; in other
    words,
    \begin{equation*}
      \bz \in \eM_0 \iff \mu(\bz) = 0~.
    \end{equation*}
  \end{enumerate}
  
  Since we have identified $\CC^N$ as the space of vacuum expectation
  values of the dynamical scalar fields, the action of $G$ on the
  fields induces an action of $G$ on $\CC^N$:
  \begin{equation*}
    \bz \mapsto e^X \bz~,
  \end{equation*}
  where $X \in \fg$ is an antihermitian matrix.

  \begin{enumerate}
  \item[4.] Show that $\eM_0$ is preserved by the action of $G$,
    so that if $\bz \in \eM_0$ then so does $e^X \bz$ for all
    $X\in\fg$.
  \end{enumerate}
  
  Since in a gauge theory field configurations which are related by a
  gauge transformations are physically indistinguishable, we have to
  identify gauge related vacua $\bz \in \eM_0$.  This means that the
  \emph{moduli space of vacua} is the quotient
  \begin{equation*}
    \eM := \eM_0 / G~,
  \end{equation*}
  which by the above result is well-defined.  It can be shown that
  $\eM$ admits a natural Kähler metric.  With this metric, $\eM$ is
  called the \emph{Kähler quotient} of $\CC^N$ by $G$.  It is
  often denoted $\CC^N/\!/G$.
  \begin{scholium}
    One of the beautiful things about supersymmetry is that it allows
    us to understand this fact in physical terms.  At low energies,
    only the lightest states will contribute to the dynamics.  The
    scalar content of the low-energy effective theory is in fact a
    sigma model on the moduli space of vacua.  We will see in the next
    lecture that since the potential vanishes on the space of vacua,
    supersymmetry is unbroken.  This means that the low-energy
    effective theory is supersymmetric; but by
    \probref{pr:chiralmodels} we know that the supersymmetric sigma
    models are defined on manifolds admitting Kähler metrics.
    Therefore $\eM$ must have a Kähler metric.  In fact, it is
    possible to work out the form of this metric exactly at least in
    one simple, but important, example: the $\CC P^N$ model, the
    Kähler quotient of $\CC^{N+1}$ by $\U(1)$.
  \end{scholium}
  Let us take $N+1$ chiral superfields $\bPhi = (\Phi^I)$ for
  $I=0,1,\dots,N$ and gauge the natural $\U(1)$ action
  \begin{equation*}
    \bPhi \mapsto e^{i\vartheta} \bPhi~,
  \end{equation*}
  with $\vartheta\in\RR$.  To simplify matters, let us take
  $2g=\kappa=1$.  We have one vector superfield $V = \bar V$.  The
  lagrangian is given by
  \begin{equation*}
    \int d^2\theta\, d^2\bar\theta\, \left( \bar\bPhi e^V \bPhi - V
    \right) + \left[ \int d^2\theta\, \Tr \tfrac14 W^\alpha W_\alpha +
    \text{c.c.} \right]~.
  \end{equation*}
  The space $\eM_0$ of minima of the potential is the unit sphere in
  $\CC^{N+1}$:
  \begin{equation*}
    \bar \bz \bz = 1~.
  \end{equation*}
  The moduli space of vacua is obtained by identifying each $\bz$ on
  the unit sphere with $e^{i\vartheta} \bz$ for any $\vartheta\in\RR$.
  The resulting space is a compact smooth manifold, denoted $\CC P^N$
  and called the complex projective space.  It is the space of complex
  lines through the origin in $\CC^{N+1}$.  The natural Kähler metric
  on $\CC P^N$ is the so-called Fubini--Study metric.  Let us see how
  supersymmetry gives rise to this metric.

  \begin{enumerate}
  \item[5.] Choose a point in $\eM_0$ and expanding around that point,
    show that the $\U(1)$ gauge symmetry is broken and that the photon
    acquires a mass.
  \end{enumerate}

  Since supersymmetry is not broken (see the next lecture) its
  superpartner, the photino, also acquires a mass.  For energies lower 
  than the mass of these fields, we can disregard their dynamics.  The 
  low-energy effective action becomes then
  \begin{equation*}
    \int d^2\theta\, d^2\bar\theta\, \left( \bar\bPhi e^V \bPhi - V
    \right)~.
  \end{equation*}

  \begin{enumerate}
  \item[6.] Eliminate $V$ using its (algebraic) equations of motion to 
    obtain the following action:
    \begin{equation*}
      \int d^2\theta\, d^2\bar\theta\, \log(\bar\bPhi \bPhi)~.
    \end{equation*}
  \item[7.] Show that this action is still invariant under the abelian
    gauge symmetry $\bPhi \mapsto e^{i\Lambda} \bPhi$, with $\Lambda$
    a chiral superfield.
  \item[8.] Use the gauge symmetry to fix, $\Phi^0 = 1$, say, and
    arrive at the following action in terms of the remaining chiral
    superfields $\Phi^I$, $I=1,\dots,N$:
    \begin{equation*}
      \int d^2\theta\, d^2\bar\theta\, \log(1 + \sum_{I=1}^N
      \Phi^I\bar\Phi_I)~.
    \end{equation*}
    \begin{caveat}
      This is only possible at those points where $\phi^0$ is
      different from zero.  This simply reflects the fact that $\CC
      P^N$, like most manifolds, does not have global coordinates.
    \end{caveat}
  \item[9.] Expand the action in components to obtain
  \begin{equation*}
    - g_{I\bar J}(\phi,\bar\phi) \d_\mu \phi^I \d^\mu \bar\phi^{\bar
      J} + \cdots
  \end{equation*}
  where $g_{I\bar J}$ is the Fubini--Study metric for $\CC P^N$.  Find
  the metric explicitly.
  \end{enumerate}
\end{tut}


\section{Spontaneous supersymmetry breaking}
\label{lec:SB}

In the previous lecture we have learned how to write down
renormalisable supersymmetric models in four dimensions.  However if
supersymmetry is a symmetry of nature, it must be broken, since we do
not observe the mass degeneracy between bosons and fermions that
unbroken supersymmetry demands.  There are three common ways to break
supersymmetry:
\begin{itemize}
\item Introducing symmetry breaking terms explicitly in the action
  (\emph{soft});
\item Breaking tree-level supersymmetry by quantum effects, either
  perturbatively or nonperturbatively (\emph{dynamical}); and
\item Breaking supersymmetry due to a choice of non-invariant vacuum
  (\emph{spontaneous}).
\end{itemize}
We will not discuss dynamical supersymmetry breaking in these lectures,
except to note that nonrenormalisation theorems usually forbid the
perturbative dynamical breaking of supersymmetry.  Neither will we
discuss soft supersymmetry breaking, except to say that this means that
the supersymmetric current is no longer conserved, and this forbids
the coupling to (super)gravity.  We will concentrate instead on
spontaneous supersymmetry breaking.

\begin{caveat}
  I should emphasise, however, that from the point of view of
  supersymmetric field theories (that is, ignoring (super)gravity) the
  most realistic models do involve soft breaking terms.  These terms
  are the low-energy manifestation of the spontaneous breaking (at
  some high energy scale) of local supersymmetry, in which the
  gravitino acquires a mass via the super-Higgs mechanism.
\end{caveat}

\subsection{Supersymmetry breaking and vacuum energy}

\begin{figure}[h!]
\centering
\def\jNO{$\times$}
\def\jSI{$\aok$}
\mbox{%
  \subfigure[SUSY~\jNO\qquad Gauge~\jNO]
  {\includegraphics[width=0.45\textwidth]{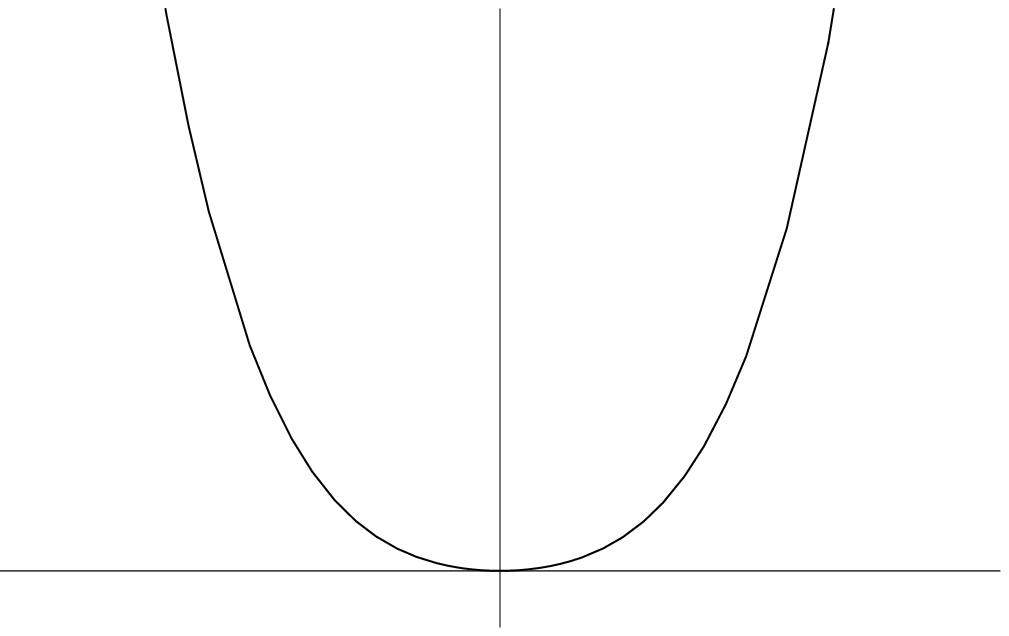}}
\qquad
\subfigure[SUSY~\jSI\qquad Gauge~\jNO]
{\includegraphics[width=0.45\textwidth]{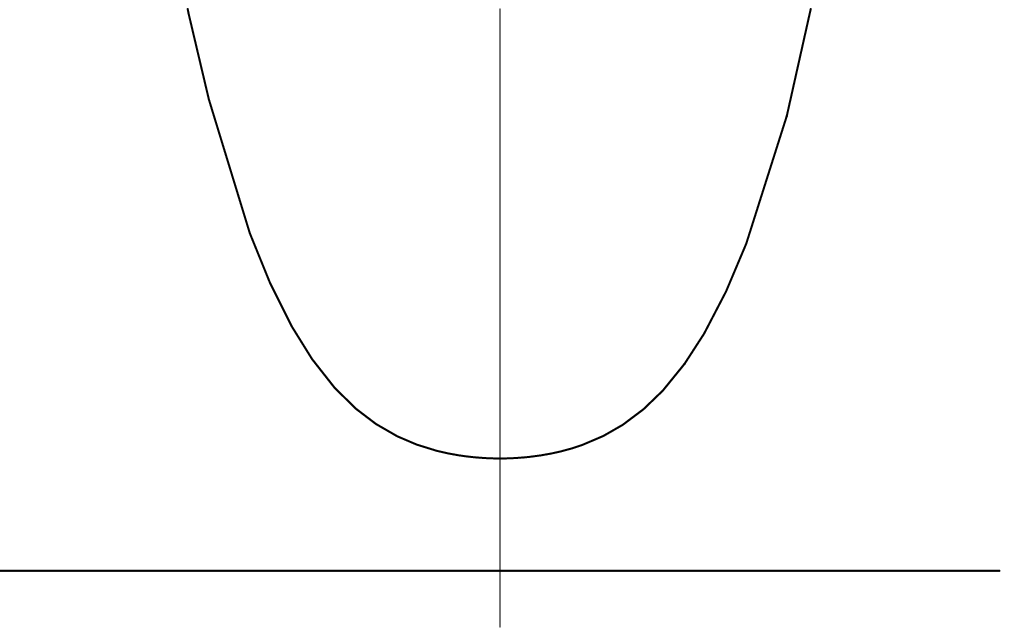}}
}\\
\mbox{%
  \subfigure[SUSY~\jNO\qquad Gauge~\jSI]
{\includegraphics[width=0.45\textwidth]{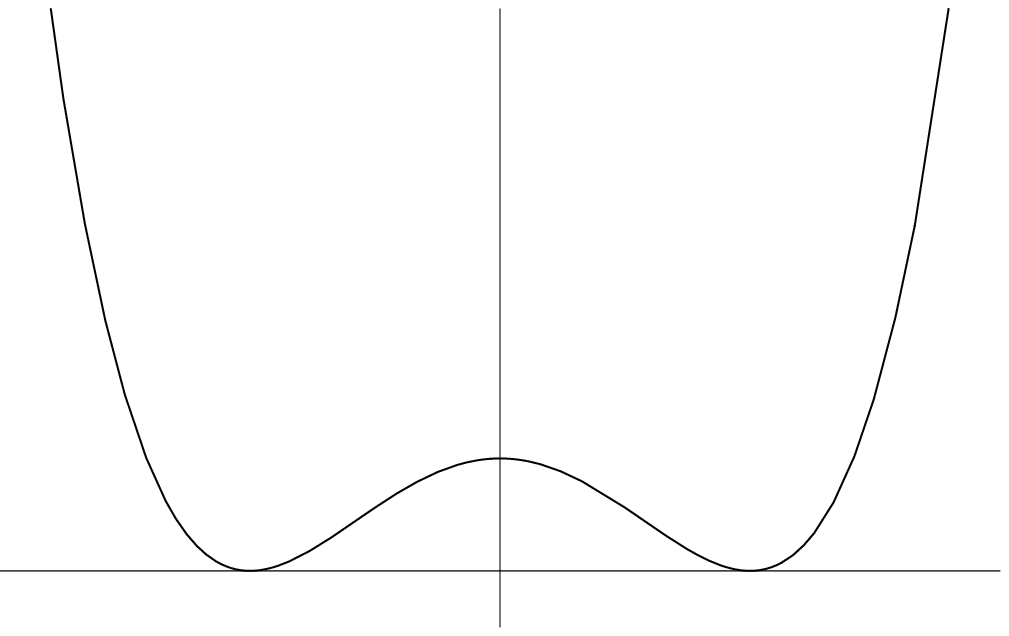}}
\qquad
\subfigure[SUSY~\jSI\qquad Gauge~\jSI]
{\includegraphics[width=0.45\textwidth]{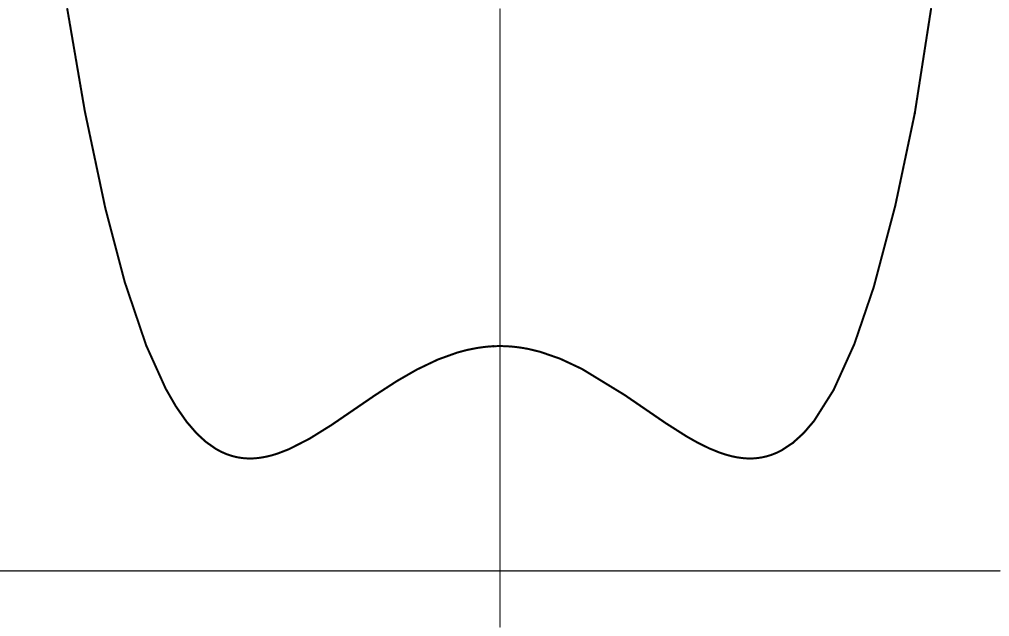}}
}
\caption{Generic forms of scalar potentials, indicating which symmetry 
  is broken (denoted by a $\aok$) for each potential.}
\label{fig:potentials}
\end{figure}

We saw in Lecture~\ref{lec:Reps} the remarkable fact that in
supersymmetric theories the energy is positive-semidefinite.  This
means in particular that the lowest-energy state---the vacuum, denoted
$\vac$---has non-negative energy.  Indeed, applying the hamiltonian to
the vacuum and using \eqref{eq:Hsusy}, we obtain
\begin{multline*}
  \cav H \vac\\
  = \tfrac14 \left( \|\sQ_1\vac\|^2 + \|\sQ_1^\dagger\vac\|^2 +
  \|\sQ_2\vac\|^2 + \|\sQ_2^\dagger\vac\|^2 \right)~,
\end{multline*}
from where we deduce that the vacuum has zero energy if and only if it
is supersymmetric, that is, if and only if it is annihilated by the
supercharges.  This gives an elegant restatement of the condition for
the spontaneous breaking of supersymmetry: \emph{supersymmetry is
spontaneously broken if and only if the vacuum energy is positive}.
This is to be contrasted with the spontaneous breaking of gauge
symmetries, which is governed by the shape of the potential of the
dynamical scalar fields.  Spontaneous breaking of supersymmetry is
impervious to the shape of the potential, but only to the minimum
value of the energy.  Figure~\ref{fig:potentials} illustrates this
point.  Whereas only potentials (b) and (d) break supersymmetry, the
potentials breaking gauge symmetry are (c) and (d).

\begin{caveat}
  You may ask whether one cannot simply shift the zero point energy in
  order to make it be precisely zero at the minimum of the potential.
  In contrast with nonsupersymmetric theories, the energy is now
  dictated by the symmetry, since the hamiltonian appears in the
  supersymmetry algebra.
\end{caveat}

\subsection{Supersymmetry breaking and VEVs}

Another criterion of spontaneous supersymmetry breaking can be given
in terms of vacuum expectation values of auxiliary fields.

We start with the observation that supersymmetry is spontaneously
broken if and only if there is some field $\varphi$ whose
supersymmetry variation has a nontrivial vacuum expectation value:
\begin{equation}
  \label{eq:vev}
  \cav \, \delta_\varepsilon \varphi\, \vac \neq 0~.
\end{equation}
Indeed, notice that $\delta_\varepsilon \varphi =
-\left[\varepsilon\sQ + \bar\varepsilon\bar\sQ, \varphi\right]$ as
quantum operators, hence
\begin{equation*}
  \cav\, \delta_\varepsilon \varphi \, \vac  = -\varepsilon^\alpha
  \cav\, [\sQ_\alpha, \varphi] \, \vac - \bar\varepsilon^{\dot\alpha}
  \cav\, [(\sQ_\alpha)^\dagger, \varphi] \, \vac~.
\end{equation*}
Because Lorentz invariance is sacred, no field which transform
nontrivially under the Lorentz group is allowed to have a nonzero
vacuum expectation value.  Since supersymmetry exchanges bosons with
fermions, and fermions always transform nontrivially under the Lorentz
group, it means that the field $\varphi$ in equation \eqref{eq:vev}
must be fermionic.  By examining the supersymmetry transformation
laws for the fermionic fields in the different superfields we have met
thus far, we can relate the spontaneous breaking of supersymmetry to
the vacuum expectation values of auxiliary fields.  This illustrates
the importance of auxiliary fields beyond merely ensuring the
off-shell closure of the supersymmetry algebra.

Let's start with the chiral superfields.  Equation
\eqref{eq:susychiral} describes how the fermions in the chiral
superfield transform under supersymmetry.  Only the dynamical scalar
and the auxiliary field can have vacuum expectation values, and only
the vacuum expectation value of the auxiliary field can give a nonzero 
contribution to equation \eqref{eq:vev}.  This sort of supersymmetry
breaking is known as $F$-term (or O'Raifeartaigh) supersymmetry
breaking.

In the case of the vector superfields, the transformation law of the
fermion is now given by equation \eqref{eq:susyNAvectorcomps}.  Only
the auxiliary field can have a nonzero vacuum expectation value and
hence give a nonzero contribution to \eqref{eq:vev}.  This sort of
supersymmetry breaking is known as $D$-term supersymmetry breaking and
will be discussed in more detail below.  Notice however that giving a
nonzero vacuum expectation value to $D$ breaks gauge invariance unless
$D$, which is Lie algebra valued, happens to belong to the center;
that is, to have vanishing Lie brackets with all other elements in the
Lie algebra.  This requires the gauge group to have abelian factors.

\begin{scholium}
  Notice that when either the $F$ or $D$ auxiliary fields acquire
  nonzero vacuum expectation values, the transformation law of some
  fermion contains an inhomogeneous term:
  \begin{equation*}
    \delta_\varepsilon \lambda_\alpha = - 2 \varepsilon_\alpha
    \left<D\right> + \cdots\qquad\text{and}\qquad
    \delta_\varepsilon \chi_\alpha = - 2 \varepsilon_\alpha
    \left<F\right> + \cdots
  \end{equation*}
  Such a fermion is called a \emph{Goldstone fermion}, by analogy with 
  the Goldstone boson which appears whenever a global continuous
  symmetry is spontaneously broken.  Just like in the standard Higgs
  mechanism, wherein a vector boson ``eats'' the Goldstone boson to
  acquire mass, in a supergravity theory the gravitino acquires a mass 
  by eating the Goldstone fermion, in a process known as the
  super-Higgs mechanism.
\end{scholium}

\subsection{The O'Raifeartaigh model}

We now consider a model which breaks supersymmetry spontaneously
because of a nonzero vacuum expectation value of the $F$ field.
Consider a theory of chiral superfields $\{\Phi^i\}$.  The most
general renormalisable lagrangian was worked out in
\probref{pr:chiralmodels}.  It consists of a positive-definite kinetic 
term
\begin{equation*}
  \int d^2\theta\, d^2\bar\theta\, \sum_i \Phi^i \bar\Phi_i 
\end{equation*}
and a superpotential term
\begin{equation*}
  \int d^2\theta\, W(\bPhi) + \text{c.c.}~,
\end{equation*}
where $W(\bPhi)$ is a cubic polynomial (for renormalisability)
\begin{equation*}
  W(\bPhi) = a_i \Phi^i + \half m_{ij} \Phi^i \Phi^j + \tfrac13
  \lambda_{ijk} \Phi^i \Phi^j \Phi^k~.
\end{equation*}

In \probref{pr:chiralmodels} we found the component expression for the
above lagrangian.  From this one can read off the equations of motion
of the auxiliary fields:
\begin{equation*}
  \bar F_i = - \frac{\d W(\phi)}{\d \phi^i} = - \left( a_i + m_{ij}
  \phi^j + \lambda_{ijk} \phi^j \phi^k \right)~.
\end{equation*}
Substituting this back into the lagrangian, one gets the potential
energy term:
\begin{equation*}
  \eV = \sum_i \bar F_i F^i = \sum_i \left| -\frac{\d W(\phi)}{\d
  \phi^i} \right|^2 = \sum_i \left| a_i + m_{ij} \phi^j +
  \lambda_{ijk} \phi^j \phi^k \right|^2~.
\end{equation*}
This potential is positive-semidefinite.  It breaks supersymmetry if
and only if there exist no vacuum expectation values
$\left<\phi^i\right>$ such that $\left< F^i \right> = 0$ for all $i$.
Notice that if $a_i=0$, then $\left<\phi^i\right>=0$ always works, so
that supersymmetry is not broken unless $a_i\neq 0$.  Can we find
superpotentials $W(\bPhi)$ for which this is the case?

It turns out that one cannot find any interesting (i.e., interacting)
such theories with less than three chiral superfields.

\begin{exercise}
  Prove that if there is only one chiral superfield $\Phi$, then the
  only cubic superpotential which breaks supersymmetry consists is
  $W(\Phi) = a \Phi$, so that the theory is free.
\end{exercise}

In fact the same is true for two chiral superfields, although the
proof is more involved.  The simplest model needs three chiral
superfields $\Phi_0$, $\Phi_1$ and $\Phi_2$.  This is the
\emph{O'Raifeartaigh model} and is described by the following
superpotential:
\begin{equation*}
  W(\bPhi) = \mu \Phi_1 \Phi_2 + \lambda \Phi_0 \left(\Phi_1^2 -
  \alpha^2\right)~,
\end{equation*}
where $\alpha$, $\mu$ and $\lambda$ can be chosen to be real and
positive by changing, if necessary, the overall phases of the chiral
superfields and of $W$.

\begin{exercise}
  Show that this superpotential is determined uniquely by the
  requirements of renormalisability, invariance under the R-symmetry
  \begin{equation*}
    \sR \cdot \Phi_0 = \Phi_0 \qquad
    \sR \cdot \Phi_1 = 0 \qquad
    \sR \cdot \Phi_2 = \Phi_2~,
  \end{equation*}
  and invariance under the discrete $\ZZ_2$ symmetry
  \begin{equation*}
    \Phi_0 \mapsto \Phi_0 \qquad 
    \Phi_1 \mapsto -\Phi_1 \qquad 
    \Phi_2 \mapsto -\Phi_2~.
  \end{equation*}
\end{exercise}

The equations of motion of the auxiliary fields are given by
\begin{equation*}
  \begin{aligned}
    \bar F_0 &= - \lambda \left( \phi_1^2 - \alpha^2
    \right)\\
    \bar F_1 &= - \left( \mu \phi_2 - 2 \lambda \phi_0 \phi_1\right)\\ 
    \bar F_2 &= - \mu \phi_1~.
\end{aligned}
\end{equation*}

\begin{exercise}
  Show that the above superpotential breaks supersymmetry
  spontaneously provided that $\lambda$, $\mu$ and $\alpha$ are
  nonzero.
\end{exercise}

Let us introduce complex coordinates $z_i = \left<\phi_i\right>$.  The 
potential defines a function $\eV: \CC^3 \to \RR$, which is actually
positive:
\begin{equation*}
  \eV = \lambda^2 |z_1^2-\alpha^2|^2 + \mu^2 |z_1|^2 + |\mu z_2^2 - 2
  \lambda z_0 z_1|^2~.
\end{equation*}
To minimise the potential, notice that provided that $\mu\neq 0$, we
can always set $z_2$ such that the last term vanishes for any values
of $z_0$ or $z_1$.  The other two terms depend only on $z_1$, hence
the potential will have a flat direction along $z_0$.

\begin{exercise}
  Show that provided $\mu^2 \geq 2 \lambda^2 \alpha^2$, the minimum of
  the potential $\eV$ is at $z_1=z_2=0$ and arbitrary values of $z_0$.
  Compute the spectrum of masses in this case and show that there is a
  massless fermion, which can be identified with the
  Goldstone fermion.\\
  {\rm (Hint: The masses will depend on $z_0$, but the fact there
    exists a massless fermion has to do with the vanishing of the
    determinant of the fermion mass matrix, and this is the case for
    all $z_0$.)}
\end{exercise}

Notice that the existence of the Goldstone fermion was inferred from
the vanishing of the determinant of the fermion mass matrix.  This
comes from the superpotential term and is protected from quantum
corrections.  But even if this were not the case, it is clear that
under radiative corrections the condition that the vacuum energy is
positive is stable under deformations, in the sense that this
condition is preserved under small perturbations.  In the language of
(point set) topology, one would say that this is an \emph{open}
condition: meaning that in the relevant space of deformation
parameters, every point for which the vacuum energy is positive has a
neighbourhood consisting of points which share this property.  This is
illustrated in Figure~\ref{fig:open} below, where the dashed lines
indicate deformations of the potential, drawn with a solid line.

\begin{figure}[h!]
  \centering
  \includegraphics[width=0.40\textwidth]{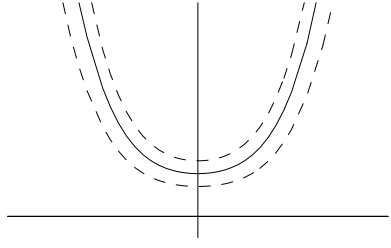}
  \caption{Potentials with positive vacuum energy are stable under
    deformations.}
\label{fig:open}
\end{figure}

\begin{scholium}
  How about chiral superfields coupled to gauge fields?  Ignoring for
  the moment the Fayet--Iliopoulos terms, which will be the subject of
  the next section, let me just mention that it is possible to show
  that in the absence of Fayet--Iliopoulos terms, it is the
  O'Raifeartaigh mechanism again which governs the spontaneous
  breaking of supersymmetry, in the sense that if the $F$ equations of
  motion ($F^i = 0$) are satisfied for some scalar vacuum expectation
  values, then it is possible to use the ``global'' gauge symmetry,
  which is a symmetry of the superpotential and hence of the $F$
  equations of motion, in order to find (possibly different) vacuum
  expectation values such that the $D$-equations of motion ($D^i = 0$)
  are also satisfied.
\end{scholium}

\subsection{Fayet--Iliopoulos terms}

The O'Raifeartaigh model breaks supersymmetry because of the linear
term in the superpotential (the $F$ term), which gives a nonzero
vacuum expectation value to the auxiliary field in the chiral
superfield.  It is also possible to break supersymmetry by giving a
nonzero vacuum expectation value to the auxiliary field in the vector
superfield.  This is possible by adding a Fayet--Iliopoulos term to
the action.  Gauge invariance requires that the Fayet--Iliopoulos term
belong to the center of the Lie algebra $\fg$ of the gauge group.
Since the gauge group is compact, its Lie algebra is the direct
product of a semisimple Lie algebra and an abelian Lie algebra.
Semisimple Lie algebras have no center, hence for the
Fayet--Iliopoulos term to exist, there has to be a nontrivial abelian
factor.  In other words, the gauge group must have at least one
$\U(1)$ factor.  To illustrate this phenomenon, we will actually
consider an abelian Yang--Mills theory with gauge group $\U(1)$:
supersymmetric QED, with superspace lagrangian \eqref{eq:sQED}, except 
that we also add a Fayet--Iliopoulos term $\kappa V$ to the
superspace lagrangian:
\begin{multline*}
    \int d^2 \theta d^2\bar\theta\, \left( \bar\Phi_+ e^{2eV} \Phi_+ +
      \bar\Phi_- e^{-2eV} \Phi_- + \kappa V\right)\\
    + \left[ \int d^2 \theta\, \left( \tfrac14 W^\alpha W_\alpha + m
      \Phi_+\Phi_-\right) + \text{c.c.} \right]~.
\end{multline*}

The potential energy terms are
\begin{multline*}
  2D^2 + \kappa D + 2eD \left(|\phi_+|^2 - |\phi_-|^2\right)\\
  + |F_+|^2 + |F_-|^2 + m \left( F_+ \phi_- + F_- \phi_+ + \bar F_+
    \bar \phi_- + \bar F_- \bar \phi_+ \right)~.
\end{multline*}
Eliminating the auxiliary fields via their equations of motion
\begin{equation*}
  \begin{aligned}
    F_\pm &= - m \bar\phi_\mp\\
    D &= -\tfrac14 \left( \kappa + 2 e \left( |\phi_+|^2 - |\phi_-|^2
    \right) \right)
  \end{aligned}
\end{equation*}
we obtain the potential energy
\begin{equation*}
  \eV = \tfrac18 \left( \kappa + 2 e  \left( |\phi_+|^2 - |\phi_-|^2
  \right) \right)^2 + m^2 \left( |\phi_-|^2 + |\phi_+|^2\right)~.
\end{equation*}
Notice that for nonzero $\kappa$ supersymmetry is spontaneously
broken, since it is impossible to choose vacuum expectation values for
the scalars such that $\left<F_\pm\right> = \left< D\right> = 0$.

Expanding the potential
\begin{multline*}
  \eV = \tfrac18 \kappa^2 + (m^2 - \half e\kappa) |\phi_-|^2 + (m^2 +
  \half e\kappa) |\phi_+|^2 + \half e^2 \left( |\phi_+|^2 -
    |\phi_-|^2\right)^2
\end{multline*}
we notice that there are two regimes with different qualitative
behaviours.

If $m^2 > \half e\kappa$ the minimum of the potential occurs for
$\left<\phi_+\right> = \left<\phi_-\right> = 0$ and the model
describes two complex scalars with masses $m_\mp^2 = m^2 \pm \half
e\kappa$.  The electron mass $m$ does not change, and the photon
and photino remain massless.  Hence supersymmetry is spontaneously
broken---the photino playing the rôle of the Goldstone fermion---and
the gauge symmetry is unbroken.  This is the situation depicted by
the potential of the type (b) in Figure~\ref{fig:potentials}.

On the other hand if $m^2 < \half e\kappa$, the minimum of the
potential is no longer at $\left<\phi_+\right> = \left<\phi_-\right> = 0$.
Instead we see that the minimum happens at $\left<\phi_+\right>=0$ but 
at $\left<\phi_-\right> = z$ where
\begin{equation*}
  |z|^2 = \left( \frac{\kappa}{2e} - \frac{m^2}{e^2}\right)~.
\end{equation*}
There is a circle of solutions corresponding to the phase of $z$.  We
can always choose the global phase so that $z$ is real and positive:
\begin{equation*}
  z = \sqrt{\frac{\kappa}{2e} - \frac{m^2}{e^2}}~.
\end{equation*}

\begin{exercise}
  Expand around $\left<\phi_+\right> = 0$ and $\left<\phi_-\right> =
  z$ and compute the mass spectrum.  Show that the photon acquires a
  mass, signalling the spontaneous breaking of the $\U(1)$ gauge
  symmetry, but that there is a massless fermion in the spectrum,
  signalling the spontaneous breaking of supersymmetry.
\end{exercise}

The situation is now the one depicted by the potential of type (d) in
Figure~\ref{fig:potentials}.

\subsection{The Witten index}

Finally let us introduce an extremely important concept in the
determination of supersymmetry breaking.  In theories with complicated
vacuum structure it is often nontrivial to determine whether
supersymmetry is broken.  The Witten index is a quantity which can
help determine when supersymmetry is \emph{not} broken, provided that
one can actually compute it.  Its computation is facilitated by the
fact that it is in a certain sense a ``topological'' invariant.

Suppose that we have a supersymmetric theory, by which we mean a
unitary representation of the Poincaré superalgebra on some Hilbert
space $\eH$.  We will furthermore assume that $\eH$ decomposes as a
direct sum (or more generally a direct integral) of energy eigenstates
\begin{equation*}
  \eH = \bigoplus_{E\geq 0}\, \eH_E~,
\end{equation*}
with each $\eH_E$ finite-dimensional.  (In practice the extension to
the general case is usually straightforward.)

Let $\beta$ be a positive real number and consider the following
quantity
\begin{equation*}
  I(\beta) = \STr_\eH e^{-\beta H} = \Tr_\eH\, (-1)^F \, e^{-\beta
  H}~,
\end{equation*}
which defines the \emph{supertrace} $\STr$, and where
$H$ is the hamiltonian and $F$ is the fermion number operator.  In
particular, this means that $(-1)^F$ is $+1$ on a bosonic state and
$-1$ on a fermionic state.  We will show that $I(\beta)$ is actually
independent of $\beta$---the resulting integer is called the
\emph{Witten index} of the representation $\eH$.

The crucial observation is that in a supersymmetric theory there are
an equal number of bosonic and fermionic states with any given
positive energy.  Hence the Witten index only receives contributions
from the zero energy states, if any.  This means in particular that a
nonzero value of the Witten index signals the existence of some zero
energy state which, by the discussion at the start of this lecture,
implies that supersymmetry is \emph{not} broken.  In contrast, a zero
value for the Witten index does not allow us to conclude anything,
since all this says is that there is an equal number of bosonic and
fermionic zero energy states, but this number could either be zero
(broken supersymmetry) or nonzero (unbroken supersymmetry).

By definition,
\begin{equation*}
  I(\beta) = \sum_{E\geq 0} e^{\beta E}\, \Tr_{\eH_E}\, (-1)^F =
  \sum_{E\geq 0} e^{\beta E} \, n(E),
\end{equation*}
where
\begin{equation*}
   n(E) = \Tr_{\eH_E}\, (-1)^F = n_+(E) - n_-(E)
\end{equation*}
is the difference between the number of bosonic states with energy $E$ 
and the number of fermionic states with the same energy.  It is here
that we make use of the assumption that $\eH_E$ is finite-dimensional: 
so that $n_\pm(E)$, and hence their difference, are well-defined.

\begin{exercise}
  Show that for $E\neq 0$, $n(E)=0$.\\
  {\rm (Hint: You may find of use the expression \eqref{eq:Hsusy} for
  the hamiltonian in terms of the supercharges.)}
\end{exercise}

\begin{scholium}
  Alternatively, one can prove the $\beta$ independence of $I(\beta)$
  by taking the derivative of $I(\beta)$ and showing that the result
  vanishes as a consequence of the expression \eqref{eq:Hsusy} for the
  Hamiltonian of a supersymmetric theory, the fact that $H$ commutes
  with the supercharges, and that the supertrace of an
  (anti)commutator vanishes.  This last result (which you are
  encouraged to prove) is the super-analogue of the well-known fact
  that the trace of a commutator vanishes.
\end{scholium}

This result implies that
\begin{equation*}
  I(\beta) = \Tr_{\eH_0}\, (-1)^F = n_+(0) - n_-(0)~,
\end{equation*}
is independent of $\beta$.  This means that it can be computed for any 
value of $\beta$, for example in the limit as $\beta \to \infty$,
where the calculation may simplify enormously.  In fact, the Witten
index is a ``topological'' invariant of the supersymmetric theory.  As 
such it does not depend on parameters, here illustrated by the
independence on $\beta$.  This means that one can take couplings to
desired values, put the theory in a finite volume and other
simplifications.

\begin{scholium}
  The Witten index is defined in principle for any supersymmetric
  theory. As we saw in \probref{pr:chiralmodels}, there are
  supersymmetric theories whose data is geometric and it is to be
  expected that the Witten index should have some geometric meaning in
  this case.  In fact, the dimensional reduction to one dimension of
  the supersymmetric sigma model discussed in
  \probref{pr:chiralmodels} yields a supersymmetric quantum mechanical
  model whose Witten index equals the Euler characteristic.  More is
  true, however, and the computation of the Witten index gives a proof
  of the well-known Gauss--Bonnet theorem relating the Euler
  characteristic of the manifold to the curvature.  In fact, the
  Witten index underlies many of the topological applications of
  supersymmetry and in particular the simplest known proof of the
  Atiyah--Singer index theorem relating the analytic index of an
  elliptic operator on a manifold to the topology of that manifold.
  
  There are many deep and beautiful connections like that one between
  supersymmetry and mathematics.  Indeed, whatever the final verdict
  might be for the existence of supersymmetry (albeit broken) in
  nature, the impact of supersymmetry in mathematics will be felt for
  many years to come.
\end{scholium}

\begin{tut}[\textsc{The Higgs mechanism}]\indent\par
  In supersymmetric theories the issue of gauge symmetry breaking
  (Higgs mechanism) and supersymmetric breaking are intimately
  related.  Although the topic of this lecture has been supersymmetry
  breaking, in this tutorial you are asked to study a simple
  example of Higgs mechanism which preserves supersymmetry.  The model
  in question is an $\SU(5)$ gauge theory coupled to adjoint matter in
  the form of chiral superfields.  In other words, the model consists
  of a nonabelian vector superfield $V = V^i (iT_i)$ and an adjoint
  chiral superfield $\bPhi = \Phi^i T_i$, where $T_i$ are $5\times 5$
  traceless antihermitian matrices.  Notice that $\Phi^i$ are chiral
  superfields, hence complex, and $V^i$ are vector superfields, hence
  real.

  The superspace lagrangian has the form
  \begin{multline*}
    \int d^2 \theta\, d^2 \bar\theta\, \Tr \bar\bPhi e^{2g\,\ad V}
    \bPhi\\
    + \left[ \int d^2 \theta \left( \tfrac14 \Tr W^\alpha
        W_\alpha + W(\bPhi) \right) + \text{c.c.} \right]~,
  \end{multline*}
  where we are treating the $\bPhi$ as matrices in the fundamental
  representation, hence $V$ acts on $\bPhi$ via the matrix commutator
  (denoted $\ad V$) and $\Tr$ is the matrix trace.  Since $\SU(5)$ is
  a simple group, there is no Fayet--Iliopoulos term in this model.
  Notice that since the $T_i$ are antihermitian, the trace form $\Tr
  T_i T_j = - K_{ij}$ where $K_{ij}$ is positive-definite.

  \begin{enumerate}
  \item Show that the most general renormalisable gauge-invariant
    superpotential takes the form
    \begin{equation*}
      W(\bPhi) = \half m \Tr \bPhi^2 + \tfrac13 \lambda \Tr \bPhi^3~,
    \end{equation*}
    and argue that $m$ and $\lambda$ can be taken to be real by
    changing, if necessary, the overall phases of $W$ and of $\bPhi$.
  \item Expanding the superspace action in components and eliminating
    the auxiliary fields $F$ and $D$, show that the scalar potential
    takes the form
    \begin{equation*}
      \eV = -\half g^2 \Tr [\bar\bphi,\bphi]^2 - \Tr \overline{\nabla
      W} \nabla W~,
    \end{equation*}
    where $\nabla W$ is defined by $\Tr\nabla W T_i = -\d
    W/\d\phi^i$.
  \end{enumerate}
  
  Let us remark that since the trace form on antihermitian matrices is
  negative-definite, the above potential is actually
  positive-semidefinite---in fact, it is a sum of squares.

  \textbf{Notation:} Let $A := \left<\bphi\right>$ be the vacuum
  expectation value of $\bphi$.  It is a $5\times 5$ traceless
  antihermitian matrix.

  \begin{enumerate}
  \item[3.] Show that $A = 0$ is a minimum of the potential $\eV$.
  \end{enumerate}

  This solution corresponds to unbroken $\SU(5)$ gauge theory and,
  since the potential is zero for this choice of $A$, unbroken
  supersymmetry.  The rest of the problem explores other
  supersymmetric minima for which $\SU(5)$ is broken down to smaller
  subgroups.  As we saw in the lecture, a vacuum is supersymmetric if
  and only if it has zero energy, hence we are interested in vacuum
  expectation values $A$ for which $\eV=0$.  These vacua will be
  degenerate, since they are acted upon by the subgroup of the gauge
  group which remains unbroken.
  
  \begin{enumerate}
  \item[4.]  Show that the minima of the potential $\eV$ correspond to 
    those matrices $A$ obeying the following two equations:
    \begin{equation*}
      [A,\bar A] = 0 \qquad\text{and}\qquad m A + \lambda \left( A^2 - 
        \tfrac15 \Tr A^2\right) = 0~,
    \end{equation*}
    where $\bar A$ is the hermitian conjugate of $A$.
  \item[5.] Conclude from the first equation that $A$ can be
    diagonalised by a matrix in $\SU(5)$, hence we can assume that $A$ 
    takes the form
    \begin{equation*}
      A = 
      \begin{pmatrix}
        \mu_1 & & & &\\
        & \mu_2 & & &\\
        & & \mu_3 & &\\
        & & & \mu_4 &\\
        & & & & \mu_5
      \end{pmatrix}
    \end{equation*}
    for complex numbers $\mu_i$ obeying $\sum_i \mu_i = 0$.
  \item[6.] Assume that $\lambda\neq 0$ and show that both
    \begin{equation*}
      \frac{3m}{\lambda}
      \begin{pmatrix}
        1 & & & &\\
        & 1 & & &\\
        & & 1 & &\\
        & & & 1 &\\
        & & & & -4
      \end{pmatrix}
      \quad\text{and}\quad
      \frac{2m}{\lambda} 
      \begin{pmatrix}
        1 & & & &\\
        & 1 & & &\\
        & & 1 & &\\
        & & & -\frac32 &\\
        & & & & -\frac32
      \end{pmatrix}
    \end{equation*}
    are possible choices for $A$ which solve the equations.  Which
    subgroup of $\SU(5)$ remains unbroken in each case?\\
    {\small (Answers: The groups are $\mathrm{S}\left(\U(4) \times
        \U(1)\right)$ and $\mathrm{S}\left(\U(3) \times
        \U(2)\right)$, which are locally isomorphic to $\SU(4) \times
      \U(1)$ and $\SU(3) \times \SU(2) \times \U(1)$,
      respectively; but you have to show this!)}
  \end{enumerate}

  It is possible to show that up to gauge transformations these are
  the only three minima of $\eV$.  Hence the situation in this problem
  corresponds to a potential which is a mixture of types (a) and (c)
  in Figure~\ref{fig:potentials}, and roughly sketched below:
  \begin{figure}[h!]
    \centering
    \includegraphics[width=0.65\textwidth]{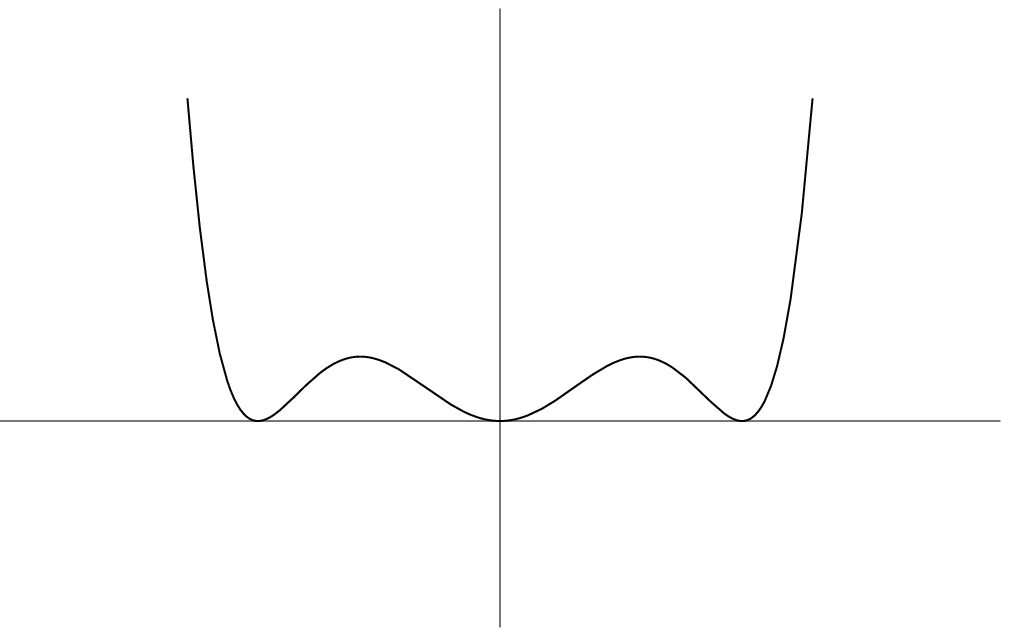}
  \end{figure}
\end{tut}


%
\appendix
\setcounter{equation}{0}
\renewcommand{\theequation}{\Alph{section}-\arabic{equation}}
\section{Basic definitions and conventions}
\label{app:conventions}

This appendix collects the basic definitions used in the lecture
concerning Lie (super)algebras, Minkowski space, the Poincaré group,
the Clifford algebra, the spin group and the different types of
spinors.  More importantly it also contains our spinor conventions.  I
learned supersymmetry from Peter van Nieuwenhuizen and these
conventions agree mostly with his.  I am however solely responsible
for any inconsistencies.

\begin{exercise}
  Find any inconsistencies and let me know!
\end{exercise}

\subsection{Lie algebras}

We now summarise the basic notions of Lie algebras and Lie
superalgebras used in the lectures.

A Lie algebra consists of a vector space $\fg$ and an antisymmetric
bilinear map
\begin{equation*}
  [-,-]: \fg \times \fg \to \fg~,
\end{equation*}
called the Lie bracket, which satisfying the Jacobi identity
\begin{equation*}
  [X,[Y,Z]] = [[X,Y],Z] + [Y,[X,Z]] \qquad\text{for all
  $X,Y,Z\in\fg$.}
\end{equation*}
Fixing a basis $\{T_i\}$ for $\fg$, the Lie bracket is specified by
the structure constants $f_{ij}{}^k = - f_{ji}{}^k$ defined by
\begin{equation*}
  [T_i,T_j] = f_{ij}{}^k T_k~.
\end{equation*}
All Lie algebras considered in these lectures are real; in other
words, $\fg$ is a real vector space and the structure constants are
real.  This means, in particular, that in a unitary representation
they are realised as antihermitian matrices.

Most Lie algebras of interest possess an invariant inner product,
denoted $\Tr$, since it can often be taken to be the trace in some
faithful representation.  Relative to a basis, the inner product is
specified by a real symmetric matrix $G_{ij} = G_{ji} = \Tr T_i T_j$.
Invariance means that
\begin{equation*}
  \Tr [T_i,T_j] T_k = \Tr T_i [T_j,T_k]
\end{equation*}
which is equivalent to $f_{ijk} := f_{ij}{}^\ell G_{\ell k}$ being
totally antisymmetric.  For a compact Lie group, one can always choose
a basis for the Lie algebra such that $G_{ij} = - \delta_{ij}$.
Notice that it is negative-definite.

The exponential of a matrix is defined in terms of the Taylor series
of the exponential function:
\begin{equation*}
  e^\sA := \1 + \sA + \half \sA^2 + \cdots~.
\end{equation*}
Suppose we are given a linear representation of a Lie algebra $\fg$.
Every element $X\in\fg$ is represented by a matrix $\sX$, and hence we 
can define the exponential $\exp(X)$ in the representation as the
exponential of the corresponding matrix $\exp(\sX)$.  Given $X,Y\in
\fg$ with corresponding matrices $\sX,\sY$ and consider the product of
their exponentials $\exp(\sX) \exp(\sY)$.  It turns out that this is
the exponential of a third matrix $\sZ$:
\begin{equation*}
  e^{\sX} e^{\sY} = e^{\sZ} \quad\text{where $\sZ = \sX + \sY +
  \cdots$},
\end{equation*}
where the omitted terms consists of nested commutators of $\sX$ and
$\sY$. This implies that there is an element $Z\in\fg$ which is
represented by $\sZ$.  The dependence of $Z$ on $X$ and $Y$ is quite
complicated, and is given by the celebrated Baker--Campbell--Hausdorff
formula.  For our purposes it will be sufficient to notice that
\begin{equation*}
  Z = X + \left( \frac{-\ad X}{e^{-\ad X} - \1} \right) \cdot Y +
  \cdots
\end{equation*}
where the omitted terms are at least quadratic in $Y$.  In this
formula, $\ad X$ is defined by $\ad X \cdot Y = [X,Y]$ and the
expression in parenthesis is defined by its Taylor series.  Keeping
only those terms at most linear in $Y$, $Z$ takes the form
\begin{equation}
  \label{eq:BCH}
  Z = X + Y + \half [X,Y] + \tfrac1{12} [X,[X,Y]] + \sum_{k\geq 2} c_k
  (\ad X)^{2k}\cdot Y~,
\end{equation}
where the $c_k$ are rational coefficients.  Notice that the sum has
only even powers of $\ad X$.

\subsection{Lie superalgebras}

The notion of a Lie superalgebra is a natural extension of the notion
of a Lie algebra.  By definition, a Lie superalgebra consists of a
$\ZZ_2$-graded vector space $\fg = \fg_0 \oplus \fg_1$ and a bilinear
operation to be defined presently.  In practice we will only consider
homogeneous elements; that is, elements in either $\fg_0$ or $\fg_1$.
For $X$ a homogeneous element the following are equivalent:
\begin{equation*}
  \begin{aligned}
    |X|= 0 & \iff X \in \fg_0 \iff \text{$X$ is even,}\\
    |X|= 1 & \iff X \in \fg_1 \iff \text{$X$ is odd,}
  \end{aligned}
\end{equation*}
which defines what we mean by even and odd.  The Lie bracket is now
$\ZZ_2$-graded
\begin{equation*}
  [-,-]: \fg_i \times \fg_j \to \fg_{i+j}
\end{equation*}
where $i+j$ are added modulo $2$. It is again bilinear and obeys
\begin{equation*}
  [X,Y] = -(-1)^{|X||Y|} [Y,X]
\end{equation*}
and
\begin{equation*}
  [X,[Y,Z]] = [[X,Y],Z] + (-1)^{|X||Y|} [Y,[X,Z]]
\end{equation*}
for all homogeneous elements $X,Y,Z\in\fg$.  We use the \emph{same}
notation $[-,-]$ for the bracket of any two elements in a
superalgebra.  We should remember however that it is symmetric if both
elements are odd and antisymmetric otherwise.  Furthermore, in a
linear representation, the bracket of two odd elements is realised as
the anticommutator of the corresponding matrices, whereas it is
realised as the commutator in all other cases.

We introduce a useful categorical concept.  Given a Lie superalgebra
defined by some brackets, by the \emph{opposite superalgebra} we will
mean the Lie superalgebra defined by multiplying the brackets by $-1$.
Clearly any Lie superalgebra is isomorphic to its opposite, by sending
each generator $X$ to $-X$.  We are only introducing this notion for
notation: I find it more convenient conceptually to think in terms of
representations of the opposite algebra than in terms of
antirepresentations of an algebra, and in these lectures we will have
to deal with both.

It is a general fact, following trivially from the axioms, that the
even subspace of a Lie superalgebra forms a Lie algebra of which the
odd subspace is a (real, in the cases of interest) representation.  It
follows in particular that a Lie algebra is a Lie superalgebra which
has no odd elements.  Hence the theory of Lie superalgebras contains
the theory of Lie algebras, and extends it in a nontrivial way.  From
a kinematic point of view, supersymmetry is all about finding field
theoretical realisations of Lie superalgebras whose even subspace
contains a Lie subalgebra isomorphic to either the Poincaré or
conformal algebras.

\subsection{Minkowski space and the Poincaré group}

Minkowski space is the four-dimensional real vector space with
``mostly plus'' metric
\begin{equation*}
  \eta_{\mu\nu}=
  \begin{pmatrix}
    -1 & & & \\
    & +1 & & \\
    & & +1 & \\
    & & & +1
  \end{pmatrix}~.
\end{equation*}
We fix an orientation $\epsilon_{\mu\nu\rho\sigma}$ by
\begin{equation*}
  \epsilon^{0123} = - \epsilon_{0123} = + 1~.
\end{equation*}

The group of isometries of Minkowski space is called the Poincaré
group.  The subgroup of isometries which preserve the origin is called
the Lorentz group.  The Poincaré group is the semidirect product of
the Lorentz group and the translation group.  Its Lie algebra is
therefore also the semidirect product of the Lorentz algebra and the
translation algebra.  Let $\sM_{\mu\nu} = - \sM_{\nu\mu}$ be a basis
for the Lorentz algebra and let $\sP_\mu$ be a basis for the
translation algebra.  They satisfy the following brackets:
  \begin{equation}
    \label{eq:poincare}
    \begin{aligned}
      \left[\sP_\mu,\sP_\nu\right] &= 0\\
      \left[\sM_{\mu\nu},\sP_\rho\right] &= \eta_{\nu\rho} \sP_\mu -
      \eta_{\mu\rho} \sP_\nu\\
      \left[\sM_{\mu\nu},\sM_{\rho\sigma}\right] &= \eta_{\nu\rho}
      \sM_{\mu\sigma} - \eta_{\mu\rho} \sM_{\nu\sigma} -
      \eta_{\nu\sigma} \sM_{\mu\rho} + \eta_{\mu\sigma}
      \sM_{\nu\rho}~.
  \end{aligned}
\end{equation}

The Poincaré group acts transitively on Minkowski space: any point can 
be reached from the origin by a Poincaré transformation.  This
transformation is not unique, since there are some transformations
which leave the origin fixed: the Lorentz transformations.  Therefore
Minkowski space (with a choice of origin) can be identified with the
space of right cosets of the Lorentz group.  Each such coset has a
unique representative which is a translation.  This allows us to
assign a unique element of the Poincaré group to each point in
Minkowski space:
\begin{equation*}
  \begin{array}{cr}
    x^\mu & \in \text{Minkowski space}\\
    \updownarrow & \\
    \exp(x^\mu \sP_\mu) & \in \text{Poincaré group},
  \end{array}
\end{equation*}
which in turn allows us to realise the action of the Poincaré group in 
Minkowski space as left multiplication in the group.

Indeed, a translation $\exp(\tau^\mu \sP_\mu)$ acts as
\begin{equation*}
  \exp(\tau^\mu \sP_\mu) \exp(x^\mu \sP_\mu) = \exp((x^\mu + \tau^\mu) 
  \sP_\mu)~,
\end{equation*}
whence $x^\mu \mapsto x^\mu + \tau^\mu$.  Similarly a Lorentz
transformation acts as
\begin{equation*}
  \exp(\half \lambda^{\mu\nu} \sM_{\mu\nu}) \exp(x^\mu \sP_\mu)
  = \exp(x^\mu \Lambda_\mu{}^\nu \sP_\nu) \exp(\half \lambda^{\mu\nu}
  \sM_{\mu\nu})~,
\end{equation*}
where $\Lambda_\mu{}^\nu$ is the adjoint matrix defined by
\begin{equation*}
  \Lambda_\mu{}^\nu \sP_\nu = \exp(\half \lambda^{\mu\nu}
  \sM_{\mu\nu}) \sP_\mu \exp(- \half \lambda^{\mu\nu} \sM_{\mu\nu})~.
\end{equation*}
Therefore the effect of a Poincaré transformation $\exp(\tau \cdot
\sP) \exp(\lambda \cdot \sM)$ is
\begin{equation*}
  x^\mu \mapsto x^\nu \Lambda_\nu{}^\mu + \tau^\mu~.
\end{equation*}
Let us call this transformation $P(\Lambda,\tau)$.  Notice that
acting on points the order of the transformations is reversed:
\begin{equation*}
  P(\Lambda_1,\tau_1) P(\Lambda_2,\tau_2) = P(\Lambda_2\Lambda_1,
  \Lambda_1\tau_2 + \tau_1)~.
\end{equation*}

Similarly, we can work out the action of the Lie algebra by
considering infinitesimal transformations:
\begin{equation*}
  \delta_\tau x^\mu = \tau^\mu \quad\text{and}\quad
  \delta_\lambda x^\mu = x^\nu \lambda_\nu{}^\mu~,
\end{equation*}
whence we see that $\sP_\mu$ and $\sM_{\mu\nu}$ are realised in terms
of vector fields
\begin{equation*}
  \sP_\mu \rightsquigarrow \d_\mu \quad\text{and}\quad
  \sM_{\mu\nu} \rightsquigarrow x_\mu \d_\nu - x_\nu \d_\mu~.
\end{equation*}
Again notice that these vector fields obey the opposite algebra.

\subsection{The Clifford algebra and its spinors}

The Lorentz group has four connected components.  The component
containing the identity consists of those Lorentz transformations
which preserve the space and time orientations, the proper
orthochronous Lorentz transformations.  This component is not simply
connected, but rather admits a simply-connected double cover (the spin
cover) which is isomorphic to the group $\SL(2,\CC)$ of $2\times 2$
complex matrices with unit determinant.  The spinorial representations
of the Lorentz group are actually representations of $\SL(2,\CC)$.

A convenient way to study the spinorial representations is via the
Clifford algebra of Minkowski space
\begin{equation*}
  \gamma_\mu \gamma_\nu + \gamma_\nu \gamma_\mu = + 2 \eta_{\mu\nu}
  \1~.
\end{equation*}
The reason is that the spin group is actually contained in the
Clifford algebra as exponentials of (linear combinations of)
\begin{equation*}
  \Sigma_{\mu\nu} = \tfrac14 \left( \gamma_\mu \gamma_\nu - \gamma_\nu 
  \gamma_\mu \right)~.
\end{equation*}
Notice that under the Clifford commutator these elements represent the 
Lorentz algebra (cf. the last equation in \eqref{eq:poincare})
\begin{equation*}
  \left[\Sigma_{\mu\nu},\Sigma_{\rho\sigma}\right] = 
  \eta_{\nu\rho} \Sigma_{\mu\sigma} - \eta_{\mu\rho}
  \Sigma_{\nu\sigma} - \eta_{\nu\sigma} \Sigma_{\mu\rho} +
  \eta_{\mu\sigma} \Sigma_{\nu\rho}~.
\end{equation*}

As an associative algebra, the Clifford algebra is isomorphic to the
algebra of $4\times 4$ real matrices.  This means that it has a unique
irreducible representation which is real and four-dimensional.  These
are the \emph{Majorana spinors}.

It is often convenient to work with the complexified Clifford algebra,
that is to say, one is allowed to take linear combination of the Dirac
$\gamma$ matrices.  The complexified Clifford algebra has a unique
irreducible representation which is complex and four-dimensional.
These are the \emph{Dirac spinors}.

We can always choose the inner product of spinors in such a way that
the Dirac matrices are unitary.  The Clifford algebra then implies
that $\gamma_0$ is antihermitian and $\gamma_i$ are hermitian.  These
conditions can be summarised succinctly as
\begin{equation*}
  \gamma_\mu^\dagger \gamma_0 = - \gamma_0 \gamma_\mu~.
\end{equation*}

One recovers the Majorana spinors as those Dirac spinors for which its 
Dirac $\bar\psi_D= \psi^\dagger i \gamma^0$ and Majorana
$\bar\psi_M=\psi^t C$ conjugates agree:
\begin{equation}
  \label{eq:majorana}
  \bar \psi := \bar\psi_D = \bar \psi_M ~,
\end{equation}
where $C$ is the charge conjugation matrix.  This implies a reality
condition on the Dirac spinor:
\begin{equation*}
  \psi^* = i C \gamma^0 \psi~.
\end{equation*}
I find it easier to work with the Majorana conjugate, since this
avoids having to complex conjugate the spinor.

Its historical name notwithstanding, $C$ is \emph{not} a matrix, since
under a change of basis it does not transform like a $\gamma$ matrix.
Introducing spinor indices $\psi^a$, the $\gamma$ matrices have indices
$(\gamma_\mu)^a{}_b$ whereas $C$ has indices $C_{ab}$.  In other
words, whereas the $\gamma$ matrices are linear transformations, the
charge conjugation matrix is a bilinear form.  We will always use $C$
to raise and lower spinor indices.

The charge conjugation matrix obeys the following properties:
\begin{equation}
  \label{eq:chargeconjugation}
  C^t = - C \qquad\text{and}\qquad C \gamma_\mu = - \gamma_\mu^t C~.
\end{equation}
Writing the indices explicitly the first of these equations becomes
\begin{equation*}
  C_{ab} = - C_{ba}~,
\end{equation*}
so that $C$ is antisymmetric.  This means that care has to be taken to
choose a consistent way to raise and lower indices.  We will raise
and lower indices using the \emph{North-West} and \emph{South-East}
conventions, respectively.  More precisely,
\begin{equation*}
  \psi^a = C^{ab} \psi_b \qquad\text{and}\qquad \psi_a = \psi^b C_{ba}~.
\end{equation*}
This implies that the inner product of Majorana spinors takes the form
\begin{equation*}
  \bar\varepsilon \psi := \varepsilon_a \psi^a = \varepsilon^b C_{ba}
  \psi^a = - \varepsilon^b \psi^a C_{ab} = - \varepsilon^b \psi_b~.
\end{equation*}
The second identity in equation \eqref{eq:chargeconjugation} can then
be written as a symmetry condition:
\begin{equation*}
  (\gamma_\mu)_{ab} = (\gamma_\mu)_{ba}~,
\end{equation*}
where
\begin{equation*}
  (\gamma^\mu)_{ab} = (\gamma^\mu)^c{}_b C_{ca} = - C_{ac}
  (\gamma^\mu)^c{}_b~.  
\end{equation*}

We will employ the following useful notation
$\gamma_{\mu\nu\dots\rho}$ for the totally antisymmetrised product of
$\gamma$ matrices.  More precisely we define
\begin{equation}
  \label{eq:antisymmetrisation}
  \gamma_{\mu_1\mu_2\dots \mu_n} := \frac{1}{n!} \sum_{\sigma\in\fS_n}
  \sign(\sigma) \gamma_{\mu_{\sigma(1)}} \gamma_{\mu_{\sigma(2)}}
  \cdots \gamma_{\mu_{\sigma(n)}}~,
\end{equation}
where the sum is over all the permutations of the set
$\{1,2,\dots,n\}$.  Notice the factorial prefactor.  For example, for
$n=2$ this formula unpacks into
\begin{equation*}
  \gamma_{\mu\nu} = \half \left( \gamma_\mu \gamma_\nu - \gamma_\nu
  \gamma_\mu \right)~.
\end{equation*}

The following identity is very convenient for computations
\begin{multline*}
  \gamma_{\mu_1\mu_2\dots\mu_n} \gamma_\nu =
  \gamma_{\mu_1\mu_2\dots\mu_n\nu} + \eta_{\nu\mu_n}
  \gamma_{\mu_1\mu_2\dots\mu_{n-1}} - \eta_{\nu\mu_{n-1}}
  \gamma_{\mu_1\mu_2\dots\widehat{\mu_{n-1}}\mu_n}\\
  + \eta_{\nu\mu_{n-2}}
    \gamma_{\mu_1\dots\widehat{\mu_{n-2}}\mu_{n-1}\mu_n} - \dots +
    (-1)^{n-1} \eta_{\nu\mu_1} \gamma_{\mu_2\mu_3\dots\mu_3}~,
\end{multline*}
where a hat over an index indicates its omission.  For example,
\begin{equation}
  \label{eq:gammagammagamma}
  \gamma_{\mu\nu} \gamma_\rho = \gamma_{\mu\nu\rho} + \eta_{\nu\rho}
  \gamma_\mu - \eta_{\mu\rho}\gamma_\nu~.
\end{equation}
As an immediate corollary, we have the following useful identities:
\begin{equation}
  \label{eq:useful}
  \gamma^\rho \gamma_\mu \gamma_\rho = - 2 \gamma_\mu
  \qquad\text{and}\qquad
  \gamma^\rho \gamma_{\mu\nu} \gamma_\rho = 0~.
\end{equation}

The Clifford algebra is isomorphic \emph{as a vector space} to the
exterior algebra of Minkowski space.  The above antisymmetrisation
provides the isomorphism.  This makes it easy to 
list a basis for the Clifford algebra
\begin{equation*}
  \1 \qquad \gamma_\mu \qquad \gamma_{\mu\nu} \qquad
  \gamma_{\mu\nu\rho} \qquad \gamma_{\mu\nu\rho\sigma}~.
\end{equation*}
There are $1 + 4 + 6 + 4 + 1 = 16$ elements which are clearly linearly 
independent.

Define $\gamma_5$ as
\begin{equation*}
  \gamma_5 = \tfrac{1}{4!} \epsilon^{\mu\nu\rho\sigma}
  \gamma_{\mu\nu\rho\sigma} = \gamma_0 \gamma_1 \gamma_2 \gamma_3~.
\end{equation*}
It satisfies the following properties:
\begin{equation*}
  \gamma_\mu\gamma_5 = - \gamma_5 \gamma_\mu \qquad \gamma_5^2 = -\1
  \qquad \gamma_5^\dagger = - \gamma_5 \qquad \gamma_5^t C = C
  \gamma_5~.
\end{equation*}
This last identity can be rewritten as the antisymmetry condition
\begin{equation*}
  (\gamma_5)_{ab} = - (\gamma_5)_{ba}~.
\end{equation*}
Using $\gamma_5$ we never need to consider antisymmetric products of
more than two $\gamma$ matrices.  Indeed, one has the following
identities:
\begin{equation*}
  \begin{aligned}
    \gamma_{\mu\nu}\gamma_5 &= - \half \epsilon_{\mu\nu\rho\sigma}
    \gamma^{\rho\sigma}\\
    \gamma_{\mu\nu\rho} &= \epsilon_{\mu\nu\rho\sigma} \gamma^\sigma
    \gamma_5\\
    \gamma_{\mu\nu\rho\sigma} &= - \epsilon_{\mu\nu\rho\sigma}
    \gamma_5~.
  \end{aligned}
\end{equation*}
Thus an equally good basis for the Clifford algebra is given by
\begin{equation}
  \label{eq:basis}
  \1 \qquad \gamma_5 \qquad \gamma_\mu \qquad \gamma_\mu\gamma_5
  \qquad \gamma_{\mu\nu}~.
\end{equation}
Lowering indices with $C$ we find that $\1$, $\gamma_5$ and
$\gamma_\mu\gamma_5$ becomes antisymmetric, whereas $\gamma_\mu$ and
$\gamma_{\mu\nu}$ become symmetric.

Let $\varepsilon_1$ and $\varepsilon_2$ be anticommuting spinors, and
let $\varepsilon_1 \bar\varepsilon_2$ denote the linear transformation
which, acting on a spinor $\psi$, yields
\begin{equation*}
  \varepsilon_1 \bar\varepsilon_2 \, \psi = (\bar\varepsilon_2\psi) \, 
    \varepsilon_1~.
\end{equation*}
Since the Clifford algebra is the algebra of linear transformations in 
the space of spinors, the basis \eqref{eq:basis} is also a basis of
this space and we can expand $\varepsilon_1 \bar\varepsilon_2$ in
terms of it.  The resulting identity is the celebrated \emph{Fierz
identity}:
\begin{multline}
  \label{eq:fierz}
  \varepsilon_1 \bar\varepsilon_2 = -\tfrac14 (\bar\varepsilon_2
  \varepsilon_1)\, \1 + \tfrac14 (\bar\varepsilon_2 \gamma_5
  \varepsilon_1)\, \gamma_5 - \tfrac14 (\bar\varepsilon_2 \gamma^\mu
  \varepsilon_1)\,  \gamma_\mu\\
  + \tfrac14 (\bar\varepsilon_2 \gamma^{\mu}\gamma_5 \varepsilon_1) \,
  \gamma_\mu\gamma_5 + \tfrac18 (\bar\varepsilon_2
  \gamma^{\mu\nu}\varepsilon_1)\, \gamma_{\mu\nu}~,
\end{multline}
whose importance in supersymmetry calculations can hardly be
overemphasised.  (For commuting spinors there is an overall minus sign
in the right-hand side.)  The Fierz identity can be proven by tracing
with the elements of the basis \eqref{eq:basis} and noticing that
$\gamma_5$, $\gamma_\mu$, $\gamma_\mu\gamma_5$ and $\gamma_{\mu\nu}$
are traceless.  An important special case of the Fierz identity is
\begin{equation}
  \label{eq:specialfierz}
  \varepsilon_1 \bar\varepsilon_2 - \varepsilon_2 \bar\varepsilon_1 =
  \half (\bar\varepsilon_1 \gamma^\mu \varepsilon_2)\,  \gamma_\mu -
    \tfrac14 (\bar\varepsilon_1 \gamma^{\mu\nu}\varepsilon_2)\,
    \gamma_{\mu\nu}~,
\end{equation}
which comes in handy when computing the commutator of two
supersymmetries.

Closely related to the Fierz identity are the following identities
involving powers of an anticommuting Majorana spinor $\theta$:
\begin{equation}
  \label{eq:thetapowers}
  \begin{aligned}
    \theta_a \theta_b &= \tfrac14 \left( \bar\theta\theta\, C_{ab} +
      \bar\theta\gamma_5 \theta\, (\gamma_5)_{ab} + \bar\theta
      \gamma^\mu \gamma_5 \theta\, (\gamma^\mu\gamma_5)_{ab}\right)\\
    \theta_a\theta_b\theta_c &= \half \bar\theta\theta\, \left( C_{ab}
      \theta_c + C_{ca} \theta_b + C_{bc} \theta_a \right)\\
    \theta_a\theta_b\theta_c\theta_d &= \tfrac18 \bar\theta\theta\,
    \bar\theta\theta\, \left( C_{ab} C_{cd} - C_{ac}C_{bd} + C_{ad}
    C_{bc} \right)~,
  \end{aligned}
\end{equation}
with all other powers vanishing.  These identities are extremely
useful in expanding superfields.

\subsection{The spin group}

The spin group is isomorphic to $\SL(2,\CC)$ and hence has a natural
two-dimensional complex representation, which we shall call $\WW$.
More precisely, $\WW$ is the vector space $\CC^2$ with the natural
action of $\SL(2,\CC)$.  If $w\in\WW$ has components $w^\alpha =
(w^1,w^2)$ relative to some fixed basis, and $M \in \SL(2,\CC)$, the
action of $M$ on $w$ is defined simply by $(M\,w)^\alpha =
M^\alpha{}_\beta w^\beta$.

This is not the only possible action of $\SL(2,\CC)$ on $\CC^2$, though.
We could also define an action by using instead of the matrix $M$, its
complex conjugate $\bar M$, its inverse transpose $(M^t)^{-1}$ or
its inverse hermitian adjoint $(M^\dagger)^{-1}$, since they all obey
the same group multiplication law.  These choices correspond,
respectively to the \emph{conjugate} representation $\overline\WW$,
the \emph{dual} representation $\WW^*$, and the \emph{conjugate dual}
representation $\overline\WW^*$.

We will adopt the following notation: if $w^\alpha \in\WW$, then $\bar
w^{\dot\alpha} \in \overline\WW$, $w_\alpha\in\WW^*$ and $\bar
w_{\dot\alpha}\in\overline\WW^*$.  These representations are not all
inequivalent, since we can raise and lower indices in an
$\SL(2,\CC)$-equivariant manner with the antisymmetric invariant
tensors $\epsilon_{\alpha\beta}$ and
$\bar\epsilon_{\dot\alpha\dot\beta}$.  (The $\SL(2,\CC)$-invariance of
these tensors is the statement that matrices in $\SL(2,\CC)$ have unit
determinant.)  Notice that we raise and lower also using the
North-West and South-East conventions:
\begin{equation*}
  w_\alpha = w^\beta \epsilon_{\beta\alpha} \qquad\text{and}\qquad
  w^\alpha = \epsilon^{\alpha\beta} w_\beta~,
\end{equation*}
and similarly for the conjugate spinors:
\begin{equation*}
  \bar w_{\dot\alpha} = \bar w^{\dot \beta}
  \bar\epsilon_{\dot\beta\dot\alpha} \qquad\text{and}\qquad
  \bar w^{\dot\alpha} = \bar\epsilon^{\dot\alpha\dot\beta} \bar
  w_{\dot\beta}~.
\end{equation*}
We choose the perhaps unusual normalisations:
\begin{equation*}
  \epsilon_{12} =1=\epsilon^{12} \quad\text{and}\quad
  \bar\epsilon_{\dot1\dot2} = -1 =  \bar\epsilon^{\dot1\dot2}~.
\end{equation*}

Because both the Lie algebra $\fsl(2,\CC)$ (when viewed as a real Lie
algebra) and $\su(2)\times \su(2)$ are real forms of the same complex
Lie algebra, one often employs the notation $(j,j')$ for
representations of $\SL(2,\CC)$, where $j$ and $j'$ are the spins of the 
two $\su(2)$'s.  In this notation the trivial one dimensional
representation is denoted $(0,0)$, whereas $\WW = (\half,0)$.  The two
$\su(2)$'s are actually not independent but are related by complex
conjugation, hence $\overline\WW = (0,\half)$.  In general, complex
conjugation will interchange the labels.  If a representation is
preserved by complex conjugation, then it makes sense to restrict to
the subrepresentation which is fixed by complex conjugation.  For
example, the Dirac spinors transform like $(\half,0) \oplus
(0,\half)$.  The subrepresentation fixed by complex conjugation are
precisely the Majorana spinors.

Another example is the representation $(\half,\half)$.  The
real subrepresentation coincides with the defining representation of
the Lorentz group---that is, the vector representation.  To see this
notice that any $4$-vector $p_\mu=(p_0,\boldsymbol{p})$ can be turned
into a bispinor as follows:
\begin{equation*}
\sigma\cdot p \equiv \sigma^\mu p_\mu = 
\begin{pmatrix}
p_0 + p_3 & p_1 - i p_2\\
p_1 + i p_2 & p_0 - p_3
\end{pmatrix}
\end{equation*}
where $\sigma^\mu = (\1, \boldsymbol{\sigma})$ with
$\boldsymbol{\sigma}$ the Pauli matrices:
\begin{equation}
  \label{eq:Pauli}
  \begin{array}{ccc}
    \sigma^1 = \begin{pmatrix}0&1\\1&0\end{pmatrix}
    &
    \sigma^2 = \begin{pmatrix}0&-i\\i&0\end{pmatrix}
    &
    \sigma^3 = \begin{pmatrix}1&0\\0&-1\end{pmatrix}~.
  \end{array}
\end{equation}
Since the Pauli matrices are hermitian, so will be $\sigma\cdot p$
provided $p_\mu$ is real.  The Pauli matrices have indices
$(\sigma^\mu)^{\alpha\dot\alpha}$, which shows how $\SL(2,\CC)$ acts
on this space.  If $M\in \SL(2,\CC)$, then the action of $M$ on such
matrices is given by $\sigma\cdot p \mapsto M\, \sigma\cdot p\,
M^\dagger$.  This action is linear and preserves both the hermiticity
of $\sigma\cdot p$ and the determinant $\det(\sigma\cdot p) = - p^2 =
p_0^2 - \boldsymbol{p}\cdot\boldsymbol{p}$, whence it is a Lorentz
transformation. Notice that both $M$ and $-M$ act the same way on
bispinors, which reiterates the fact that the spin group is the double
cover of the Lorentz group.

\subsection{Weyl spinors}

Although the Dirac spinors form an irreducible representation of the
(complexified) Clifford algebra, they are not an irreducible
representation of the spin group.  Indeed, since $\gamma_5$
anticommutes with $\gamma_\mu$, it follows that it commutes with
$\Sigma_{\mu\nu}$ and is not a multiple of the identity.  Schur's
lemma implies that the Dirac spinors are reducible under the spin
group.  In fact, they decompose into two irreducible two-dimensional
representations, corresponding to the eigenspaces of $\gamma_5$.
Since $(\gamma_5)^2 = -\1$, its eigenvalues are $\pm i$ and the
eigenspaces form a complex conjugate pair.  They are the \emph{Weyl
spinors}.

We now relate the Weyl spinors and the two-dimensional representations
of $\SL(2,\CC)$ discussed above.  To this effect we will use the
following convenient realisation of the Clifford algebra
\begin{equation}
  \label{eq:realisation}
  \gamma^\mu = 
  \begin{pmatrix}
    0 & -i \sigma^\mu\\
    i \bar\sigma^\mu & 0
  \end{pmatrix}~,
  \qquad \text{where $\bar\sigma^\mu =
  (-\1,\boldsymbol{\sigma})$.}
\end{equation}
Notice that $\bar\sigma^\mu$ is obtained from $\sigma^\mu$ by lowering
indices:
\begin{equation}
  \label{eq:barPauli}
  (\bar\sigma^\mu)_{\dot\alpha\alpha} =
  (\sigma^\mu)^{\beta\dot\beta} \epsilon_{\beta\alpha}
  \bar\epsilon_{\dot\beta\dot\alpha}~.
\end{equation}

Notice that the indices in $\gamma^\mu$ are such that it acts
naturally on objects of the form
\begin{equation}
  \label{eq:Dirac2}
  \psi^a =
  \begin{pmatrix}
    \chi^\alpha\\ \bar\zeta_{\dot\alpha}
  \end{pmatrix}~,
\end{equation}
whence we see that a Dirac spinor indeed breaks up into a pair of
two-component spinors.  To see that these two-component spinors are
precisely the Weyl spinors defined above, notice that in this
realisation $\gamma_5$ becomes
\begin{equation*}
  \gamma_5 = 
  \begin{pmatrix}
    -i \1^\alpha{}_\beta & 0 \\
    0 & i \1_{\dot\alpha}{}^{\dot\beta}
  \end{pmatrix}~,
\end{equation*}
so that $\WW$ and $\bar\WW$ are indeed complex conjugate eigenspaces
of $\gamma_5$.

In this realisation the generators of the spin algebra
$\Sigma_{\mu\nu}$ become block diagonal
\begin{equation*}
  \Sigma_{\mu\nu} = 
  \begin{pmatrix}
    \half \sigma_{\mu\nu} & 0 \\
    0 & \half \bar\sigma_{\mu\nu}
  \end{pmatrix}~,
\end{equation*}
where
\begin{equation*}
  \begin{aligned}
    (\sigma_{\mu\nu})^\alpha{}_\beta &= \half
    (\sigma_\mu\bar\sigma_\nu -
    \sigma_\nu \bar\sigma_\mu)^\alpha{}_\beta\\
    (\bar\sigma_{\mu\nu})_{\dot\alpha}{}^{\dot\beta} &= \half
    (\bar\sigma_\mu\sigma_\nu - \bar\sigma_\nu
    \sigma_\mu)_{\dot\alpha}{}^{\dot\beta}~.
  \end{aligned}
\end{equation*}
Notice that $\sigma_{\mu\nu}$ and $\bar\sigma_{\mu\nu}$ with both
(spinor) indices up or down are symmetric matrices.

We collect here some useful identities involving the Pauli matrices:
\begin{equation}
  \label{eq:PauliIds}
  \begin{aligned}
    (\sigma^\mu)^{\alpha\dot\beta} (\bar\sigma^\nu)_{\dot\beta\gamma}
    &= \eta^{\mu\nu} \delta_\gamma^\alpha +
    (\sigma^{\mu\nu})^\alpha{}_\gamma\\
    (\bar\sigma^\mu)_{\dot\beta\alpha} (\sigma^\nu)^{\alpha\dot\gamma} 
    &= \eta^{\mu\nu} \delta_{\dot\beta}^{\dot\gamma} +
    (\bar\sigma_{\mu\nu})_{\dot\beta}{}^{\dot\gamma}\\
    (\sigma^\mu)^{\alpha\dot\beta} (\sigma_\mu)^{\gamma\dot\delta} &=
    2 \epsilon^{\alpha\gamma}\bar\epsilon^{\dot\beta\dot\delta}.
  \end{aligned}
\end{equation}

Using the relation between the $\gamma$ matrices and the Pauli
matrices, it is possible to prove the following set of identities:
\begin{equation}
  \label{eq:PauliIdsMore}
  \begin{aligned}
    \sigma^\mu \bar\sigma^\nu \sigma^\rho &= i
    \epsilon^{\mu\nu\rho\tau}\sigma_\tau + \eta^{\nu\rho}\sigma^\mu 
    - \eta^{\mu\rho}\sigma^\nu + \eta^{\mu\nu} \sigma^\rho\\
    \bar\sigma^\mu \sigma^\nu \bar\sigma^\rho &= - i
    \epsilon^{\mu\nu\rho\tau}\bar\sigma_\tau +
    \eta^{\nu\rho}\bar\sigma^\mu - \eta^{\mu\rho}\bar\sigma^\nu +
    \eta^{\mu\nu} \bar\sigma^\rho\\
    \sigma^{\mu\nu}\sigma^\rho &= \eta^{\nu\rho}\sigma^\mu -
    \eta^{\mu\rho}\sigma^\nu + i \epsilon^{\mu\nu\rho\tau}\sigma_\tau\\
    \bar\sigma^{\mu\nu}\bar\sigma^\rho &= \eta^{\nu\rho}\bar\sigma^\mu -
    \eta^{\mu\rho}\bar\sigma^\nu - i
    \epsilon^{\mu\nu\rho\tau}\bar\sigma_\tau\\
    \half \epsilon_{\mu\nu\rho\tau} \sigma^{\rho\tau} &= + i
    \sigma_{\mu\nu}\\
    \half \epsilon_{\mu\nu\rho\tau} \bar\sigma^{\rho\tau} &= - i
    \bar\sigma_{\mu\nu}\\
    \Tr \left(\sigma^{\mu\nu} \sigma^{\rho\tau}\right) &= 2 \left(
    \eta^{\nu\rho}  \eta^{\mu\tau} - \eta^{\mu\rho} \eta^{\nu\tau} + i 
    \epsilon^{\mu\nu\rho\tau}\right)~.
  \end{aligned}
\end{equation}

In this realisation, a Majorana spinor takes the form
\begin{equation}
  \label{eq:Majorana2}
  \psi^a = 
  \begin{pmatrix}
    \psi^\alpha \\ \bar\psi_{\dot\alpha}
  \end{pmatrix}~,
\end{equation}
which is the same as saying that the charge conjugation matrix takes
the form
\begin{equation}
  \label{eq:C2}
  C_{ab} = 
  \begin{pmatrix}
    \epsilon_{\alpha\beta} & 0\\
    0 & \bar\epsilon^{\dot\alpha\dot\beta}
  \end{pmatrix}~.
\end{equation}
In particular the (Majorana) conjugate spinor is given by
\begin{equation*}
  \bar\psi_a = \psi^b C_{ba} = (\psi_\alpha,
  -\bar\psi^{\dot\alpha})~.
\end{equation*}
The passage from Majorana to Weyl spinor inner products is given by:
\begin{equation}
  \label{eq:4to2innerproduct}
  \bar\chi\psi = \chi_a \psi^a = \chi_\alpha \psi^\alpha -
  \bar\chi^{\dot\alpha} \bar\psi_{\dot\alpha} = - (\chi^\alpha
  \psi_\alpha + \bar\chi^{\dot\alpha} \bar\psi_{\dot\alpha})~.
\end{equation}
where the spinors on the left are four-component Majorana and those on 
the right are two-component Weyl.

\subsection{Two-component Fierz identities}

One of the advantages of the two-component formalism is that Fierz
identities simplify considerably; although there are more of them.
For example, suppose that $\varepsilon$ and $\theta$ are two
anticommuting spinors, then we have the following Fierz identities:
\begin{equation}
  \label{eq:fierz2}
  \begin{aligned}
    \varepsilon_\alpha \theta_\beta &= - \half \varepsilon\theta\,
    \epsilon_{\alpha\beta} - \tfrac18 \varepsilon \sigma^{\mu\nu}
    \theta \, (\sigma_{\mu\nu})_{\alpha\beta}\\
    \bar\varepsilon_{\dot\alpha} \bar\theta_{\dot\beta} &= - \half
    \bar\varepsilon\bar\theta\, \bar\epsilon_{\dot\alpha\dot\beta} -
    \tfrac18 \bar \varepsilon \bar \sigma^{\mu\nu} \bar\theta \, (\bar 
    \sigma_{\mu\nu})_{\dot\alpha\dot\beta}\\
    \varepsilon_\alpha \bar\theta_{\dot\beta} &= + \half
    \varepsilon\sigma^\mu \bar \theta\,
    (\bar\sigma_\mu)_{\dot\beta\alpha}~,
  \end{aligned}
\end{equation}
where we have used the following contractions
\begin{equation}
  \label{eq:contractions}
  \begin{aligned}
    \varepsilon\theta &= \varepsilon^\alpha \theta_\alpha\\
    \bar\varepsilon\bar\theta &= \bar\varepsilon^{\dot\alpha}
    \bar\theta_{\dot\alpha}\\
    \varepsilon\sigma^\mu\bar\theta &= \varepsilon_\alpha
    (\sigma^\mu)^{\alpha\dot\beta} \bar\theta_{\dot\beta}\\
    \bar\varepsilon\bar\sigma^\mu\theta &=
    \bar\varepsilon^{\dot\alpha} (\bar\sigma^\mu)_{\dot\alpha\beta}
    \theta^\beta\\
    \varepsilon \sigma^{\mu\nu} \theta &= \varepsilon_\alpha
    (\sigma^{\mu\nu})^\alpha{}_\beta \theta^\beta\\
    \bar\varepsilon \bar\sigma^{\mu\nu} \bar\theta &=
    \bar\varepsilon^{\dot\alpha}
    (\bar\sigma^{\mu\nu})_{\dot\alpha}{}^{\dot\beta}
    \bar\theta_{\dot\beta}~.
\end{aligned}
\end{equation}
These contractions satisfy the following (anti)symmetry properties:
\begin{equation}
  \label{eq:(anti)symmetry}
  \begin{aligned}
    \varepsilon\theta &= + \theta\varepsilon\\
    \bar\varepsilon\bar\theta &= + \bar\theta\bar\varepsilon\\
    \bar\varepsilon\bar\sigma^\mu\theta &= - \theta \sigma^\mu \bar
    \varepsilon\\
    \varepsilon \sigma^{\mu\nu} \theta &= - \theta
    \sigma^{\mu\nu}\varepsilon\\
    \bar\varepsilon \bar\sigma^{\mu\nu} \bar\theta &= -
    \bar\theta \bar\sigma^{\mu\nu} \bar\varepsilon~.
  \end{aligned}
\end{equation}
(For commuting spinors, all the signs change.)

These Fierz identities allow us to prove a variety of useful
identities simply by contracting indices and using equations
\eqref{eq:PauliIds} and \eqref{eq:PauliIdsMore}.  For example,
\begin{equation}
  \label{eq:usefultoo}
  \bar\theta\bar\sigma^\mu\theta\, \bar\theta\bar\sigma^\nu\theta = 
    - \half \theta^2 \bar\theta^2\, \eta^{\mu\nu}
\end{equation}
and
\begin{equation}
  \label{eq:useful3}
  \begin{aligned}
    \theta\psi\, \theta\sigma^\mu\bar\xi &= - \half \theta^2 \,
    \psi\sigma^\mu\bar\xi\\
    \bar\theta\bar\psi\, \bar\theta\bar\sigma^\mu\xi &= - \half
    \bar\theta^2 \, \bar\psi\bar\sigma^\mu\xi~.
  \end{aligned}
\end{equation}
These and similar identities come in handy when working out component
expansions of superfields.

\subsection{Complex conjugation}

Finally we come to complex conjugation.  By definition, complex
conjugation is always an involution, so that $(O^*)^* = O$ for any
object $O$.  For spinorial objects, we have that
\begin{equation*}
  (\theta_\alpha)^* = \bar\theta_{\dot\alpha}~,
\end{equation*}
which, because of our sign conventions, implies
\begin{equation*}
(\theta^\alpha)^* = - \bar\theta^{\dot\alpha}~.
\end{equation*}

Complex conjugation \emph{always} reverses the order of anticommuting
objects.  For example,
\begin{equation*}
  (\theta_\alpha \theta_\beta)^* = \bar\theta_{\dot\beta}
  \bar\theta_{\dot\alpha} \qquad\text{and}\qquad
  (\theta^\alpha \theta^\beta \theta^\gamma)^* = -
  \bar\theta^{\dot\gamma} \bar\theta^{\dot\beta} \bar\theta^{\dot\alpha}~.
\end{equation*}
In so doing, it does \emph{not} give rise to a sign.  This is
\emph{not} in conflict with the fact that the objects are
anticommuting, since conjugation actually changes the objects being
conjugated.

Hermiticity of the Pauli matrices means that
\begin{equation*}
  \begin{aligned}
    \left((\sigma^\mu)^{\alpha\dot\alpha}\right)^* &=
    (\bar\sigma^\mu)^{\dot\alpha\alpha}\\
    \left( (\sigma_{\mu\nu})^\alpha{}_\beta \right)^* &= -
    (\bar\sigma_{\mu\nu})_{\dot\beta}{}^{\dot\alpha}\\
    \left( (\sigma_{\mu\nu})^{\alpha\beta} \right)^* &= +
    (\bar\sigma_{\mu\nu})^{\dot\alpha\dot\beta}~.
  \end{aligned}
\end{equation*}
The last two equations show that complex conjugation indeed exchanges
the two kinds of Weyl spinors.

In particular, notice that
\begin{equation*}
  \begin{aligned}
    (\varepsilon\theta)^* &= + \bar\theta\bar\varepsilon = +
    \bar\varepsilon\bar\theta\\
    (\varepsilon\sigma^\mu\bar\theta)^* &= - \bar\varepsilon
    \bar\sigma^\mu \theta = + \theta\sigma^\mu
    \bar\varepsilon\\
    \left(\varepsilon\sigma^{\mu\nu}\theta \right)^* &= + \bar\theta
    \bar\sigma^{\mu\nu}\bar\varepsilon = - \bar\varepsilon
    \bar\sigma^{\mu\nu} \bar\theta~.
  \end{aligned}
\end{equation*}

This rule applies also to conjugating derivatives with respect to
anticommuting coordinates.  This guarantees that spinorial derivatives of
scalars are indeed spinors. For example,
\begin{equation*}
  (\d_\alpha)^* = \bar\d_{\dot\alpha} \qquad\text{and}\qquad
  (\d^\alpha)^* = - \bar\d^{\dot\alpha}~.
\end{equation*}
More generally, the rule applies to spinorial indices, as in
\begin{equation*}
  (\epsilon_{\alpha\beta})^* = \bar\epsilon_{\dot\beta\dot\alpha}~.
\end{equation*}

A useful ``reality check'' is to make sure that any result involving
bar'd objects agrees with the complex conjugate of the corresponding
result with unbar'd objects.  This simple procedure catches many a
wayward sign.


\twocolumn
\section{Formulas}
\label{app:formulas}
\scriptsize


\begin{fminipage}{0.85\columnwidth}
  \begin{equation*}
    \begin{aligned}
      \eta_{\mu\nu}&=
      \begin{pmatrix}
        -1 & & & \\
        & +1 & & \\
        & & +1 & \\
        & & & +1
      \end{pmatrix}\\
      \epsilon^{0123} &= - \epsilon_{0123} = + 1
    \end{aligned}
  \end{equation*}
\end{fminipage}

\vspace{10pt}



\begin{fminipage}{0.85\columnwidth}
  \begin{equation*}
    \begin{gathered}
      \gamma_\mu \gamma_\nu + \gamma_\nu \gamma_\mu = + 2
      \eta_{\mu\nu}
      \1\\
      \gamma_{\mu\nu} := \half (\gamma_\mu \gamma_\nu -
      \gamma_\nu\gamma_\mu)\\
      \gamma_{\mu\nu} \gamma_\rho = \gamma_{\mu\nu\rho} +
      \eta_{\nu\rho}
      \gamma_\mu - \eta_{\mu\rho}\gamma_\nu\\
      \gamma^\rho \gamma_\mu \gamma_\rho = - 2 \gamma_\mu\\
      \gamma^\rho \gamma_{\mu\nu} \gamma_\rho = 0
    \end{gathered}
  \end{equation*}
\end{fminipage}

\vspace{10pt}


\begin{fminipage}{0.85\columnwidth}
  \begin{gather*}
      \gamma_5 := \tfrac{1}{4!} \epsilon^{\mu\nu\rho\sigma}
      \gamma_{\mu\nu\rho\sigma} = \gamma_0 \gamma_1 \gamma_2
      \gamma_3\\
    \begin{aligned}
      \gamma_5^2 &= -\1\\
      \gamma_\mu\gamma_5 &= - \gamma_5 \gamma_\mu\\
      \gamma_{\mu\nu}\gamma_5 &= - \half \epsilon_{\mu\nu\rho\sigma}
      \gamma^{\rho\sigma}\\
      \gamma_{\mu\nu\rho} &= \epsilon_{\mu\nu\rho\sigma} \gamma^\sigma
      \gamma_5\\
      \gamma_{\mu\nu\rho\sigma} &= - \epsilon_{\mu\nu\rho\sigma}
      \gamma_5
  \end{aligned}
\end{gather*}
\end{fminipage}

\vspace{10pt}


\begin{fminipage}{0.85\columnwidth}
  \begin{equation*}
    \begin{aligned}
      C^t &= - C \\
      C \gamma_\mu &= - \gamma_\mu^t C\\
      C \gamma_5 &= + \gamma_5^t C\\
      C \gamma_{\mu\nu} &= - \gamma_{\mu\nu}^t C\\
      \bar \psi_M &:= \psi^t C
    \end{aligned}
  \end{equation*}
\end{fminipage}

\vspace{10pt}


\begin{fminipage}{0.85\columnwidth}
  \begin{equation*}
    \begin{aligned}
      C_{ab} &= - C_{ba}\\
      (\gamma_\mu)_{ab} &:= (\gamma_\mu)^c{}_b C_{ca} =
      (\gamma_\mu)_{ba}\\
      (\gamma_{\mu\nu})_{ab} &= (\gamma_{\mu\nu})_{ba}\\
      (\gamma_\mu\gamma_5)_{ab} &= - (\gamma_\mu\gamma_5)_{ba}\\
      (\gamma_5)_{ab} &= - (\gamma_5)_{ba}\\[3pt]
      \psi^a &= C^{ab} \psi_b\\
      \psi_a &= \psi^b C_{ba}\\
      \bar\varepsilon \psi &:= \varepsilon_a \psi^a = - \varepsilon^b
      \psi_b
    \end{aligned}
  \end{equation*}
\end{fminipage}

\vspace{10pt}


\begin{fminipage}{0.85\columnwidth}
  \begin{gather*}
    \begin{aligned}
      \gamma_\mu^\dagger &= \gamma_0 \gamma_\mu \gamma_0\\
      \gamma_5^\dagger &= - \gamma_5\\
      \gamma_{\mu\nu}^\dagger &= \gamma_0  \gamma_{\mu\nu} \gamma_0\\
      \bar\psi_D &:= \psi^\dagger i \gamma^0
    \end{aligned}\\
    \bar\psi_D = \bar\psi_M \iff \psi^* = i C \gamma^0 \psi    
  \end{gather*}
\end{fminipage}

\vspace{10pt}


\begin{fminipage}{0.85\columnwidth}
  \begin{equation*}
    \begin{aligned}
      \varepsilon_1 \bar\varepsilon_2 = & {} -\tfrac14
      (\bar\varepsilon_2 \varepsilon_1)\, \1\\
      & {} + \tfrac14 (\bar\varepsilon_2 \gamma_5
      \varepsilon_1)\, \gamma_5\\
      & {} - \tfrac14 (\bar\varepsilon_2 \gamma^\mu \varepsilon_1)\,
      \gamma_\mu\\
      & {} + \tfrac14 (\bar\varepsilon_2 \gamma^{\mu}\gamma_5
      \varepsilon_1) \, \gamma_\mu\gamma_5\\
      & {} + \tfrac18 (\bar\varepsilon_2
      \gamma^{\mu\nu}\varepsilon_1)\,
      \gamma_{\mu\nu}\\
      \varepsilon_1 \bar\varepsilon_2 - \varepsilon_2
      \bar\varepsilon_1 = & {} + \half (\bar\varepsilon_1 \gamma^\mu
      \varepsilon_2)\, \gamma_\mu\\
      & {} - \tfrac14 (\bar\varepsilon_1
      \gamma^{\mu\nu}\varepsilon_2)\, \gamma_{\mu\nu}
    \end{aligned}
  \end{equation*}
\end{fminipage}

\vspace{10pt}


\begin{fminipage}{0.85\columnwidth}
  \begin{equation*}
    \begin{aligned}
      \theta_a \theta_b ={} & \tfrac14 \left( \bar\theta\theta\,
        C_{ab} + \bar\theta\gamma_5 \theta\, (\gamma_5)_{ab} \right.\\
      & \left. {} + \bar\theta \gamma^\mu \gamma_5 \theta\,
        (\gamma^\mu\gamma_5)_{ab}\right)\\
      \theta_a\theta_b\theta_c = {} & \half \bar\theta\theta\, \left(
        C_{ab} \theta_c + C_{ca} \theta_b \right. \\
        & \left. {} + C_{bc} \theta_a \right)\\
      \theta_a\theta_b\theta_c\theta_d = {} &\tfrac18 \bar\theta\theta\,
      \bar\theta\theta\, \left( C_{ab} C_{cd} - C_{ac}C_{bd} \right. \\
        & \left. {} + C_{ad} C_{bc} \right)~,
    \end{aligned}
  \end{equation*}
\end{fminipage}

\vspace{10pt}



\begin{fminipage}{0.85\columnwidth}
  \begin{equation*}
    \begin{gathered}
      \epsilon_{12} =1=\epsilon^{12}\\
      \bar\epsilon_{\dot1\dot2} = -1 =  \bar\epsilon^{\dot1\dot2}\\
      (\epsilon_{\alpha\beta})^* =
      \bar\epsilon_{\dot\beta\dot\alpha}\\
      w_\alpha = w^\beta \epsilon_{\beta\alpha} \qquad  w^\alpha =
      \epsilon^{\alpha\beta} w_\beta\\
      \bar w_{\dot\alpha} = \bar w^{\dot \beta}
      \bar\epsilon_{\dot\beta\dot\alpha} \qquad \bar w^{\dot\alpha} =
      \bar\epsilon^{\dot\alpha\dot\beta} \bar w_{\dot\beta}
    \end{gathered}
  \end{equation*}
\end{fminipage}

\vspace{10pt}


\begin{fminipage}{0.85\columnwidth}
  \begin{equation*}
    \begin{gathered}
      \sigma^\mu = (\1, \boldsymbol{\sigma})\qquad
      \bar\sigma^\mu = (-\1,\boldsymbol{\sigma})\\
      \boldsymbol{\sigma}:\,\begin{pmatrix}0&1\\1&0\end{pmatrix}~
      \begin{pmatrix}0&-i\\i&0\end{pmatrix}~
      \begin{pmatrix}1&0\\0&-1\end{pmatrix}\\
      (\bar\sigma^\mu)_{\dot\alpha\alpha} =
      (\sigma^\mu)^{\beta\dot\beta} \epsilon_{\beta\alpha}
      \bar\epsilon_{\dot\beta\dot\alpha}\\
      \left((\sigma^\mu)^{\alpha\dot\alpha}\right)^* =
      (\bar\sigma^\mu)^{\dot\alpha\alpha}\\
    \end{gathered}
  \end{equation*}
\end{fminipage}


\begin{fminipage}{0.85\columnwidth}
  \begin{equation*}
    \begin{aligned}
      (\sigma_{\mu\nu})^\alpha{}_\beta &:= \half
      (\sigma_\mu\bar\sigma_\nu -
      \sigma_\nu \bar\sigma_\mu)^\alpha{}_\beta\\
      (\bar\sigma_{\mu\nu})_{\dot\alpha}{}^{\dot\beta} &:= \half
      (\bar\sigma_\mu\sigma_\nu - \bar\sigma_\nu
      \sigma_\mu)_{\dot\alpha}{}^{\dot\beta}\\
      \left( (\sigma_{\mu\nu})^\alpha{}_\beta \right)^* &= -
      (\bar\sigma_{\mu\nu})_{\dot\beta}{}^{\dot\alpha}\\
      \left( (\sigma_{\mu\nu})^{\alpha\beta} \right)^* &= +
      (\bar\sigma_{\mu\nu})^{\dot\alpha\dot\beta}\\
      (\sigma_{\mu\nu})_{\alpha\beta} &=
      (\sigma_{\mu\nu})_{\beta\alpha}\\
      (\bar\sigma_{\mu\nu})_{\dot\alpha\dot\beta} &=
      (\bar\sigma_{\mu\nu})_{\dot\beta\dot\alpha}
    \end{aligned}
  \end{equation*}
\end{fminipage}

\vspace{10pt}


\begin{fminipage}{0.85\columnwidth}
  \begin{gather*}
    \begin{aligned}
      (\sigma^\mu)^{\alpha\dot\beta} (\bar\sigma^\nu)_{\dot\beta\gamma}
      &= \eta^{\mu\nu} \delta_\gamma^\alpha +
      (\sigma^{\mu\nu})^\alpha{}_\gamma\\
      (\bar\sigma^\mu)_{\dot\beta\alpha} (\sigma^\nu)^{\alpha\dot\gamma} 
      &= \eta^{\mu\nu} \delta_{\dot\beta}^{\dot\gamma} +
      (\bar\sigma_{\mu\nu})_{\dot\beta}{}^{\dot\gamma}\\
      (\sigma^\mu)^{\alpha\dot\beta} (\sigma_\mu)^{\gamma\dot\delta} &=
      2 \epsilon^{\alpha\gamma}\bar\epsilon^{\dot\beta\dot\delta}
    \end{aligned}
  \end{gather*}
\end{fminipage}

\vspace{10pt}

\begin{fminipage}{0.85\columnwidth}
  \begin{equation*}
    \begin{aligned}
      \sigma^\mu \bar\sigma^\nu \sigma^\rho = & + i
      \epsilon^{\mu\nu\rho\tau}\sigma_\tau +
      \eta^{\nu\rho}\sigma^\mu\\
      & {} - \eta^{\mu\rho}\sigma^\nu + \eta^{\mu\nu} \sigma^\rho\\
      \bar\sigma^\mu \sigma^\nu \bar\sigma^\rho = & - i
      \epsilon^{\mu\nu\rho\tau}\bar\sigma_\tau +
      \eta^{\nu\rho}\bar\sigma^\mu\\
      & {} - \eta^{\mu\rho}\bar\sigma^\nu + \eta^{\mu\nu}
      \bar\sigma^\rho\\
      \sigma^{\mu\nu}\sigma^\rho =& + \eta^{\nu\rho}\sigma^\mu -
      \eta^{\mu\rho}\sigma^\nu\\
      & {} + i \epsilon^{\mu\nu\rho\tau}\sigma_\tau\\
      \bar\sigma^{\mu\nu}\bar\sigma^\rho =& + \eta^{\nu\rho}
      \bar\sigma^\mu - \eta^{\mu\rho}\bar\sigma^\nu\\
      & {} - i \epsilon^{\mu\nu\rho\tau}\bar\sigma_\tau
    \end{aligned}
  \end{equation*}
\end{fminipage}

\vspace{10pt}

\begin{fminipage}{0.85\columnwidth}
  \begin{equation*}
    \begin{aligned}
      \half \epsilon_{\mu\nu\rho\tau} \sigma^{\rho\tau} =& + i
      \sigma_{\mu\nu}\\
      \half \epsilon_{\mu\nu\rho\tau} \bar\sigma^{\rho\tau} =& - i
      \bar\sigma_{\mu\nu}\\
      \Tr \left(\sigma^{\mu\nu} \sigma^{\rho\tau}\right) = {} & 2
      \left(\eta^{\nu\rho} \eta^{\mu\tau} - \eta^{\mu\rho}
      \eta^{\nu\tau}\right.\\
      & {} + \left.  i \epsilon^{\mu\nu\rho\tau}\right)
    \end{aligned}
  \end{equation*}
\end{fminipage}

\vspace{10pt}


\begin{fminipage}{0.85\columnwidth}
  \begin{gather*}
    \text{(Majorana)}\quad \psi^a = \begin{pmatrix}
      \psi^\alpha \\[3pt] \bar\psi_{\dot\alpha}
    \end{pmatrix}\\[3pt]
    C = 
    \begin{pmatrix}
      \epsilon_{\alpha\beta} & 0\\
      0 & \bar\epsilon^{\dot\alpha\dot\beta}
    \end{pmatrix}\\[3pt]
    \bar\psi_a = \psi^b C_{ba} = (\psi_\alpha,
    -\bar\psi^{\dot\alpha})\\
    \bar\chi\psi = \chi_a \psi^a = - (\chi\psi + \bar\chi
    \bar\psi)\\[3pt]
    (\psi_\alpha)^* = \bar\psi_{\dot\alpha}\\
    (\psi^\alpha)^* = - \bar\psi^{\dot\alpha}
  \end{gather*}
\end{fminipage}

\vspace{10pt}

\begin{fminipage}{0.85\columnwidth}
  \begin{gather*}
    \gamma^\mu = 
    \begin{pmatrix}
      0 & -i \sigma^\mu\\
      i \bar\sigma^\mu & 0
    \end{pmatrix}\\
    \gamma_5 = 
    \begin{pmatrix}
      -i \1^\alpha{}_\beta & 0 \\
      0 & i \1_{\dot\alpha}{}^{\dot\beta}
    \end{pmatrix}\\
    \Sigma_{\mu\nu} = 
    \begin{pmatrix}
      \half \sigma_{\mu\nu} & 0 \\
      0 & \half \bar\sigma_{\mu\nu}
    \end{pmatrix}\\[3pt]
  \end{gather*}
\end{fminipage}

\vspace{10pt}


\begin{fminipage}{0.85\columnwidth}
  \begin{equation*}
    \begin{aligned}
      \varepsilon\theta &:= \varepsilon^\alpha \theta_\alpha\\
      \bar\varepsilon\bar\theta &:= \bar\varepsilon^{\dot\alpha}
      \bar\theta_{\dot\alpha}\\
      \varepsilon\sigma^\mu\bar\theta &:= \varepsilon_\alpha
      (\sigma^\mu)^{\alpha\dot\beta} \bar\theta_{\dot\beta}\\
      \bar\varepsilon\bar\sigma^\mu\theta &:=
      \bar\varepsilon^{\dot\alpha} (\bar\sigma^\mu)_{\dot\alpha\beta}
      \theta^\beta\\
      \varepsilon \sigma^{\mu\nu} \theta &:= \varepsilon_\alpha
      (\sigma^{\mu\nu})^\alpha{}_\beta \theta^\beta\\
      \bar\varepsilon \bar\sigma^{\mu\nu} \bar\theta &:=
      \bar\varepsilon^{\dot\alpha}
      (\bar\sigma^{\mu\nu})_{\dot\alpha}{}^{\dot\beta}
      \bar\theta_{\dot\beta}\\
      \theta\sigma^\mu\bar\sigma^\nu\varepsilon &:= \theta_\alpha
      (\sigma^\mu)^{\alpha\dot\alpha}
      (\bar\sigma^\nu)_{\dot\alpha\beta} \varepsilon^\beta\\
      \bar\theta\bar\sigma^\mu\sigma^\nu\bar\varepsilon &:=
      \bar\theta^{\dot\alpha} (\bar\sigma^\mu)_{\dot\alpha\alpha}
      (\sigma^\nu)^{\alpha\dot\beta} \bar\varepsilon_{\dot\beta}
    \end{aligned}
  \end{equation*}
\end{fminipage}

\vspace{10pt}


\begin{fminipage}{0.85\columnwidth}
\begin{gather*}
  \begin{aligned}
    \varepsilon\theta &= + \theta\varepsilon\\
    \bar\varepsilon\bar\theta &= + \bar\theta\bar\varepsilon\\
    \bar\varepsilon\bar\sigma^\mu\theta &= - \theta \sigma^\mu \bar
    \varepsilon\\
    \varepsilon \sigma^{\mu\nu} \theta &= - \theta
    \sigma^{\mu\nu}\varepsilon\\
    \bar\varepsilon \bar\sigma^{\mu\nu} \bar\theta &= - \bar\theta
    \bar\sigma^{\mu\nu} \bar\varepsilon\\
    \theta\sigma^\mu\bar\sigma^\nu\varepsilon &= - \eta^{\mu\nu}
    \theta\varepsilon + \theta \sigma^{\mu\nu}\varepsilon\\
    \bar\theta\bar\sigma^\mu\sigma^\nu\bar\varepsilon &= +
    \eta^{\mu\nu} \bar\theta\bar\varepsilon + \bar\theta
    \bar\sigma^{\mu\nu}\bar\varepsilon\\
  \end{aligned}\\
  \begin{aligned}
    (\varepsilon\theta)^* &= + \bar\theta\bar\varepsilon = +
    \bar\varepsilon\bar\theta\\
    (\varepsilon\sigma^\mu\bar\theta)^* &= - \bar\varepsilon
    \bar\sigma^\mu \theta = + \theta\sigma^\mu
    \bar\varepsilon\\
    \left(\varepsilon\sigma^{\mu\nu}\theta \right)^* &= + \bar\theta
    \bar\sigma^{\mu\nu}\bar\varepsilon = - \bar\varepsilon
    \bar\sigma^{\mu\nu} \bar\theta
  \end{aligned}
\end{gather*}
\end{fminipage}

\vspace{10pt}


\begin{fminipage}{0.85\columnwidth}
  \begin{equation*}
    \begin{aligned}
      \varepsilon_\alpha \theta_\beta &= - \half \varepsilon\theta\,
      \epsilon_{\alpha\beta} - \tfrac18 \varepsilon \sigma^{\mu\nu}
      \theta \, (\sigma_{\mu\nu})_{\alpha\beta}\\
      \bar\varepsilon_{\dot\alpha} \bar\theta_{\dot\beta} &= - \half
      \bar\varepsilon\bar\theta\, \bar\epsilon_{\dot\alpha\dot\beta} -
      \tfrac18 \bar \varepsilon \bar \sigma^{\mu\nu} \bar\theta \, (\bar
      \sigma_{\mu\nu})_{\dot\alpha\dot\beta}\\
      \varepsilon_\alpha \bar\theta_{\dot\beta} &= + \half
      \varepsilon\sigma^\mu \bar \theta\,
      (\bar\sigma_\mu)_{\dot\beta\alpha}
    \end{aligned}
  \end{equation*}
\end{fminipage}

\vspace{10pt}


\begin{fminipage}{0.85\columnwidth}
  \begin{equation*}
    \begin{aligned}
      \theta_\alpha \theta_\beta &= - \half \theta^2 \,
      \epsilon_{\alpha\beta}\\
      \bar\theta_{\dot\alpha} \bar\theta_{\dot\beta} &= - \half
      \bar\theta^2\, \bar\epsilon_{\dot\alpha\dot\beta}\\
      \theta_\alpha \bar\theta_{\dot\beta} &= - \half
      \bar\theta\bar\sigma^\mu \theta\,
      (\bar\sigma_\mu)_{\dot\beta\alpha}
    \end{aligned}
  \end{equation*}
\end{fminipage}

\vspace{10pt}


\begin{fminipage}{0.85\columnwidth}
  \begin{equation*}
    \begin{aligned}
      \theta\psi\, \theta\varepsilon &= -\half \theta^2 \psi\varepsilon\\
      \bar\theta\bar\psi\, \bar\theta\bar\varepsilon &= -\half
      \bar\theta^2 \bar\psi\bar\varepsilon\\
      \theta\psi\, \bar\theta\bar\varepsilon &= + \half
      \theta\sigma_\mu\bar\theta\, \bar\varepsilon\bar\sigma^\mu\psi\\      
      \theta\psi\, \theta\sigma^\mu\bar\xi &= - \half \theta^2 \,
      \psi\sigma^\mu\bar\xi\\
      \bar\theta\bar\psi\, \bar\theta\bar\sigma^\mu\xi &= - \half
      \bar\theta^2 \, \bar\psi\bar\sigma^\mu\xi\\
      \bar\theta\bar\sigma^\mu\theta\, \theta \sigma^\nu
      \bar\varepsilon &= + \half \theta^2 \bar\theta\bar\varepsilon
      \eta^{\mu\nu} + \half \theta^2 \bar\theta \bar\sigma^{\mu\nu}
      \bar\varepsilon\\
      \bar\theta\bar\sigma^\mu\theta\, \bar\theta \bar\sigma^\nu
      \varepsilon &= - \half \bar\theta^2 \theta\varepsilon
      \eta^{\mu\nu} + \half \bar\theta^2 \theta \sigma^{\mu\nu}
      \varepsilon
    \end{aligned}
  \end{equation*}
\end{fminipage}

\vspace{10pt}



\begin{fminipage}{0.85\columnwidth}
  \begin{align*}
    \left[\sM_{\mu\nu},\sP_\rho\right] = {} & + \eta_{\nu\rho} \sP_\mu -
    \eta_{\mu\rho} \sP_\nu\\
    \left[\sM_{\mu\nu},\sM_{\rho\sigma}\right] = {} & + \eta_{\nu\rho}
    \sM_{\mu\sigma} - \eta_{\mu\rho} \sM_{\nu\sigma}\\
    & {} - \eta_{\nu\sigma} \sM_{\mu\rho} +
    \eta_{\mu\sigma}\sM_{\nu\rho}\\[3pt]
    \left[\sP_\mu,\sD\right] = {} & + \sP_\mu\\
    \left[\sK_\mu,\sD\right] = {} & -\sK_\mu\\
    \left[\sP_\mu,\sK_\nu\right] = {} & + 2 \eta_{\mu\nu} \sD - 2
    \sM_{\mu\nu}\\
    \left[\sM_{\mu\nu}, \sK_\rho\right] = {} & + \eta_{\nu\rho} \sK_\mu -
    \eta_{\mu\rho} \sK_\nu\\[3pt]
    \left[\sM_{\mu\nu}, \sQ_a\right] = {} & -
    \left(\Sigma_{\mu\nu}\right)_a{}^b \sQ_b\\
    \left[\sQ_a, \sQ_b\right] = {} & + 2 \left(\gamma^\mu\right)_{ab}
    \sP_\mu\\[3pt]
    \left[\sK_\mu,\sQ_a\right] = {} & + (\gamma_\mu)_a{}^b \sS_b\\
    \left[\sM_{\mu\nu}, \sS_a\right] = {} & -
    \left(\Sigma_{\mu\nu}\right)_a{}^b \sS_b\\
    \left[\sP_\mu,\sS_a\right] = {} & -(\gamma_\mu)_a{}^b \sQ_b\\
    \left[\sS_a,\sS_b\right] = {} & - 2 (\gamma^\mu)_{ab} \sK_\mu\\
    \left[\sQ_a,\sS_b\right] = {} & + 2 C_{ab} \sD - 2 (\gamma_5)_{ab}
    \sR\\
    & {} + (\gamma^{\mu\nu})_{ab} \sM_{\mu\nu}\\
    \left[\sR, \sQ_a\right] = {} & + \half (\gamma_5)_a{}^b \sQ_b\\
    \left[\sR, \sS_a\right] = {} & - \half (\gamma_5)_a{}^b \sS_b\\
    \left[\sD, \sQ_a\right] = {} & - \half \sQ_a\\
    \left[\sD, \sS_a\right] = {} & + \half \sS_a
  \end{align*}
\end{fminipage}

\newpage


\begin{fminipage}{0.85\columnwidth}
  \begin{align*}
    \left[\sM_{\mu\nu}, \sQ_\alpha\right] = {} & - \half
    \left(\sigma_{\mu\nu}\right)_\alpha{}^\beta \sQ_\beta\\
    \left[\sM_{\mu\nu}, \bar\sQ_{\dot\alpha}\right] = {} & +
    \half \left(\sigma_{\mu\nu}\right)_{\dot\alpha}{}^{\dot\beta}
    \bar\sQ_{\dot\beta}\\
    \left[\sQ_\alpha, \bar\sQ_{\dot\beta}\right] = {} & + 2 i
    \left(\bar\sigma^\mu\right)_{\dot\beta\alpha}
    \sP_\mu
  \end{align*}
\end{fminipage}

\vspace{10pt}


\begin{fminipage}{0.85\columnwidth}
  \begin{gather*}
    e^X \, e^Y = e^Z\\
    \begin{aligned}
      Z = {} & X + Y + \half [X,Y]\\
      & {} + \tfrac1{12} [X,[X,Y]] + \cdots      
    \end{aligned}
  \end{gather*}
\end{fminipage}

\vspace{10pt}



\begin{fminipage}{0.85\columnwidth}
  \begin{equation*}
    \begin{aligned}
      \d_\alpha \theta^\beta &= \delta_\alpha{}^\beta\qquad
      \bar\d_{\dot\alpha} \bar\theta^{\dot\beta} &=
      \delta_{\dot\alpha}{}^{\dot\beta}\\
      \d_\alpha \theta_\beta &= \epsilon_{\alpha\beta}\qquad
      \bar\d_{\dot\alpha} \bar\theta_{\dot\beta} &=
      \bar\epsilon_{\dot\alpha\dot\beta}\\
      \d_\alpha \theta^2 &= 2\theta_\alpha\qquad
      \bar\d_{\dot\alpha} \bar\theta^2 &= 2\bar\theta_{\dot\alpha}\\
      \d^2 \theta^2 &= -4\qquad \bar\d^2 \bar\theta^2 &= -4
    \end{aligned}
  \end{equation*}
\end{fminipage}

\vspace{10pt}


\begin{fminipage}{0.85\columnwidth}
  \begin{equation*}
    \begin{aligned}
      Q_\alpha &:= \d_\alpha + i (\sigma^\mu)_{\alpha\dot\alpha}
      \bar\theta^{\dot\alpha} \d_\mu\\
      \bar Q_{\dot\alpha} &:= \bar\d_{\dot\alpha} + i
      (\bar\sigma^\mu)_{\dot\alpha\alpha} \theta^\alpha \d_\mu\\
      [Q_\alpha, \bar Q_{\dot\alpha}] &=
      + 2i(\bar\sigma^\mu)_{\dot\alpha\alpha} \d_\mu\\[3pt]
      D_\alpha &:= \d_\alpha - i (\sigma^\mu)_{\alpha\dot\alpha}
      \bar\theta^{\dot\alpha} \d_\mu\\
      \bar D_{\dot\alpha} &:= \bar\d_{\dot\alpha} - i
      (\bar\sigma^\mu)_{\dot\alpha\alpha} \theta^\alpha \d_\mu\\
      [D_\alpha, \bar D_{\dot\alpha}] &=
      -2i(\bar\sigma^\mu)_{\dot\alpha\alpha} \d_\mu
    \end{aligned}
  \end{equation*}
\end{fminipage}

\vspace{10pt}


\begin{fminipage}{0.85\columnwidth}
  \begin{gather*}
      U := \theta \sigma^\mu \bar \theta \d_\mu = -
      \bar\theta\bar\sigma^\mu \theta \d_\mu\\
    \begin{aligned}
      D_\alpha &= e^{iU} \d_\alpha e^{-iU}\\
      \bar D_{\dot\alpha} &= e^{-iU} \bar\d_{\dot\alpha} e^{iU}\\
      Q_\alpha &= e^{-iU} \d_\alpha e^{iU}\\
      \bar Q_{\dot\alpha} &= e^{iU} \bar\d_{\dot\alpha} e^{-iU}
    \end{aligned}
  \end{gather*}
\end{fminipage}

\newpage



\begin{fminipage}{0.85\columnwidth}
  \begin{gather*}
    \bar D_{\dot\alpha} \Phi = 0\\
    \Phi = e^{-iU} \left[\phi + \theta\chi + \theta^2 F\right]\\
    \begin{aligned}
      \Phi = {} & \phi + \theta\chi + \theta^2 F + i
      \bar\theta\bar\sigma^\mu\theta \d_\mu\phi\\
      & {} - \tfrac{i}{2} \theta^2 \bar\theta \bar\sigma^\mu
      \d_\mu\chi + \tfrac14 \theta^2 \bar\theta^2 \dalem\phi
    \end{aligned}
  \end{gather*}
\end{fminipage}

\vspace{10pt}


\begin{fminipage}{0.85\columnwidth}
  \begin{gather*}
    \bar V = V\\
    e^V \mapsto e^{-\bar\Lambda} e^V e^{-\Lambda}\\
    \bar D_{\dot\alpha} \Lambda = 0 \quad 
    D_\alpha \bar\Lambda = 0
  \end{gather*}
  In WZ gauge:
  \begin{gather*}
    V = \bar\theta\bar\sigma^\mu\theta v_\mu + \bar\theta^2
    \theta\lambda +  \theta^2 \bar\theta\bar\lambda +
    \theta^2\bar\theta^2 D\\
    \begin{aligned}
      W_\alpha & := -\frac{1}{8g} \bar D^2 e^{-2g\, V} D_\alpha e^{2g\,V}\\
      \overline W_{\dot\alpha} & := -\frac{1}{8g} D^2 e^{-2g\,V} \bar
      D_{\dot\alpha} e^{2g\,V}
    \end{aligned}\\
    \bar D_{\dot\alpha} W_\alpha = 0 \qquad D_\alpha \overline
    W_{\dot\alpha} = 0\\
    D^\alpha W_\alpha = \bar D^{\dot\alpha} \overline
    W_{\dot\alpha}
\end{gather*}
\end{fminipage}

\vspace{10pt}


\begin{fminipage}{0.85\columnwidth}
  \begin{gather*}
    \begin{aligned}
      \eL = {} & \int d^2 \theta d^2 \bar\theta\, \left( \bar\bPhi e^{2g\,V}
      \bPhi +  \Tr \mu V \right)\\
      & {} + \left[ \int d^2\theta\,  \Tr \tfrac14 W^\alpha W_\alpha +
        \text{c.c.}\right]\\
      & {} + \left[ \int d^2\theta \, W(\bPhi) + \text{c.c.}\right]
    \end{aligned}\\
    \begin{aligned}
      W(\bPhi) ={} & a_I \Phi^I + \half m_{IJ} \Phi^I \Phi^J\\
      & {} + \tfrac13 \lambda_{IJK} \Phi^I \Phi^J \Phi^K
  \end{aligned}
\end{gather*}
\end{fminipage}


%
%
%
\end{document}